\documentclass[nofootinbib,amsmath,prd,aps,superscriptaddress,tightenlines,12pt]{revtex4}
\usepackage[hypertex]{hyperref}
\usepackage{graphicx}
\usepackage{epstopdf}
\usepackage{bm}
\usepackage{epsfig}
\usepackage{graphics}
\usepackage{xspace}
\usepackage{pstricks}

\def\Dslash{D\!\!\!\!\slash}

\def\OMIT#1{}

\newcommand{\nn}{\nonumber}

\newcommand{\bea}{\begin{eqnarray}}
\newcommand{\eea}{\end{eqnarray}}

\newcommand{\be}{\begin{equation}}
\newcommand{\ee}{\end{equation}}

\begin{document}
\setlength\baselineskip{17pt}


\preprint{ \vbox{  \hbox{CALT-68-}}}

\title{\bf Implications of a Scalar Dark Force for Terrestrial Experiments}

\author{Sean M. Carroll}
\email[]{seancarroll@gmail.com}
\affiliation{California Institute of Technology, Pasadena, CA 91125}
\author{Sonny Mantry}
\email[]{mantry@wisc.edu}

\affiliation{University of Wisconsin-Madison, Madison, WI 53706}
\author{Michael J. Ramsey-Musolf}
\email[]{mjrm@physics.wisc.edu}
\affiliation{University of Wisconsin-Madison, Madison, WI 53706}
\affiliation{California Institute of Technology, Pasadena, CA 91125}


\begin{abstract}
  \vspace*{0.3cm}
A long range Weak Equivalence Principle  (WEP) violating force between Dark Matter   (DM) particles, mediated by an ultralight scalar, is tightly constrained by galactic dynamics and large scale structure formation. We examine the implications of such a \lq\lq dark force" for  several terrestrial experiments, including  E\"otv\"os tests of the WEP, direct-detection DM searches, and collider studies.  The presence of a dark force implies a non-vanishing effect in E\"otv\"os tests that could be probed by current and future experiments depending on the DM model. For scalar singlet DM scenarios, a dark force of astrophysically relevant magnitude is ruled out  in large regions of parameter space by the DM relic density and WEP constraints. WEP tests also imply constraints on the Higgs-exchange contributions to the spin-independent (SI) DM-nucleus direct detection cross-section. For WIMP scenarios, these considerations constrain Higgs-exchange contributions to the SI cross-section to be subleading compared to gauge-boson mediated contributions. In multicomponent DM scenarios, a dark force would preclude large shifts in the rate for Higgs decay to two photons associated with DM-multiplet loops that might otherwise lead to measurable deviations at the LHC or a future linear collider. The
combination of observations from galactic dynamics, large scale structure formation, E\"otv\"os experiments, DM-direct-detection experiments, and colliders can  further constrain the size of new long range forces in the dark sector.


\OMIT{
A long range Weak Equivalence Principle  (WEP) violating force between Dark Matter   (DM) particles, mediated by an ultralight scalar, is tightly constrained by galactic dynamics and large scale structure formation. 
We examine the implications of such a \textit{dark force} for  several terrestrial experiments. Depending on the DM model, the presence of a dark force implies constraints for E\"otv\"os experiments, DM-nucleus cross-sections, and collider signatures.
We illustrate this with minimal gauge singlet and WIMP DM models. For scalar singlet and real triplet DM coupled to a dark force, there are constraints on the Higgs-DM interactions from WEP tests.  For scalar singlet DM, tight constraints on the Higgs-DM interactions lead to a suppression on of the DM annihilation rate leading to a closed universe ruling out a dark force in large regions of parameter space. The DM-Higgs-interaction constraints also lead to upper bounds on DM-nucleus cross-sections for singlet and real triplet scalar DM. In most cases these upper bounds are below current sensitivities in direct detection.
In multicomponent DM scenarios, involving light WIMPs with a non-zero coupling to the Higgs and a dark force, there are testable constraints on Higgs decay to two photons at colliders.   Such analyses can be generalized to other DM models where  new types of constraints in terrestrial experiments might emerge.  The
synergy of observations from galactic dynamics, large scale structure formation, E\"otv\"os experiments, DM-direct-detection experiments, and colliders can be a powerful probe of dark forces.    If the ultralight scalar plays the role of dark energy, such analyses can complement existing constraints on interactions between dark matter and dark energy.}

\end{abstract}

\maketitle

\newpage
\tableofcontents

\section{Introduction}

There is now compelling evidence for the $\Lambda$CDM model or the `Standard Model' of cosmology according to which the energy of the universe is about 74\% dark energy, 22\% Dark Matter  (DM), and 4\% baryonic matter. There have been independent confirmations of the dark energy component of the universe from observations of high redshift Type Ia supernovae \cite{Perlmutter:1998np, Riess:1998cb, Schmidt:1998ys, Garnavich:1998th, Knop:2003iy}. The evidence for DM is even more compelling from the study of galactic rotation curves~\cite{Faber:1979pp, Bosma:1981zz, Rubin:1985ze}, acoustic oscillations in the cosmic microwave background~\cite{Hu:1994uz,Hu:1994jd, Jungman:1995bz,Zaldarriaga:1997ch}, large scale structure formation~\cite{Eisenstein:1997ik, Eisenstein:2005su}, and gravitational lensing~\cite{Clowe:2006eq, Zhang:2007nk}. In spite of such strong  evidence for the existence of dark energy and DM, almost nothing is known about their properties. The simplest explanation of dark energy is a {small} but non-zero cosmological constant. The DM properties such as its mass, quantum numbers, and interactions with the Standard Model  (SM) remain unknown. Furthermore, it remains to be seen if there is only one type of DM particle responsible for all of the observational evidence, or if  there exists a rich spectrum of DM particles analogous to the complexity seen in the visible sector.
Many experiments are underway to detect DM and determine its properties. Ground based direct detection experiments~\cite{Angle:2007uj,Ahmed:2008eu} put limits on the DM mass and the strength of its interaction with baryonic matter from observations of recoiling nuclei. Experiments~\cite{Adriani:2008zr, Barwick:1997ig,Beatty:2004cy,Aguilar:2007yf} studying cosmic rays from the galactic halo 
have recently seen indications of an electron/positron excess, which could be interpreted as evidence for DM annihilation, and can constrain the DM mass and interactions. There has also been a recent proposal to observe a possible DM magnetic moment via the gyromagnetic Faraday effect~\cite{Gardner:2006za, Gardner:2008yn}.

Another set of experiments are devoted to question of whether DM violates the Weak Equivalence Principle (WEP) and are the focus of this paper. There exist a variety of scenarios for new interactions confined solely to the dark sector and the possibility that they might lead to WEP violation. The possibility of  gauge or Yukawa  forces confined to the dark sector have been studied in other contexts \cite{DeRujula:1989fe, Feng:2008mu, ArkaniHamed:2008qn, Ackerman:2008gi, Baumgart:2009tn, Aguirre:2001xs, Zurek:2008qg, Bai:2009it}. Short range WEP tests have  been studied in \cite{Moody:1984ba,Hoyle:2004cw,Kaplan:2000hh} for example. In this work, we focus  on a long range dark force, mediated by an ultralight scalar, and study its implications for terrestrial experiments. For this scenario, the dark force  can be communicated to ordinary matter via virtual DM loops that connect the  scalar with ordinary matter, as long as the DM candidate is not sterile. This mechanism will give rise to effects in terrestrial experiments. We investigate the resulting impact that WEP violation in the dark sector may have on  DM detection experiments, laboratory based WEP tests, or even studies of Higgs boson properties at colliders. Constraints on WEP violation in ordinary matter induced by WEP violation in the dark sector were recently studied in \cite{Bovy:2008gh, Carroll:2008ub}. In addition, a connection between direct DM detection experiments and WEP violation was shown in \cite{Bovy:2008gh}.  

Many  models that contain the interaction of an ultralight scalar with DM~\cite{Damour:1990tw,Friedman:1991dj, Gradwohl:1992ue,  Anderson:1997un, Carroll:1998zi, Amendola:2001rc, Farrar:2003uw, Gubser:2004du, Gubser:2004uh, Bertolami:2004nh, Nusser:2004qu, Bean:2007ny} have been proposed to explain features in the DM distribution and explore the possibility of DM-quintessence interactions. More recently, work with non-universal scalar-tensor theories of gravity with the Abnormally Weighting Energy (AWE) Hypothesis~\cite{Alimi:2008ee, Fuzfa:2007sv} also invoke couplings of an ultratight scalar to the dark sector as a way of explaining the observed cosmic acceleration even in the absence of a dark energy fluid. Constraints on such scenarios from big bang nucleosynthesis have also been studied~\cite{Coc:2008yu}. There are several other observational motivations,  including higher than predicted supercluster densities \cite{Einasto:2006si} and voids~\cite{Farrar:2003uw, Nusser:2004qu} ( for a summary see \cite{Kesden:2006vz,Bovy:2008gh}). The existence of a long-range attractive Yukawa force between DM particles would accelerate structure formation and could help explain some of these observations. Strong constraints on such a {dark}  force are derived from observations of DM dynamics in the tidal stream  of the Sagittarius dwarf galaxy~\cite{Kesden:2006vz,Kesden:2006zb}, which indicate a force with strength less that 20\% of gravity. However, new observational systematic errors have been recently discovered~\cite{Chou:2006ia} that could require a revision of this result, perhaps allowing for a stronger dark force. A more recent analysis~\cite{Bean:2008ac} considers the effect of a dark force on the evolution of density perturbations and the resulting impact on the CMB spectrum. This analysis constrains the strength of a dark force to be less than 5\% of gravity.

From a purely theoretical perspective, the existence of an ultralight scalar $\phi $ with mass $m_\phi < 10^{-25}$ eV, able to mediate a long-range force over scales of interest to galactic dynamics, would introduce a new hierarchy in addition to that between the weak scale $m_W \sim 100$ GeV and the Planck scale $M_P\sim 10^{19}$ GeV.  However, as we still await experimental evidence for a mechanism to explain  the the hierarchy between the weak and Planck scales, and in light of the discovery of an unnaturally small cosmological constant, we keep an open mind and do not attempt to provide an explanation for  the ultralight scalar mass. We assume the existence of a finely-tuned ultralight scalar mediating a long range force dark force and study its consequences for terrestrial experiments.  

In what follows, we amplify on our earlier work\cite{Carroll:2008ub} and that of Ref.~\cite{Bovy:2008gh}, using simple DM scenarios to illustrate the prospective implications of dark sector WEP violation for terrestrial experiments. We study three representative minimal DM scenarios to explore the range of possible implications: scalar DM that is a singlet with respect to SM gauge interactions; scalar DM that is the neutral component of a real SU(2$)_L$ triplet; and fermionic DM that lives in a vector-like representation of SU(2$)_L$. Our main conclusions are:
\begin{itemize}
\item[(i)] The presence of a dark force  implies a non-zero effect in E\"otv\"os experiments if the DM interacts with Standard Model (SM) fields. For scalar singlet DM, this effect arises from loop-induced mixing between the ultralight scalar and the SM Higgs, while for representative WIMP scenarios (scalar or fermionic) additional contributions arise from loop-induced higher-dimension operators that couple the ultralight scalar directly to matter.  We derive order-of-magnitude expectations for the minimum size of this effect for these representative scenarios as illustrated in Fig.~\ref{fermionicWIMPetabeta}  for WIMP DM and Table. \ref{relic-table} for scalar singlet DM.  For a dark force with strength roughly 20\% of gravity, one could expect a non-vanishing effect, for non-minimal WIMP DM models and in certain regions of parameter space of scalar singlet DM models, within reach of future approved E\"otv\"os experiments such as Microscope ~\cite{Lammerzahl:2001qr} able to detect anomalous accelerations to a sensitivity of $\Delta a/a \sim 10^{-15}$. The  MiniSTEP experiment\cite{Lockerbie:1998ar} with an increased sensitivity of $\Delta a/a \sim 10^{-18}$ , currently under study by NASA and the ESA, could see non-vanishing effects in minimal WIMP models which can induce a WEP violating starting at two loops.

\item[(ii)]  For scalar singlet DM, a dark force of astrophysical relevance, is already ruled out in large regions of parameter space. The bounds from E\"otv\"os experiments constrain the size of DM-Higgs interaction
which determines the relic density.
In large regions of parameter space, the bound on DM-Higgs interactions implies a suppression in the DM annihilation rate resulting in a relic density that over-closes the universe. As a result, relic density considerations in scalar singlet DM models can yield the strongest bounds on the size of a dark force.

\item[(iii)] The constraints on DM-Higgs interactions lead to upper bounds on the magnitude of Higgs-exchange contributions to DM-nucleus cross sections.  These bounds depend on the Higgs mass, implying that a combination of direct detection experiments and Higgs boson discovery could be used to test the simplest scenario for a dark force in some DM scenarios. In particular,  Higgs-exchange contributions dominate the SI scalar singlet DM-nucleus cross section, so that dark force considerations -- together with the observed DM relic density --  imply constraints on the entire cross section. In contrast, WIMP-nucleus cross sections receive contributions from electroweak gauge boson-exchange that are not constrained by the presence of a dark force. As we show below, dark force considerations and present limits from E\"otv\"os experiments imply that the Higgs exchange contributions are sub-leading compared to those from gauge boson-exchange. The corresponding bounds for the scalar singlet and real triplet models are illustrated in Figs.~\ref{singletWIMPetaE}. Tests of the WEP can only constrain the full DM-nucleus cross section  if the DM particles are singlets with respect to the SM gauge symmetries  (see, {\em e.g.}, \cite{Barger:2007im,He:2008qm,Barger:2008jx} and references therein) so that elastic scattering proceeds only via Higgs exchange (at least at tree level). 
\item[(iv)] In multi-component WIMP DM scenarios, where one of the  light ($\lesssim 200$ GeV) DM components has a non-zero coupling to the Higgs, the presence of a dark force -- together with tests of the WEP -- imply testable upper bounds on one-loop WIMP-induced shifts in the branching ratio for the SM Higgs to decay to two photons. These bounds  generally lie well below the prospective sensitivities of LHC studies of Br$(H\to\gamma\gamma)$ as seen in Fig.~\ref{a2shiftrate}. The observation of a significant shift in this branching ratio would likely preclude this scenario for a dark force.
\item[(v)] The existence of an observable long range dark force which requires $m_\phi < 10^{-25}$ eV, implies restrictions in the space of finite renormalized parameters in addition to the usual fine tuning of radiative corrections that are sensitive to the cutoff. We discuss these regions in parameter space and their implications for the observation of a dark force.
\end{itemize}

In arriving at these conclusions, we emphasize we have drawn upon representative cases rather than carrying out a comprehensive study. We expect that our conclusions will generalize to other DM scenarios, but do not preclude the possibility of exceptions in some cases. We also note that our analysis and conclusions differ  from those of Ref.~\cite{Bovy:2008gh}, who first observed that bounds on WEP and the presence of an astrophysically relevant dark force could imply constraints on DM-nucleus cross sections. The bounds obtained in that work lie well below the reach of future direct detection experiments. In what follows, we argue that an effective operator analysis consistent with the fine-tuning needed to maintain a vanishingly small scalar mass implies considerably weaker bounds  than given in Ref.~\cite{Bovy:2008gh}.  

The outline of the paper is as follows. In section \ref{section2} we review the phenomenology of experimental WEP tests and establish notation. In section \ref{eft} we review the derivation of the ultralight scalar coupling to macroscopic objects in terms of its couplings to the Standard Model  (SM) particles. In section \ref{section4} we discuss in a model independent way the mechanisms by which the ultralight scalar can couple to the SM. In sections \ref{section5} and \ref{sectionVI} we examine the experimental consequences of a dark force for various minimal DM models. In section
\ref{sectionVII} we discuss the regions in parameter space where an observable dark force is possible and how they relate to our analysis. We conclude in section \ref{section8}.

\section{Fifth-Force Phenomenology}
\label{section2}

We begin by considering the force between two bodies mediated by
a scalar field $\phi$ with mass $m_\phi$.  In the non-relativistic limit, the Yukawa potential
between a test body $i$ and a source $s$ separated by a distance $r$ is given 
 (in units where $\hbar=c=1$) by
\bea
  V_\phi = -\xi_i \xi_s \frac{Q_i Q_s}{4\pi r}e^{-m_\phi r}\,,
\eea
where $Q_{i,s}$ denote the charges of the test and source objects under the force mediated by $\phi$. The parameters $\xi_{i,s}$ are\footnote{The t-channel $\phi$ exchange amplitude is accompanied by an extra factor of $2m_{i,s}$ for fermions relative to scalars. This is due to the  fermionic spinor normalization $\bar{u}_{i,s} u_{i,s}=2m_{i,s}$ in the non-relativistic limit. These factors are absorbed by switching to states with normalization $\langle \textbf{p} | \textbf{q} \rangle =  (2\pi)^3 \delta^{ (3)} (\textbf{p} - \textbf{q}) $ in order to compare with the non-relativistic Born amplitude. For  scalars we are then left  with an additional factor of $\frac{1}{2m_{i,s}}$ in the potential relative to fermions. }
\bea
\label{xi}
\xi_{i,s} = \begin{cases} 
1&\text{for fermionic objects},\\ 
\frac{1}{2m_{i,s}}&\text{for scalar objects. } \end{cases}
\eea
Note that  the charges $Q_{i,s}$  are of mass  dimension one and zero for scalar and fermionic objects respectively, so that the equation is dimensionally consistent. These mass dimensions will become apparent when we study  specific models. 
The Newtonian gravitational potential between a body with mass $M_i$
and a source with mass $M_s$ is 
\be
  V_G = -\frac{GM_iM_s}{r}\,,
\ee
where $G$ is Newton's constant.  It is therefore convenient to write the total potential as
\be
  V = -\frac{GM_iM_s}{r}\left (1 + \alpha_{is}e^{-m_\phi r}\right)\,,
  \label{totalpotential}
\ee
where
\be
  \alpha_{is} = \frac{1}{4\pi G}\frac{q_iq_s}{\mu_i\mu_s} \hat{\xi}_i \hat{\xi} _s,
  \label{alpha}
\ee
is a dimensionless parameter characterizing the strength of the new force relative to
gravity, expressed in terms of the charge-to-mass ratio $q/\mu = Q/M$, where $\mu$
is the mass in atomic mass units. The parameters $\hat{\xi}_{i,s}$ are

\bea
\hat{\xi}_{i,s} = \begin{cases} 
1&\text{for fermionic objects},\\ 
\frac{1}{2\mu_{i,s}}&\text{for scalar objects. } \end{cases}
\eea
The parameter $\alpha_{is}$ is not  universal  and in general
depend on the composition of the macroscopic bodies acting as sources for $\phi$.

E\"otv\"os experiments look for violations of the equivalence principle by measuring the difference in acceleration of two test bodies of different compositions in the presence of a common source. Experimental constraints on new long-range composition-dependent forces are
typically expressed in terms of the E\"otv\"os parameter,
\be
  \eta = 2\frac{|a_1 - a_2|}{|a_1+a_2|}\simeq \Big | \frac{\Delta a}{a}\Big |\,,
  \label{eotvos1}
\ee
where $a_i$ is the total acceleration of object $i=1,2$, $\Delta a\equiv a_1-a_2$, and $a$ is the universal gravitational acceleration in the absence of any new long range forces. The last approximation made above is valid when the fifth force is weaker than gravity. From  (\ref{totalpotential}),
the acceleration of object $i$ due to the source $s$ is
\be
  a_i = \frac{GM_s}{r^2}\left[1+\alpha_{is} (1+m_\phi r)e^{-m_\phi r}\right]\,.
\ee
We are interested in forces that are considerably weaker than gravity, and
distances less than the Compton wavelength of the scalar, $r \ll m_\phi^{-1}$.
The E\"otv\"os parameter is then
\be
  \eta _s^{1,2}= \frac{1}{4\pi G}\left|\frac{q_1 \hat{\xi}_1}{\mu_1} - \frac{q_2 \hat{\xi}_2}{\mu_2}\right| \>\left|\frac{q_s \hat{\xi}_s}{\mu_s}\right|\,.
  \label{eotvos}
\ee
Currently, the strongest limits on violations of the weak equivalence  come from
torsion balance E\"otv\"os experiments~\cite{Schlamminger:2007ht} which give the constraints 
\bea
\label{bounds}
 \eta _{_\text{E}}^{{\text{Be,Ti}}} <  (0.3 \pm 1.8) \times 10^{-13}, \qquad  \eta _{_\text{DM}}^{{\text{Be,Ti}}} <  (4\pm 7) \times 10^{-5}.
\eea
The E\"otv\"os parameters $\eta _{_\text{E}}^{{\text{Be,Ti}}} $ and $\eta _{_\text{DM}}^{{\text{Be,Ti}}}$ measure differential acceleration of laboratory test samples of Beryllium and Titanium with the Earth and galactic dark matter as the source bodies respectively.

Future experiments, currently being studied, are expected to further improve the the bound on the E\"otv\"os parameter by several orders of magnitude as shown in Table.~\ref{eotvos-future}
The MiniSTEP experiment \cite{Lockerbie:1998ar}, currently under study, would use test objects of different composition orbiting earth in free fall and new technology to reduce thermal noise. If approved, this experiment is expected to achieve the highest sensitivity of $\eta \sim 10^{-18}$. The Microscope experiment, which has been approved, uses the same principle but is expected to reach a sensitivity of $\eta \sim  10^{-15}$.
In the method of Lunar Laser Ranging (LLR) used by the APOLLO collaboration\cite{Williams:2003wu}, the differential acceleration of the Earth and Moon is measured in the presence of a source like the Sun or galatic dark matter. The APOLLO collaboration, which is currently underway, is expected to 
achieve a sensitivity of $\eta \sim 10^{-14}$ improving the current bound on $\eta$ by an order of magnitude.
Methods using atom interferometry \cite{Dimopoulos:2006nk} could reach a sensitivity of $\eta \sim 10^{-17}$.

These experiments are also sensitive to WEP violation in the dark sector, if the DM has interactions with the SM. Through quantum effects involving virtual DM, WEP violation in the dark sector will be communicated to ordinary matter  and these effects can be tested in E\"otv\"os experiments. WEP violation in the dark sector is already constrained from an analysis of the tidal disruption in satellite galaxies \cite{Kesden:2006vz}. This study constrains the coupling of $\phi$ to DM particles by putting bounds on the parameter $\beta$
\bea
\label{eq:beta}
\beta = \frac{M_P}{\sqrt{4 \pi}} \frac{|g_\chi|}{M_\chi} \xi_\chi ,
\eea 
where $g_\chi$ denotes the DM charge under the fifth force, $M_\chi$ denotes the DM mass, $M_P =1/\sqrt{G}$ is the Planck mass, and $\xi_{\chi}$ is as defined in Eq.~(\ref{xi}).
The coupling $g_\chi$ appears in the Lagrangian via interaction terms for fermionic\footnote{For simplicity we assume that the fermionic DM is in a vector-like gauge representation so that $\bar{\chi}\chi\phi$ is gauge invariant. In the more general case the coupling of $\phi$ to fermionic DM may arise from higher dimension operators.} and scalar DM of the form
\bea
\label{eq:chicoup}
\delta {\cal L} =  \begin{cases} 
g_\chi \bar{\chi}  \chi \phi  ,&\text{fermionic DM},\\ 
g_\chi \chi^\dagger \chi \phi,&\text{scalar DM,} \end{cases}
\eea
\begin{table}
\begin{tabular}{|c | c |}
\hline 
Experiment \qquad\qquad & \qquad Expected Future Sensitivity in $\eta$ \\
\hline 
MiniSTEP\cite{Lockerbie:1998ar} & $10^{-18}$ \\
Microscope\cite{Lammerzahl:2001qr} & $10^{-15}$ \\
Apollo (LLR)\cite{Williams:2003wu} & $10^{-14}$ \\
\hline
\end{tabular}
\caption{Expected sensitivities for the E\"otv\"os parameter in future experiments testing the WEP. The MiniSTEP experiment is currently under study by NASA and the ESA. Microscope has been approved and the Apollo(LLR) experiment is underway. }
\label{eotvos-future}
\end{table}
Thus, we see that for fermionic DM, $g_\chi$ is dimensionless and for scalar DM it has dimension one. From the analysis of tidal streams in the Saggiatrius galaxy, Kamionkowski and Kesden\cite{Kesden:2006vz} obtained the approximate upper bound of
\bea
\beta \lesssim 0.2.
\eea
Newly discovered systematic errors~\cite{Chou:2006ia} could lead to a revision of this bound and more recently, the work of \cite{Peebles:2009th} showed the possibility of $\beta \sim 1$ consistent with observations of galactic dynamics. A more recent analysis~\cite{Bean:2008ac} of the CMB and large scale structure formation gives a tighter bound of $\beta < 0.05$.  In this paper we use  $\beta = 0.2$ as a reference value for most discussions, and our results be straightforwardly translated to other values of $\beta$.


\section{Light Scalar Coupling to Macroscopic Objects}\label{eft}

The charge to mass ratio under a fifth force for an elementary particle is straightforward to obtain in terms of the Lagrangian parameters. For example, the charge to mass ratio for elementary fermionic or scalar DM $\chi$ is given by
\bea
\Big  (\frac{q}{\mu}\Big)_\chi = \frac{g_\chi}{M_\chi}.
\eea
This charge to mass ratio is obtained by computing the tree level $\phi$ exchange diagram between two DM particles and taking the non-relativistic limit to compare with Eq.~(\ref{totalpotential}).

For composite materials the calculation of the charge to mass ratio is more complicated, as one has to take into 
account hadronic, nuclear, and atomic matrix elements of various operators containing SM fields that couple to $\phi$ as well as the effects of binding energy. In particular, one needs the charge to mass ratio for the various types of atoms that make up the laboratory test materials.  We compute these ratios using an effective field theory valid near the nucleon mass scale that involves the light  quarks $q=\{u,d,s\}$, gluons, the charged leptons $\ell =\{ e, \mu \}$, the photon, and the light scalar $\phi $. All other heavier degrees of freedom have been integrated out. The interaction terms in this effective Lagrangian take the form:
\bea
\label{eftlag}
{\cal L}_{\phi} = \sum_{q}\frac{g_q}{m_p} m_q\> \bar{q}q \phi + \sum_{\ell }\frac{g_\ell}{m_p} m_\ell \>\bar{\ell}\ell \phi + c_g\> \phi \> G_{\mu \nu}^a G^{\mu \nu}_a +c_\gamma \> \phi F_{\mu \nu} F^{\mu \nu}.
\eea
 As we discuss below, the effects of the $\phi$ coupling to heavy quarks, the tau lepton, massive gauge bosons, and $\chi$  that have been integrated out  are encoded in the operator coefficients $g_{q,\ell}$ and $c_{g,\gamma}$. We assume that the couplings of $\phi$ to the SM fermions are linearly proportional to the fermion mass. This will make the analysis simpler, as we will see, by allowing us to exploit the scale invariance of the energy momentum tensor. This assumption is realized in several types of DM models. The couplings $c_g$ and $c_\gamma$ can be straightforwardly computed in any given  model. To illustrate, consider a model in which $\phi$ couples to the SM fermions at the electroweak scale as
\bea
\label{phi-fermions}
{\cal L}_{\phi f{\bar f}} = \sum _q \frac{g_q}{m_p} m_q\> \bar{q}q \phi +  \sum_{\ell }\frac{g_\ell}{m_p} m_\ell \>\bar{\ell}\ell \phi + \frac{g_\tau}{m_p} m_\tau \>\bar{\tau}\tau \phi+  \sum _Q \frac{g_Q}{m_p} m_Q\> \bar{Q}Q \phi,
\eea
where the sum over $Q$ denotes a sum over the heavy $b,c,t$ quarks and all the couplings $g_{\ell, \tau, q,Q}$ above are independent of the SM fermion masses. One can then integrate out the heavy quarks to obtain~\cite{Shifman:1978zn} the renormlaization group invariant relation 
\bea
\label{integrateQ}
m_Q \bar{Q}Q = -\frac{\alpha_s}{12 \pi} G_{\mu \nu}^a G^{\mu \nu}_a  -\frac{\alpha }{16 \pi}  F_{\mu \nu} F^{\mu \nu}, \qquad m_\tau \bar{\tau}\tau = -\frac{\alpha }{16 \pi}  F_{\mu \nu} F^{\mu \nu}
\eea
to leading order in the heavy quark expansion and perturbation theory. Note that the RHS above is independent of the heavy quark mass. In this case the couplings $c_g$ and $c_\gamma$
in Eq.~(\ref{eftlag}) are given by 
\bea
\label{cg-cgamma}
c_g = -\frac{1}{m_p}  ( \sum_Q g_Q ) \frac{\alpha_s}{12\pi}, \qquad c_\gamma = -\frac{1}{m_p} (\sum _Q g_Q + g_\tau)  \frac{\alpha}{48\pi}, 
\eea
at leading order. The mass operators on the LHS of the equations in Eq.~(\ref{integrateQ}) appear in the QCD$+$QED energy momentum tensor and are scale invariant, allowing us to evaluate $\alpha_s$ and $\alpha$ in Eq.~(\ref{cg-cgamma}) at the low energy scale of the effective theory (when taking the nucleon matrix element). Because they do not run below the electroweak scale,   couplings $g_{Q,\tau}$  are evaluated at that scale.

We now evaluate the coupling of $\phi$ to an atom~\cite{Kaplan:2000hh}  of type `A'. Doing this allows us to determine the charge to mass ratio $q_A \hat{\xi}_A/\mu_A$ needed for  E\"otv\"os parameters, as seen from Eq.(\ref{eotvos}), if the test or source bodies are made up of atoms of type  `A'. We  define the effective atomic coupling as
\bea
\label{atomic-coup}
{\cal L}_{AA\phi} = \begin{cases} g_{_A} \bar{A}A\phi,&\text{fermionic atoms}, \\
g_{A} A^\dagger A \phi, &\text{scalar atoms},\end{cases}
\eea
where the $A$ is the  field that destroys the atomic state and again $g_A$ is dimensionless for a spin 1/2 atom and has dimension one for a spin zero atom.
We  determine $g_A$ by a  matching calculation
\bea
\label{atomic-eff}
\langle A| {\cal L}_{AA\phi} | A \phi \rangle = g_{_A} \xi_A = \langle A | {\cal L}_{\phi} | A \phi \rangle,
\eea
where we have used a non-relativistic normalization for the atomic states $\langle A  (p) | A  (q) \rangle =  (2\pi)^3 \delta ^3  (\vec{p} - \vec{q}\>)$, $\xi_A$ is the normalization factor defined in Eq.~(\ref{xi}), and ${\cal L}_\phi$ is defined in \eqref{eftlag}. From Eq.(\ref{atomic-eff}), as explained in Appendix \ref{appexA}, the general expression for the charge to mass ratio $q_A \hat{\xi}_A/\mu_A  $ is
\bea
\label{cmr-atomic}
\hat{\xi}_A \Bigg  ( \frac{q}{\mu} \Bigg )_A = \frac{g_A \xi_A}{M_A} = \frac{2 c_g g_3}{\beta_3}+\frac{1}{M_A}\Big [ Z   (\zeta_e m_e + \sum_q \zeta_q m_q \> x_{q,p}) +  (A-Z) \sum_q \zeta_q m_q\>  x_{q,n} + \omega _A \Big ], \nn \\
\eea
where the quantity $\omega_A$ is given by
\bea
\label{omegaA}
\omega_A \equiv \kappa \langle A | F^{\mu \nu}F_{\mu \nu} | A \rangle  -\sum_k \zeta_k  m_k \frac{d{\cal E}_A}{dm_k} , \ \ \ 
\eea
${\cal E}_A$ is the atomic binding energy as defined in Eq.(\ref{binding-mass}), the quantities $\zeta_k$ and $\kappa$ are given by
\bea
\label{zetakappa}
\zeta_k = \frac{g_k}{m_p} - \frac{2 g_3}{\beta_3}c_g , \qquad \kappa = c_\gamma -   \frac{g_3 \beta_e}{e \beta_3}c_g,
\eea
as in Eq.(\ref{zetakappa-appex}), and $x_{q,p}$ and $x_{q,n}$ denote the nucleon matrix elements 
\bea
x_{q,p} = \langle p | \bar{q}q | p \rangle , \qquad x_{q,n} = \langle n | \bar{q}q | n \rangle ,
\eea
which are known experimentally~\cite{Belanger:2008sj, Jungman:1995df} and given in Eq.(\ref{nuclearmelt})
of appendix~\ref{appexA}.  In Eq.(\ref{zetakappa}), $\beta_3$ and $\beta_e$ denote the QCD and QED beta functions respectively.

Using Eq.(\ref{cmr-atomic}) in Eq.(\ref{eotvos}) for  test objects made up of atoms with atomic weights $A_1$ and $A_2$, the general expression for the E\"otv\"os parameter $\eta _{_S}$ with source $S$ is
\bea
\label{etaS}
\eta_{_{S}} &=& \frac{M_P^2}{4\pi} \left|\hat{\xi}_S\Big  (\frac{q}{\mu}\Big )_{S}\right|\> \Big | \> \big  (\frac{Z_1}{M_{A_1}}-\frac{Z_2}{M_{A_2} } \big ) \big  ( \zeta_e m_e + \sum_q \zeta_q m_q \> x_{q,p} \big ) \nn \\
&&\qquad\qquad + \big  ( \frac{A_1-Z_1}{M_{A_1}}-\frac{A_2-Z_2}{M_{A_2}} \big )  \sum_q \zeta_q m_q \> x_{q,n} 
+ \big  ( \frac{\omega_{A_1}}{M_{A_1}}-  \frac{\omega_{A_2}}{M_{A_2}} \Big ) \> \Big |, \nn \\
\eea
where $\big  (\frac{q}{\mu}\big )_S$ denotes the charge to mass ratio for the source object and 
$A_{k}, Z_{k}$ ($k=1,2$) refer to the atomic weights and atomic numbers of the two laboratory samples.
For order of magnitude estimates, we follow Ref.~\cite{Kaplan:2000hh} and ignore  binding energy effects, encoded
in the quantitities $\omega_{A_{1,2}}$. Setting  $M_A \simeq A m_N$ for the atomic masses, we then  obtain the simpler expression
\bea
\label{simplified-eta}
\eta_{_{S}} &\simeq& \frac{M_P^2}{4\pi m_N} \left|\hat{\xi}_S \Big  (\frac{q}{\mu}\Big )_S\right| \left |  \> \big  (\frac{Z_1}{A_1}-\frac{Z_2}{A_2 } \big ) \big\{ \zeta_e m_e + \sum_q \zeta_q m_q \> (x_{q,p}  -  \> x_{q,n})\big\}  \> \right | \ \ \ .
\eea
From Eqs.~(\ref{cg-cgamma}) and  (\ref{zetakappa}), the parameters
$\zeta_k$ appearing above are given by
\bea
\label{zetakappa2}
\zeta_k = \frac{1}{m_p} \Big [ g_k - \frac{2}{27} \sum_Q g_Q \Big ]
\eea
at leading order. Here $g_k$ denotes the couplings of $\phi$ to the light quarks and charged leptons and $g_Q$ denotes its coupling to the heavy $(b,c,t)$ quarks. A special case that will be of particular interest in subsequent discussion occurs when the couplings to fermions are universal, apart from the fermion Yukawa couplings explicitly factored out  via the factors of $m_f$ in Eqs. (\ref{eftlag},\ref{phi-fermions}). Setting 
\be
\label{eq:univcoup}
g_k=g_Q\equiv{\bar g}
\ee
and $m_p=m_n=m_N$ leads to
\bea
\label{eq:univeta}
\eta_{_{S}}^\mathrm{univ} &\simeq& {\bar g}\, \left(\frac{M_P^2}{4\pi m_N^2}\right)\left(\frac{7}{9}\right) \left|\hat{\xi}_S \Big  (\frac{q}{\mu}\Big )_S\right| \left |  \> \big  (\frac{Z_1}{A_1}-\frac{Z_2}{A_2 } \big ) \big\{  m_e + \sum_q  m_q \> (x_{q,p}  -  \> x_{q,n})\big\}  \> \right | \ \ \ .
\eea

Typical  source objects `$S$' used in E\"otv\"os experiments
include the Earth, the Sun, and galactic DM,  and one needs to obtain their charge to mass ratio $\hat{\xi}_S \Big  (\frac{q}{\mu}\Big )_S$ that appears in Eq.(\ref{eq:univeta}). If galactic DM is made of of elementary particles, then as already discussed, the charge to mass ratio under the dark force is given by
\be
\label{eq:ctmdm}
\hat{\xi}_S \Big  (\frac{q}{\mu}\Big)_S \Big\vert_{S=DM} = \left(\frac{g_\chi}{M_\chi}\right)\, \hat{\xi}_\chi\ \ \ .
\ee
For objects like the Earth that are made up of many different types of atoms, the effective charge-to-mass ratio  is obtained by a superposition of the couplings of $\phi$ to all the different atoms present in the object. In contrast to the situation for differences in charge-to-mass ratios for test bodies, it suffices to approximate this ratio for the bulk source object by ignoring atomic binding energy effects and summing over the couplings of $\phi$ to all the neutrons, protons, and electrons present. Doing so in the case of the Earth leads to
\bea
\label{earth-phi}
\hat{\xi}_E \Big  (\frac{q}{\mu}\Big )_E \simeq \frac{g_p N_p + g_n N_n + g_e(m_e/m_N) N_e}{m_N  (N_p + N_n)+ m_e N_e},
\eea
where $g_p$ and  $g_n$ denote the couplings of $\phi$ to protons and neutrons, respectively:
\be
\label{eq:gncoup}
g_N = \langle N| {\cal L}_{\phi f{\bar f}} | N\rangle\ \ \ ,
\ee
for $N=p$ or $n$. In the limit of a universal coupling as in Eq. (\ref{eq:univcoup}), we have
\be
\label{eq:univcoupN}
g_N=  g_h {\bar g}\, \left(\frac{v}{m_N}\right)\ \ \ ,
\ee
where $v=246$ GeV is the vacuum expectation value of the neutral component of the Higgs doublet and $g_h/\sqrt{2}$ is the coupling of the physical Higgs boson to the nucleon
\bea
\label{higgs-nucleon}
g_h = \langle N| \left( \sum_{q}  \frac{m_q}{v}\bar{q}q + \sum_Q \frac{m_Q}{v}\bar{Q}Q\right) |N\rangle.
\eea
 Using similar methods to those employed to determine $g_A$ and ignoring small difference between the neutron and proton coupling, one has\cite{Cheng:1988cz, Cheng:1988im, Burgess:2000yq}  
\be
\label{eq:gh}
g_h\simeq 1.71 \times 10^{-3}\ \ \ .
\ee
The resulting expression for the Earth's charge-to-mass ratio in this case is
\bea
\label{eq:earthuniv}
\hat{\xi}_E \Big  (\frac{q}{\mu}\Big )_E \Bigg\vert_\mathrm{univ} \simeq {\bar g}\left(\frac{v}{m_N^2}\right) 
\, \frac{g_h(N_p+N_n)+ (m_e/v) N_e}{(N_p + N_n)+ (m_e/m_N) N_e} \simeq 0.0017 \> \bar{g} \left (\frac{v}{m_N^2}\right ) \ \ \ .
\eea
The number of protons, neutrons, and electrons are $N_p \simeq 1.9 \times 10^{51}, N_n\simeq 2.0\times 10^{51},$ and $N_e\simeq 1.9 \times 10^{51}$ respectively.
We will make use of Eq.~(\ref{eq:earthuniv}) in what follows.

\section{Light Scalar Coupling to the Standard Model}
\label{section4}

We now give a model-independent discussion of the coupling of ordinary matter to the ultralight  singlet scalar that mediates the long range force.  In doing so, we  will lay the groundwork for calculating the parameters $g_{f}$ ($f=q,Q,\ell$), $c_g,$ and $c_\gamma$ of Eq.~(\ref{eftlag}) and Eq.$ (\ref{phi-fermions})$ or equivalently the parameters $\zeta_k$ and $\kappa$ in Eqs.~(\ref{zetakappa}) and  (\ref{zetakappa2}). In general there are two mechanisms for a singlet scalar to couple to the SM fermions and gauge bosons. The first mechanism involves mixing between the ultralight scalar and the Higgs, which allows the ultralight scalar to couple to the SM fermions  and gauge bosons. The second mechanism entails a coupling of the ultralight scalar to the SM through higher-dimension (non-renormalizable) operators. We discuss these two mechanisms in this section and establish notation. We also address the need for fine-tuning of the ultralight scalar mass when its interactions with the SM are non-negligible, looking ahead to a similar issue when we consider its coupling to DM. 

\subsection{Coupling to the Higgs Sector}

We assume that the mediator of the dark force carries no SM charges and that it can be described by a gauge singlet $S$. There exist no renormalizable couplings of such a singlet scalar to the SM fermions or gauge bosons, but it can couple to the SM Higgs doublet with operators of mass dimension $n\leq 4$. After electroweak symmetry breaking, the $n=3$ interaction $H^\dagger H S$ will generate mixing between $S$ and the neutral component of the Higgs doublet, $h$. We will identify the ultralight force-carrying scalar $\phi$ with the lighter mass eigenstate, and the heavier eigenstate with the physical Higgs boson.  
The Lagrangian for the singlet $S$ including its renormalizable and super-renormalizable interactions is given by
\bea
\label{higgsmixlag}
{\cal L} = \frac{1}{2} \partial _\mu S\partial ^\mu S - V (H,S)\> ,
\eea
where the potential is 
\bea
\label{Higgs-S}
V (H,S) &=& -\mu^2_h H^\dagger H + \frac{\lambda}{4}  (H^\dagger H)^2  + \frac{\delta _1}{2} H^\dagger H S +
\frac{\delta _2}{2} H^\dagger H S^2 \nn \\
&-& \Big  (\frac{\delta _1 \mu^2_h}{\lambda } \Big ) S + \frac{\kappa_2}{2} S^2 + \frac{\kappa_3}{3}S^3 + \frac{\kappa_4}{4}  S^4.
\eea
We have shifted the scalar $S$ so that it has no tree level vacuum expectation value without loss of generality. We follow the  notation of Ref.~\cite{O'Connell:2006wi, Barger:2007im}, which explored the presence of such a singlet scalar in the context of collider phenomenology. The parameters $\delta_{1,2}$ may arise from a more fundamental theory of which the $S$ is a residual, low-energy degree of freedom. As we discuss below, they may also receive contributions from DM loops if the DM particles couple to the both $H$ and $S$.

After electroweak symmetry breaking  the $H^\dag H S$ interaction induces  mixing between the Higgs boson $h$ and the scalar $S$. In unitary gauge the neutral component of the Higgs doublet $H$ is given by
\bea
H^0= \frac{v+ h}{\sqrt{2}}, \qquad v=\sqrt{\frac{2\mu_h^2}{\lambda}},
\eea
and the mass terms in the potential are
\bea
\label{Vmass}
V_{\text{mass}} = \frac{1}{2}  (\mu_h^2\> h^2 + \mu_S^2\> S^2 +\mu_{hS}^2 \> h S),
\eea
where
\bea
\label{mu-kappa-delta}
\mu_h^2 = \frac{\lambda v^2}{2}, \qquad \mu_S^2 = \kappa _2 + \frac{\delta _2 v^2}{2}, \qquad \mu^2_{hS} = \delta _1 v.
\eea
The mass eigenstates $h_\pm$ in terms of $S$ and $h$ can be written in terms of a mixing angle $\theta$ as
\bea
\label{mixing}
h_- = S \cos \theta  - h \sin \theta, \qquad h_+ = S\sin \theta  + h \cos \theta, \qquad \tan \theta = \frac{x}{1+ \sqrt{1+x^2}},
\eea
with corresponding masses
\bea
\label{mpm}
m_\pm^2 = \frac{\mu_h^2+\mu_S^2}{2}\pm \frac{\mu_h^2-\mu_S^2}{2} \> \sqrt{1+x^2},\eea 
and  we have defined
\bea
x\equiv \frac{\mu_{hS}^2}{\mu_h^2-\mu_S^2}.
\eea
\begin{figure}
\centering
\includegraphics[width=6in, height=2in]{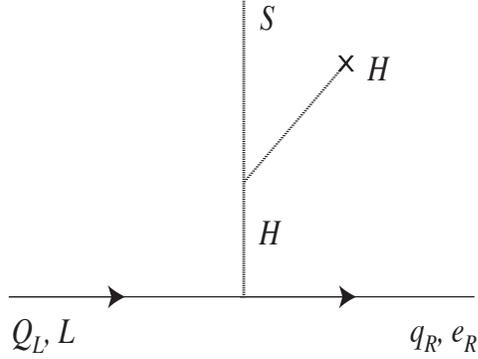}
\caption{Interaction of $S$ with SM fermions by mixing with the Higgs via the operator $SH^\dagger H$. Here, \lq\lq $X$" denotes the vacuum expectation value of the Higgs doublet.}
\label{higgsmixing}
\end{figure}

We assume that $m_-^2 \ll m_+^2$, so that the physical Higgs boson and light scalar are $h_+$ and $h_-$ respectively.  
\OMIT{From the expression for $m_-^2$ and requiring $\mu_S^2 \geq 0$ one can arrive at the constraint on the mixing angle $\theta$ and the parameter $\mu_h^2$
\bea
\frac{\tan ^2 \theta}{\tan ^2\theta -1}  \>\mu_h^2  \leq m_-^2.
\eea
Assuming that $\mu_h^2 \gg m_-^2$ we get the simpler constraint
\bea
| \tan \theta | < 1.
\eea}
The light scalar $h_-$ can couple to quarks and charged leptons through its mixing with the Higgs as shown in Eq.~(\ref{mixing}). We identify the light scalar that mediates the long range force as
\bea
\label{notation}
\phi \equiv h_-, \qquad m_\phi = m_-. 
\eea
The scalar $\phi$ couples to the SM fermions through its Higgs component, 
giving rise to the couplings $g_f$, where $f$ denotes any of the light quarks $q=u,d,s$, charged leptons $\ell = e,\mu,\tau$, or heavy quarks $Q=c,b,t$. One has
\bea
\label{gude1}
g_f = - \sin \theta \frac{m_p}{m_{{f}}} \frac{m_{f}}{v} =  - \sin \theta \frac{m_p}{v} ,
\eea
where the extra factor of $m_p/m_{f}$ after the first equality is included to be consistent with the convention in Eq.~(\ref{eftlag}).
This process is 
depicted in Fig. \ref{higgsmixing}. We see that in this mechanism the coupling of $\phi$ to ordinary matter is proportional to $\sin \theta$, with the constant of proportionality given entirely in terms of known quantities. The mixing angle $\theta$ will also receive corrections at the loop level and in the rest of the analysis we assume that $\theta$ is the renormalized mixing angle. 

For later use, we note that in the limit that $\mu_h \simeq m_h \gg \mu_S$ corresponding to a small mixing angle $\theta$, we can write\bea
\label{mphi-approx}
m_\phi^2 \simeq \mu_S^2 - \frac{\mu_{hS}^4}{4m_h^2}.
\eea
The existence of an ultralight scalar that can mediate a dark force over intergalactic distances
requires $m_\phi < 10^{-25}$ eV. In addition to the usual fine tuning of the parameters $\mu_S$
and $\mu_{hS}$ against radiative corrections sensitive to the cutoff (see section \ref{sec:finetune}), the finite renormalized parameters
$\mu_S$ and $\mu_{hS}$ are restricted in parameter space to satisfy the condition $m_\phi < 10^{-25}$ eV in Eq.(\ref{mphi-approx}). As we will discuss in section \ref{sectionVII} in more detail, this gives rise to three types of regions in parameter space. In the first region, $\mu_S$ and $\mu_{hS}$ are both individually small in which case there will be no observable dark force. In the second region,
$\mu_S$ and $\mu_{hS}$ are large enough to give rise to an observable dark force but cancel against each other in Eq.(\ref{mphi-approx}) to maintain an ultralight mass. In the third region, as will become clear in later sections, $\mu_S$ and $\mu_{hS}$ are again individually small as in the first region, but each is determined by a sum of much larger terms that cancel among each other.  The second region is phenomenologically the most interesting and is the focus of this paper.

\subsection{Non-renormalizable interactions}

If both the $S$ and $H$ couple to additional fields with masses above the electroweak scale, then these interactions will in general induce higher dimensional operators that involve both the $S$ and $H$ in a low-energy effective theory that does not contain the heavy degrees of freedom explicitly. Minimal dark matter models, for example, can require TeV-scale DM particles in order to achieve the observed relic density\cite{Cirelli:2005uq}, and these fields may generate the higher dimensional $S$-$H$ operators. At dimension five, one has 
seven independent  operators coupling $S$ to the SM fermions and gauge bosons:
\bea
\label{dim5ops}
{\cal O}_{u}^H &=& S \> \bar{Q}_L\> \epsilon H^\dagger \>C^H_{u}\> u_R + \text{h.c},\nn \\ 
{\cal O}_{d}^H &=&  S \> \bar{Q}_L \>H \>C^H_{d}\> d_R\> + \text{h.c}, \nn \\
{\cal O}_{e}^H &=& S \> \bar{L}_L\>  H \>C^H_{e}\> e_R \>  + \text{h.c},\nn \\
{\cal O}^W &=& C^WS\> \text{Tr} \big [W_{\mu \nu}W^{\mu \nu}\big ] ,\nn \\
{\cal O}^B &=& C^BS\> B_{\mu \nu}B^{\mu \nu},\nn \\
{\cal O}^G &=& C^G S\> \text{Tr} \big [G_{\mu \nu}G^{\mu \nu}\big ], \nn \\
{\cal O}^H &=& C^H S  (H^\dagger H)  (H^\dagger H).
\eea
The flavor indices on the fields $Q_L, L, u_R, e_R$ and the matrices $C^H_{u,d,e}$ are suppressed for simplicitly. Operators of the form $S \> \bar{Q}_L i\Dslash \>Q_L, S \> \bar{u}_R i\Dslash \>u_R, S \>\bar{d}_R i\Dslash \>d_R$, and $S \> \bar{e}_R i\Dslash \>e_R$ can be related to the operators  ${\cal O}_{u,d,e}^H$ by using the equations of motion
\bea
i\Dslash \> Q_L &=& \epsilon H^\dagger Y_u u_R + H Y_d d_R, \nn \\
i\Dslash \>u_R &=& \epsilon H Y_u^\dagger Q_L, \nn \\ 
i\Dslash \>d_R &=&  H^\dagger Y_d^\dagger Q_L, \nn \\ 
i\Dslash \> e_R &=& H^\dagger Y_e^\dagger L_L \ \ \ ,
\eea
where $Y_f$ denotes the matrix of SM Yukawa couplings. We have omitted operators that involve derivative or pseudoscalar couplings  of  $S$. Such couplings are spin dependent and have a negligible effect in  experiments which use unpolarized test objects. 

In general, the Wilson coefficients $C^H_{u,d,e}$ are $3\times 3$ matrices in flavor space, and can lead to flavor changing interactions of quarks and leptons with $S$. Since the couplings of $S$ to quarks and leptons are extremely small (as dictated by the WEP violation bounds) there is no danger of introducing dangerous flavor-changing neutral currents. In the specific model examples considered in subsequent sections of the paper,  find that the  $C^H_{u,d,e}$  are proportional to the Yukawa 
matrices:
\bea
C_{u,d,e}^H \equiv c_{u,d,e} Y_{u,d,e},
\eea
where $c_{u,d,e}$ are the constants of proportionality.  After expressing the fermion fields in the mass basis, in unitary gauge where  the operators ${\cal O}_{u,d,e}^H$ become flavor diagonal.  We can write
\bea
\tilde{{\cal O}}_{u}^H &=& c_u y_u^i S \> \bar{u}_L^i\> H^0 \> u_R^i + \text{h.c} \equiv \tilde{c}_u^i\, S \> \bar{u}_L^i\> H^0 \> u_R^i + \text{h.c},\nn \\ 
\tilde{{\cal O}}_{d}^H &=&  c_d y_d^i S \> \bar{d}_L^i \>H^0 \> d_R^i\> + \text{h.c}\equiv  \tilde{c}_d^i\, S \> \bar{d}_L^i \>H^0 \> d_R^i\> + \text{h.c}, \nn \\
\tilde{{\cal O}}_{e}^H &=& c_e y^i_e S \> \bar{e}_L^i\>   H^0 \> e_R^i \>  + \text{h.c}\equiv \tilde{c}_e^i \, S \> \bar{e}_L^i\>   H^0 \> e_R^i \>  + \text{h.c},
\eea
where $H^0$ is the lower component of the Higgs field $H$ in unitary gauge before electroweak symmetry breaking, the index $i=\{1,2,3\}$ runs over the three flavor generations, and we have defined $\tilde{c}^i_a = c_a y^i_a$.

After electroweak symmetry breaking, the operator ${\cal O}^H$ in Eq.~(\ref{dim5ops}) will also contribute to
$\sin \theta$. To linear order in $h$ one has
\be
\mathcal{O}^H\rightarrow C^H v^3 S h\ \ \ ,
\ee
thereby generating a contribution to the off-diagonal element of the mass-squared matrix
\be
\Delta\mu_{hS}^2= 2 \> C^H v^3\ \ \ .
\ee
We will explore the consequences of this term when discussing scalar DM models below.  For the moment we assume that this contribution has been included in $\sin \theta$.

Collecting the contributions to the couplings $g_{q,\ell,Q}$ from the higher-dimension operators ${\cal O}_{u,d,e}^H$ and mixing effects after electroweak symmetry breaking, the coupling of the ultralight scalar $\phi$ to SM fields at the electroweak scale is given by
\bea
\label{gqfinal}
g_{f} ( v) &=&  \frac{m_p}{m_{f}}\Big [\cos \theta\frac{v}{\sqrt{2}} \tilde{c}_{f}^i  ( v ) - \sin \theta \frac{m_{f}}{v} ( v)\Big ]\nn \\
& \simeq & \frac{m_p}{m_{f}}\frac{v}{\sqrt{2}} \tilde{c}_{f}  ( v ) - \sin \theta \frac{m_p}{v} ( v), 
\eea
where the last approximation is obtained from $\cos \theta \simeq 1$ since $\theta$ is constrained to be very small. We have included an extra factor of $m_p/m_{f}$ on the RHS above to be consistent with the convention in Eq.~(\ref{eftlag}). We have ignored the running between the scales $\Lambda \sim $ TeV and the electroweak scale for simplicity, but these effects can be incorporated by computing the appropriate anomalous dimension matrix and solving the corresponding renormalization group equations. 
We can now use Eq.~(\ref{gqfinal})  in Eqs.~(\ref{zetakappa}) and  (\ref{etaS}) to compute the E\"otv\"os parameters. In particular, we note that the contribution proportional to $\sin\theta$ is universal, so its contribution to the $\eta_S$ can be evaluated using Eqs.~({\ref{eq:univeta}) and (\ref{eq:earthuniv}). 

In general, the origin and parametric dependence of  $\sin \theta$ and $\tilde{c}_{f} $ are independent.  In most of the parameter space where there are no strong cancellations between the two terms in Eq.(\ref{gqfinal}), WEP violation constraints  can separately bound each of the two terms in Eq.~(\ref{gqfinal}). In the next three sections we use this feature with the representative minimal DM models, in the presence of a dark force mediated by $\phi$, and determine the implications for terrestrial experiments of direct DM-detection, E\"otv\"os experiments, and the colliders.

\subsection{Fine tuning and the light scalar mass}
\label{sec:finetune}

Before proceeding, we observe that in the absence of a symmetry that protects the light scalar mass from significant renormalization, one must resort to fine tuning to maintain the long range character of the dark force. To illustrate, we consider the contributions from the $H^\dag H S$ and $H^\dag H S^2$ in $V(H,S)$ to the singlet self energy. In the unitary gauge one has 
\bea
\label{eq:selfa}
\Sigma(p^2)_{H^\dag H S} & = & -\frac{\delta_1^2}{128\pi^2} \left[\frac{1}{\varepsilon}-\gamma+\ln 4\pi +\ln\mu^2 -F_0(m_H^2, m_H^2, p^2)\right] \\
\label{eq:selfb}
\Sigma(p^2)_{H^\dag H S^2} & = &\frac{\delta_2 m_H^2}{32\pi^2} \left[\frac{1}{\varepsilon}-\gamma+1+\ln 4\pi-\ln\frac{m_H^2}{\mu^2}\right]\ \ \ ,
\eea
where we work in $d=4-2\varepsilon$ dimensions, $\mu$ is the corresponding t'Hooft scale, and
\be
F_0(a,b,c)=\int_0^1\, dx \ln\left[(1-x)a+x b - x(1-x)c\right]\ \ \ .
\ee
Had we regulated the integrals with a momentum cutoff $\Lambda_\mathrm{CO}$ , the quadratic divergence proportional to $\delta_2 m_H^2/\varepsilon$ would be replaced by an expression proportional to $\delta_2 \Lambda^2_\mathrm{CO}$. 

For either choice of regulator, preservation of a tiny scalar mass requires a mass counterterm $\delta \mu_S^2$ to cancel  both the quadratic and logarithmically divergent contributions. In addition, as one sees using dimensional regularization, the finite,  
$\mu$-dependent contributions require a corresponding $\mu$-dependence in $\delta \mu_S^2$ as needed to maintain the scale-independence of the physical pole mass (or range of the dark force). The divergent and $\mu$-dependent finite contributions can be minimized by either choosing $\delta_{1,2}$ to be sufficiently tiny or by allowing for large cancellations between $\delta \mu_S^2(\mu)$ and the one-loop contributions. For the particular example discussed here, the finite contributions can also be minimized by choosing $\mu\approx m_H$, but these contributions will not be small at all scales unless the coefficients $\delta_{1,2}$ are tiny or there exists a large cancellation (fine-tuning) between $\delta \mu_S^2(\mu)$ and the one-loop contributions for $\mu\not= m_H$. In short, allowing for any appreciable interaction between the singlet $S$ and the Higgs sector of the SM invariably requires fine-tuning at some scale in order to ensure that the dark force mediator remains ultralight. In what follows, we will return to this point when considering the coupling of $S$ to DM. Even after allowing fine tuning, the finite renormalized parameters are restricted in parameter space in order to maintain an ultralight mass for the dark force mediator. We will discuss this  issue  in more detail in section \ref{sectionVII}.

\OMIT{
THE REMAINING PART OF THIS SECTION WILL BE REWRITTEN OR DELETED:
The Wilson coefficients $C_\alpha = \{\tilde{c}_{u,d,e}^i, C_G, C_W,C_B, C_H \}$ are first determined by matching the UV theory onto the EFT at the scale UV scale $\Lambda$. For example in models with TeV scale DM, the Wilson coefficients are determined by integrating out the DM. Next using the RG equations 
\bea
\mu \frac{d}{d\mu} C_\alpha  =  \gamma _{\alpha \beta} \> C_\beta,
\eea
one runs down to the electroweak scale $\mu \sim v$ summing logarithms of $\Lambda/ v$ in the process. Here $\gamma_{\alpha \beta}$ is the anomalous dimension matrix of the effective operators and in general is non-diagonal allowing for mixing between the operators.
Fig.~(\ref{gaugeloop}) shows examples of the one loop mixing of $O^{B,W,G}$ into $O^{H}_{u,d,e}$. }

\OMIT{
At the electroweak scale the Higgs field $H^0$ gets a vev 
which as described earlier also leads to a mixing between the Higgs field and $S$. Swicthing to the mass eigenstate basis $h$ and $\phi$ for the Higgs and light scalar respectively, the effective interaction Lagrangian describing the interactions of $\phi$ with quarks, leptons and gauge bosons are 
\bea
\label{electroweak}
{\cal L}_{\phi}^v &=&  \tilde{g}_u^i   (\mu \sim v) \bar{u}^iu^i \phi + \tilde{g}_d^i  (\mu \sim v)\bar{d}^i \>    d^i \phi  +  \tilde{g}_e^i  (\mu \sim v) \bar{e}^i e^i \phi + \tilde{C}_W (\mu \sim v) \phi W_{\mu \nu}^+ W^{- \mu \nu} + \tilde{C}_Z  (\mu \sim v)\phi Z_{\mu \nu} Z^{ \mu \nu} \nn \\
&&+ \tilde{C}_\gamma  (\mu \sim v) \phi F_{\mu \nu} F^{ \mu \nu} + C_G  (\mu \sim v) \phi \text{Tr} \big [G_{\mu \nu} G^{ \mu \nu}\big ] + {\cal V} (h,\phi), 
\eea
where the coefficients $\tilde{g}_{u,d,\ell}^i$ are given by
\bea
\tilde{g}_{u,d,e}^i  (\mu \sim v) = \frac{v}{\sqrt{2}} \tilde{c}_{u,d,e}^i  (\mu \sim v ) + \sin \theta \frac{m_{u,d,e}^i}{v} (\mu \sim v).
\eea
The second term above comes from the SM Yukawa term after the mixing of the Higgs field  with $S$. and the coefficients $\tilde{C}_{\gamma, W, Z} (\mu \sim v)$ are obtained from linear combinations of $C_{W} (\mu \sim v)$  and $C_{B} (\mu \sim v)$ after electroweak symmetry breaking where the $W^\pm$ and $Z$ bosons acquire mass.
Next we integrate out the top quarks and the $W,Z$ gauge bosons by matching at the electroweak scale and followed by running to the bottom quark mass scale
\bea
\label{bottom}
{\cal L}_\phi^{m_b} = \hat{g}_u^r   (\mu \sim m_b) \bar{u}^ru^r \phi + \hat{g}_d^i  (\mu \sim m_b)\bar{d}^i \>    d^i \phi  +  \hat{g}_e^i  (\mu \sim m_b) \bar{e}^i e^i \phi  + \hat{C}_\gamma  (\mu \sim m_b) \phi F_{\mu \nu} F^{ \mu \nu} + \hat{C}_G  (\mu \sim m_b) \phi \text{Tr} \big [G_{\mu \nu} G^{ \mu \nu}\big ], \nn \\
\eea
where the index $r=\{1,2\}$ in the up quark sector runs only over the first two generations since the top quark has been integrated out. The hatted coefficients above are obtained by matching at the scale $\mu \sim v$ followed by RG running down to the bottom mass scale. In Fig.~(\ref{?}) we show diagrams with top quarks and $W,Z$ bosons in loops which contribute to the matching calculations in determining $\hat{C}_\alpha = \{\hat{g}_{u}^r, \hat{g}_{d,e}^i, \hat{C}_{\gamma, G}\}$ at the electroweak scale. The RG equations again take a matrix equation form due to mixing between operators
\bea
\mu \frac{d}{d\mu} \hat{C}_\alpha  =  \hat{\gamma} _{\alpha \beta} \> \hat{C}_\beta,
\eea
where the anomalous dimension matrix at one loop is determined by the diagrams shown in Fig.~(\ref{?}). The running between the electroweak scale and the bottom quark mass scale sums up logarithms of $\sim m_b/v$.}
\OMIT{
\begin{figure}
\includegraphics[width=80mm]{gaugeloop.pdf}
\includegraphics[width=80mm]{gauge2loop.pdf}
\caption{Mixing of the ${\cal O}^{B,G}$ operators into ${\cal O}_{u,d,e}^{H}$ at one loop
and two loops.}
\label{gaugeloop}
\end{figure}}

\OMIT{
Next we integrate out the bottom and charm quarks and run down to the nucleon mass scale just before QCD becomes non-perturbative. The EFT takes the form of Eq.~(\ref{eftlag}) which we show again for clarity
\bea
{\cal L}_{\phi} &=& \frac{g_q}{m_p}  m_q\> \bar{q}q \phi + \frac{g_\ell}{m_p} m_\ell\> \bar{\ell}\ell\phi + c_g\> \phi \>\text{Tr}  G_{\mu \nu} G^{\mu \nu} +c_\gamma \> \> \phi F_{\mu \nu} F^{\mu \nu},\nn \\
&\equiv & c_q  (\mu \sim m_N)  \> \bar{q}q \phi + c_\ell  (\mu \sim m_N) \> \bar{\ell}\ell\phi + c_g (\mu \sim m_N)\> \phi \>\text{Tr}  G_{\mu \nu} G^{\mu \nu} + c_\gamma (\mu \sim m_N) \> \> \phi F_{\mu \nu} F^{\mu \nu}
\eea
where the index $q$ runs over the light quark flavors $u,d,s$ and the index $\ell$ runs over the charged leptons $e, \mu, \tau$ and the coefficients $c_\alpha =\{ c_q, c_\ell, c_g, c_\gamma\}$ are evaluated near the scale $\mu \sim m_N$. Once again at one loop the diagrams contributing to the matching calculation at the $m_b$ scale are shown in Fig.~(\ref{?}) with RG running equations
\bea
\mu \frac{d}{d\mu} \hat{c}_\alpha  =  \hat{\gamma} _{\alpha \beta} \> \hat{c}_\beta,
\eea
and the one loop diagrams contributing to anomalous dimension matrix are shown in Fig.~(\ref{?}).
From this effective Lagrangian the cupling of $\phi$ to atoms is determined as already discussed in section \ref{eft} . Thus, once the couplings of $\phi$ to the SM fermions and gauge bosons are are known at the high scale in any given model, the coupling of $\phi$ to atoms in ordinary matter can be determined by matching and running between a series of EFTs.}

\section{WIMP DM and E\"otv\"os Experiments }\label{DM-violation}
\label{section5}

In an earlier work \cite{Carroll:2008ub}, 
we examined constraints on the size of the coupling of an ultralight scalar to ordinary matter induced via virtual WIMP DM. The connection between constraints from galactic dynamics and E\"otv\"os experiments and the size of the ultralight scalar couplings to DM and ordinary matter were shown in Fig.~1 of \cite{Carroll:2008ub}. In analyzing the astrophysical constraints, we assumed only an upper bound $\beta<0.2$ and showed that in representative WIMP scenarios, it leads to stronger constraints on the strength of the $\phi$-WIMP coupling than do the present E\"otv\"os bounds on $\eta_{_{E,DM}}$. An improvement~\cite{Carroll:2008ub} of about eight orders of magnitude in E\"otv\"os experiments would be required to compete with the bounds from astrophysical constraints.

Here, we explore the prospective implications of a non-vanishing $\beta$. The presence of a modified, long-range dark force could help alleviate tensions in the $\Lambda$CDM paradigm (we refer the reader to Refs.~\cite{Kesden:2006vz,Bovy:2008gh} for an extensive discussion). In what follows, we show that a non-vanishing $\beta$ implies a lower bound on $\eta_{_{E,DM}}$ in simple WIMP scenarios, so that future E\"otv\"os experiments with improved sensitivity could be used to test this possibility. For purposes of illustration, we consider both scalar and fermionic WIMP DM. For fermionic WIMPs we restrict our attention to vector-like gauge representations, which simplifies the structure of the coupling of the ultralight singlet scalar $\phi$ to DM.

The Lagrangian for minimal  WIMP DM  takes the form
\bea
\label{WIMP-lag}
{\cal L} = \begin{cases} 
\bar{\chi}  (i\Dslash + M_0) \chi   , & \text{fermionic DM},\\ 
c  (D_\mu \chi)^\dagger D^\mu \chi  -  c\> M_0^2 \chi^\dagger \chi -V (\chi, H), & \text{scalar DM, } \end{cases}
\eea    
where $c=1/2$ for a real scalar and $c=1$ for a complex scalar. 
The covariant derivative depends on the $SU (2)_L$ and $U (1)_Y$ representations of $\chi$. Assuming that a single WIMP species saturates the relic density, one finds that 
 typical masses of such DM candidates are in the TeV range \cite{Cirelli:2005uq}. In general, $V (\chi,H)$ can contribute to the scalar DM mass after electroweak symmetry breaking. However, since the typical WIMP DM masses are in the TeV range, such a contribution will be much smaller than the size of the mass parameter $M_0 \sim $ TeV in the second line of Eq.(\ref{WIMP-lag}).
In what follows we assume this parameter $M_0$ to be the total DM mass since we are only interested in order of magnitude estimates. For gauge singlet scalar DM models with DM masses in the $100$ GeV range, the contribution to the mass from electroweak symmetry breaking can be important. We will consider the case of singlet scalar DM in the next section. Furthermore, since electroweak symmetry  breaking can in general induce mixing between  the scalar DM and the Higgs, we impose a  $Z_2^\chi$ symmetry $( \chi \to -\chi )$ to ensure stability of the DM particle. The interactions in $V (\chi,H)$ can also be constrained from WEP tests, and we will explore this in the next two sections for scalar DM. 
For vector-like fermionic DM  no renormalizable couplings exist between the Higgs and DM.  Such  couplings can however be present for chiral DM. 

We consider the impact of a WEP violating force in the dark sector via the interactions of DM with the ultralight scalar $\phi$ as in Eq.(\ref{eq:chicoup}).
These couplings are gauge invariant for fermionic DM only for vector-like gauge representations to which we restrict our attention. For chiral fermionic DM, the coupling to $\phi$ can only arise from higher dimension operators by gauge invariance. 
Assuming no other low-energy degrees of freedom besides those of the SM plus the $\chi$ and $\phi$, the dark sector interactions (\ref{eq:chicoup}) induce a coupling of $\phi$ to the SM fermions at two-loop order, as illustrated in  Fig.~\ref{gaugeloop}. The graph involving only virtual U(1$)_Y$ gauge bosons  [left panel of Fig.~\ref{gaugeloop}] directly  generate the operators ${\cal O}^H_{f}$ in Eq.~(\ref{dim5ops}), while the one-particle irreducible diagram involving both $W$ and $B$ bosons in the $\phi+f\to H+f$ \lq\lq Compton amplitude" [right panel of Fig.~(\ref{gaugeloop})] generates operators of the form ${\bar Q} i(\overleftarrow{/\!\!\!\! D}-\overrightarrow{/\!\!\!\! D}) Q$, etc. As noted earlier, operators of this type can be expressed in terms of ${\cal O}^H_{f}$ using the equations of motion, indicated symbolically by the presence of the $H$ field on the external leg in the right panel of Fig.~\ref{gaugeloop}. In either case,  the Wilson coefficients $C_{f}^H$ are proportional to the Yukawa matrices due to the Higgs insertions. After the neutral component of the Higgs field obtains a vev, the loop-induced operators ${\cal O}^H_{f}$ give rise to the interactions ${\bar f} f \phi$ of Eq.~(\ref{phi-fermions}). Any mixing between the ultralight scalar and the Higgs will also contribute, corresponding to the second term as usual in Eq.~(\ref{gqfinal}).

For $SU (2)_L$ triplet DM with hypercharge $Y=0$, only the SU(2$)_L$ gauge boson exchange diagrams of the right panel of Fig.~(\ref{gaugeloop}) contribute.  The resulting coupling of $\phi$ to the SM fermions are as in Eq.~(\ref{phi-fermions}), with
\bea
\label{tripletDM-fermions}
g_f=C_3 \Big  (\frac{\alpha_{em}}{\pi} \Big )^2 \frac{m_{p}}{M_\chi} g_\chi{\hat\xi}_\chi - \sin \theta \frac{m_p}{v}\ \ \ ,
\eea
where we have employed naive dimensional analysis (NDA) to estimate the first term on the RHS of Eq.~(\ref{tripletDM-fermions}). Although the precise $\mathcal{O}(1)$ coefficient $C_3$  can be obtained from a complete computation,  for our  purposes of arriving at order-of-magnitude relationships between $\beta$ and $\eta$,  the NDA expression suffices. \footnote{The subscript in $C_3$ refers to the dimension of the triplet representation of $SU(2)_L$.} We note  that the sum of all  loop graphs of the type in Fig.~(\ref{gaugeloop}) is finite because we began with only renormalizable couplings and the operators ${\cal O}^H_{f}$ have dimension $n=5$. We also observe that the coupling to different species of fermions is universal since we have factored out the explicit dependence on the Yukawa coupling in the definition of the $g_f$ in Eq.~(\ref{phi-fermions}). 


\begin{figure}
\includegraphics[height=2.1in, width= 2.5in]{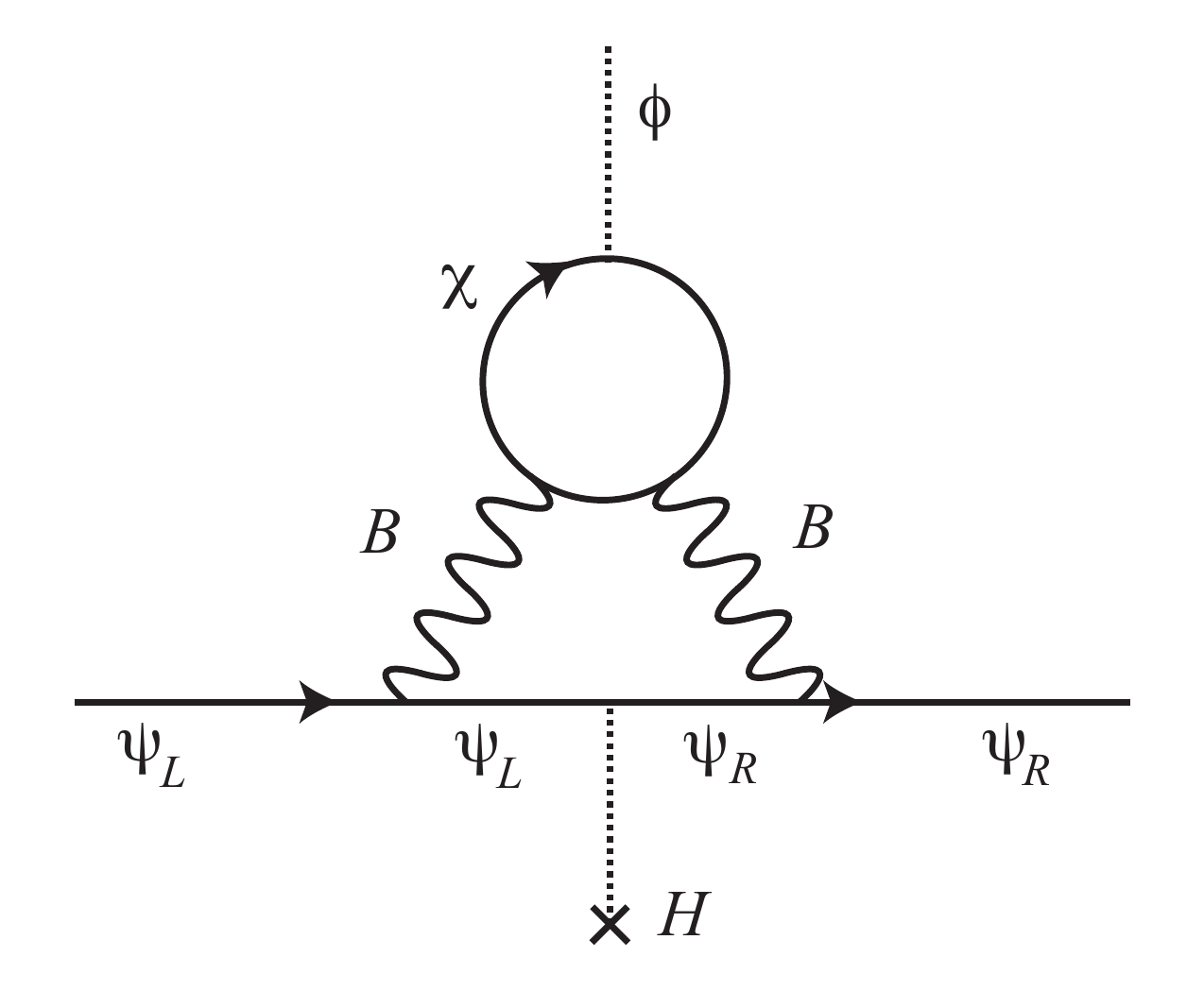}
\includegraphics[height=1.7in, width= 2.5in]{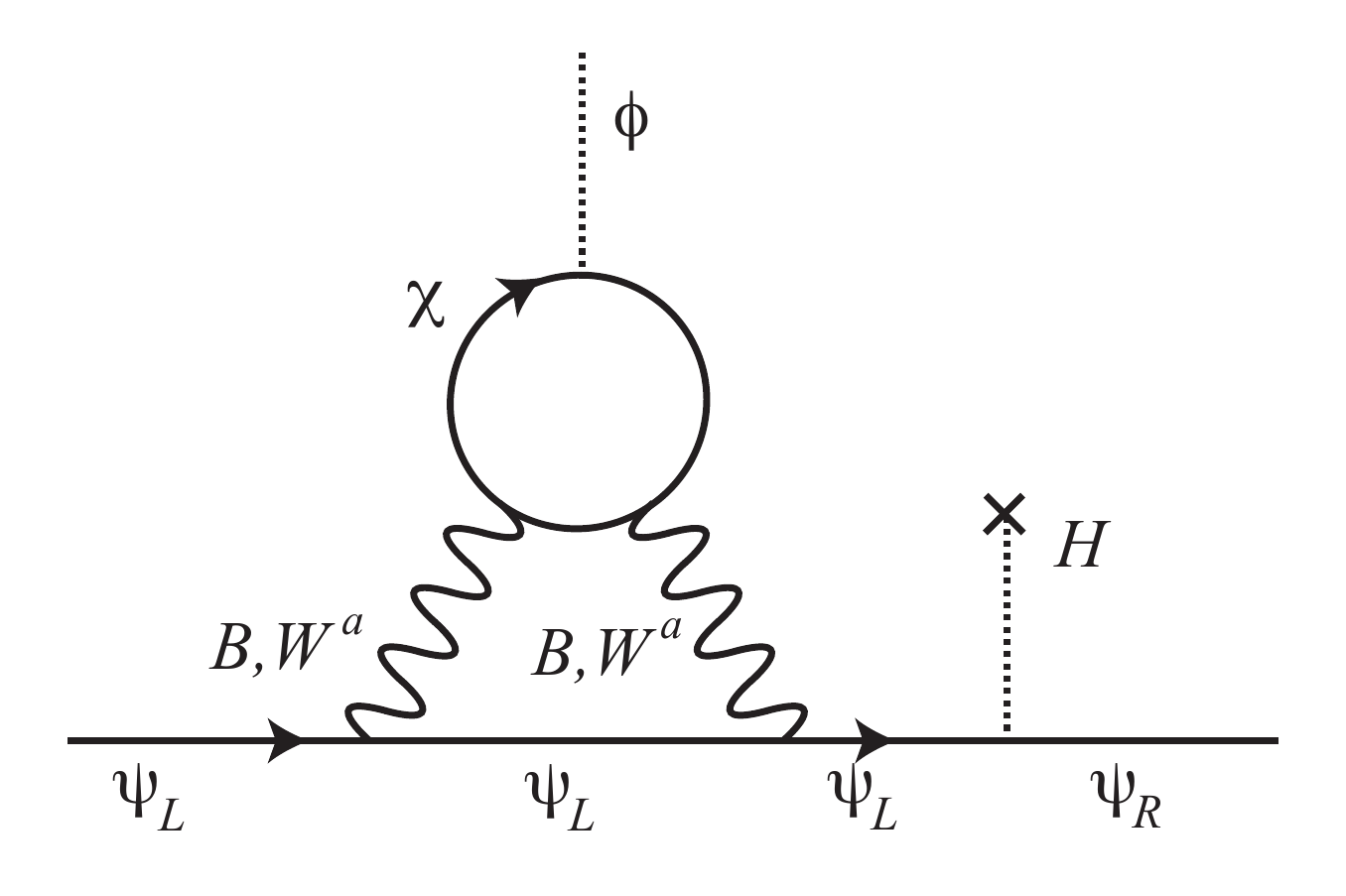}
\caption{Two-loop diagrams in WIMP DM models that generate the operators ${\cal O}^H_{f}$ in Eq.~(\ref{dim5ops}). Thus, after electroweak symmetry breaking the ultralight scalar couples to SM fermions.  }
\label{gaugeloop}
\end{figure}

For  $SU(2)_L$ multiplet DM with hypercharge $Y\neq 0$, the induced couplings 
of the ultralight scalar to DM is given by
\bea
\label{doubletDM-fermions}
g_f=C_N \Big  (\frac{\alpha_{em}}{\pi} \Big )^2 \frac{m_{p}}{M_\chi} g_\chi{\hat\xi}_\chi + C_Y  Y^2 \Big  (\frac{\alpha_{em}}{4\pi} \Big )^2\frac{m_{p}}{M_\chi} g_\chi {\hat\xi}_\chi - \sin \theta \frac{m_p}{v}\ \ \ ,
\eea
where $C_{N,Y}$ are $\mathcal{O}(1)$ coefficients that, as before, can be obtained from a complete two-loop computation. 
We observe that the first terms on the RHS of Eqs.~(\ref{tripletDM-fermions}) and (\ref{doubletDM-fermions}) are universal for different fermion species and come from the exchange of the $SU (2)_L$ gauge bosons $W^a$ in Fig.~(\ref{gaugeloop}). The last terms containing $\sin\theta$ are also universal, having been generated from the mixing between the Higgs and light scalar $\phi$. The middle term in Eq.~(\ref{doubletDM-fermions}), involving the square of the SM  hypercharge $Y$, are non-universal and are generated by  the exchange of $U (1)_Y$  gauge boson $B$. We point out that such minimal WIMP DM models with non-zero hypercharge are typically ruled out~\cite{Cirelli:2005uq} by direct detection experiments. Here we discuss these minimal DM models with non-zero hypercharge, only as illustrative examples keeping in mind that such DM could be part of a non-minimal extension which avoids the direct detection bounds.

A similar analysis can be performed for other WIMP models of DM that may involve additional degrees of freedom. In supersymmetry, for example,  the DM matter particle $\chi$
is a linear superposition of winos, binos, and higgsinos. In addition there are squark and slepton particles which give interactions of the type $\lambda \tilde{\psi} \bar{\psi} \chi + h.c.$. 
In theories with such a spectrum of particles one can induce a coupling of $\phi$ to ordinary matter via virtual DM at one loop as shown in Fig.~\ref{1-loop-squark}\footnote{Of course the presence of an ultralight scalar would introduce a new hierarchy problem which spoils the main motivation for supersymmetric theories. Here we invoke supersymmetry simply as a familiar example to illustrate the possibility of new types of interactions that can induce a coupling of the ultralight scalar to ordinary matter.}.  If the ultralight scalar $\phi$ is the scalar component of a singlet superfield $\hat{S}$, a superpotential  term of the form $(\mu + g_\chi \hat{S}) \hat{H}_u\cdot \hat{H}_d$ will lead to a coupling to fermions of the form
\bea
g_f \sim \frac{1}{16\pi ^2} \frac{m_{\tilde{\psi}} \mu \lambda^2}{M_{\text{SUSY}}^2} g_\chi.
\eea
If $\chi$ is primarily a bino, then $\lambda \simeq g_Y$, the hypercharge coupling. If $\chi$
is primarily Higgsino, the coupling of $\phi $ to the light quarks will be suppressed. The coupling
of $\phi$ will be primarily to the top quark which has order one Yukawa couplings. Thus, in such models
it is possible to induce a stronger WEP violating coupling to ordinary matter at one loop leading to bigger effects in
E\"otv\"os experiments.  For the sake of brevity, we do not consider such non-minimal scenarios and  we will only focus on minimal DM models without additional degrees of freedom such as squarks and sleptons. 
\begin{figure}
\includegraphics[height=2in, width=2in]{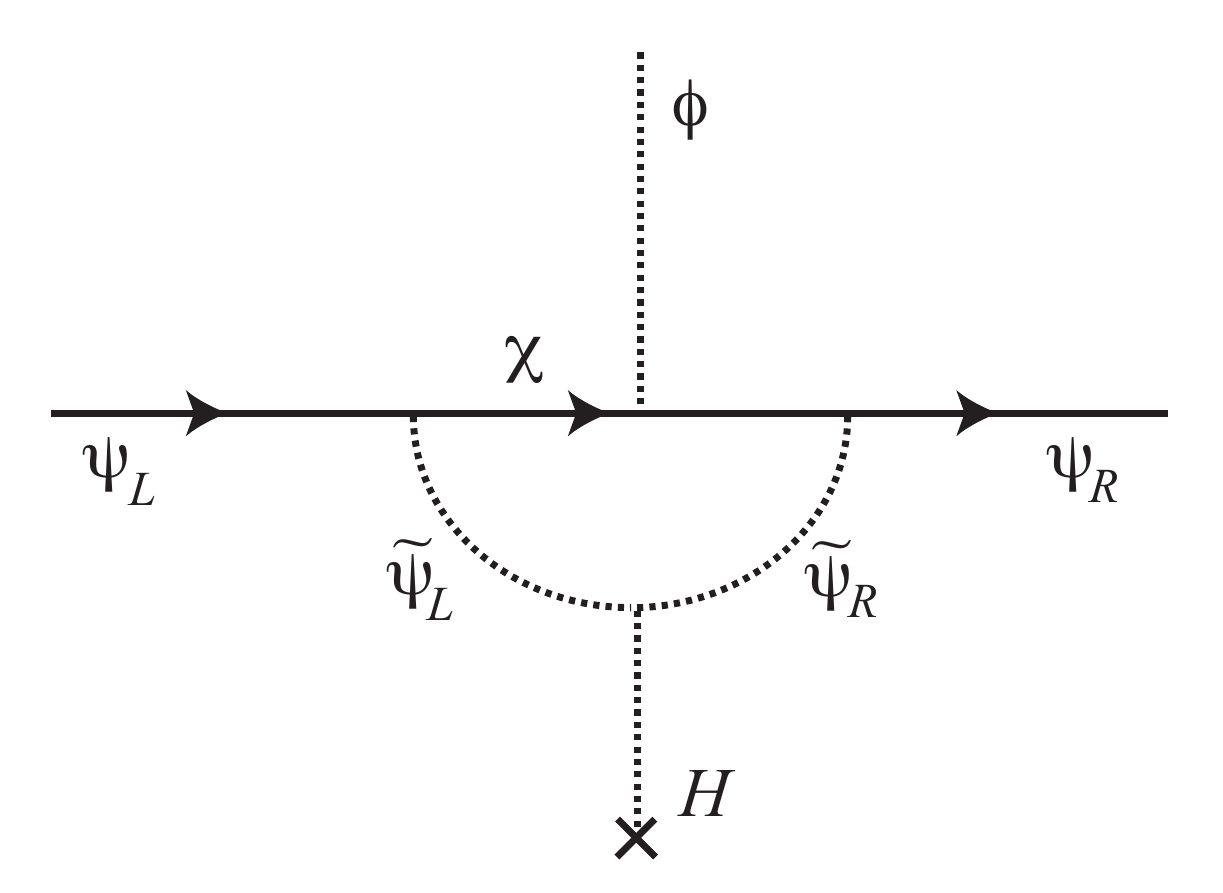}
\caption{DM-induced coupling of $\phi$ to SM fermions at one loop in the presence of additional squark and slepton like degrees of freedom. }
\label{1-loop-squark}
\end{figure}

It is possible that the  loop-induced ${\cal O}^H_{f}$ operator contributions to $g_f$ and those generated indirectly by $H$-$S$  mixing (proportional to $\sin \theta$) are individually much larger than $g_f$ yet cancel to produce a much smaller coupling. However, away from this special region of parameter space,  each contribution will be roughly no larger in magnitude than $g_f$ itself (as already discussed following Eq.(\ref{gqfinal})). In this generic case, we are able to obtain expectations for the size of  E\"otv\"os parameters in our illustrative minimal WIMP dark sector models, for a given value of  $\beta$, from the two-loop gauge contribution. We will consider the contribution of $H-S$ mixing to $g_f$ in the next section. We note  that the contributions from the SU(2$)_L$ gauge bosons are generically an order of magnitude larger than those from the hypercharge gauge bosons due to the relative sizes of their couplings [leading to the additional factor of $1/16$ in the second term of Eq.~(\ref{doubletDM-fermions})]. Consequently, for purposes of making order-of-magnitude estimates, we may employ the expressions for $\eta_S$ in the presence of universal couplings given in Eqs.~(\ref{eq:univeta}) and (\ref{eq:earthuniv}) with
\be
\label{eq:gbargauge}
{\bar g} \rightarrow \Big  (\frac{\alpha_{em}}{\pi} \Big )^2 \frac{m_{p}}{M_\chi} g_\chi{\hat\xi}_\chi\ \ \ .
\ee
\begin{figure}
\includegraphics[height=2.2in, width = 3in]{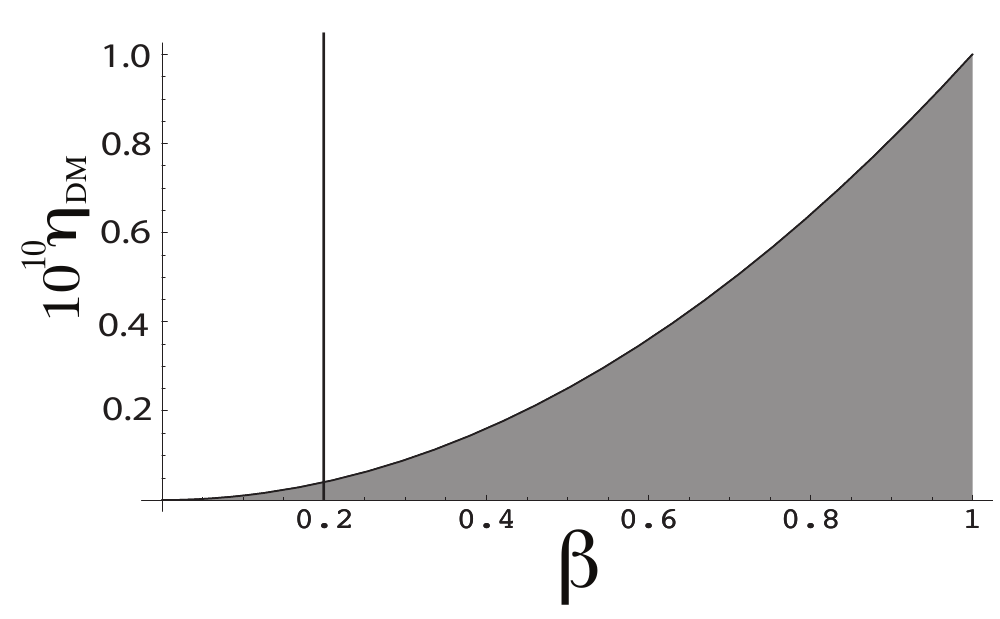} 
\hfill
\includegraphics[height=2.2in, width = 3in]{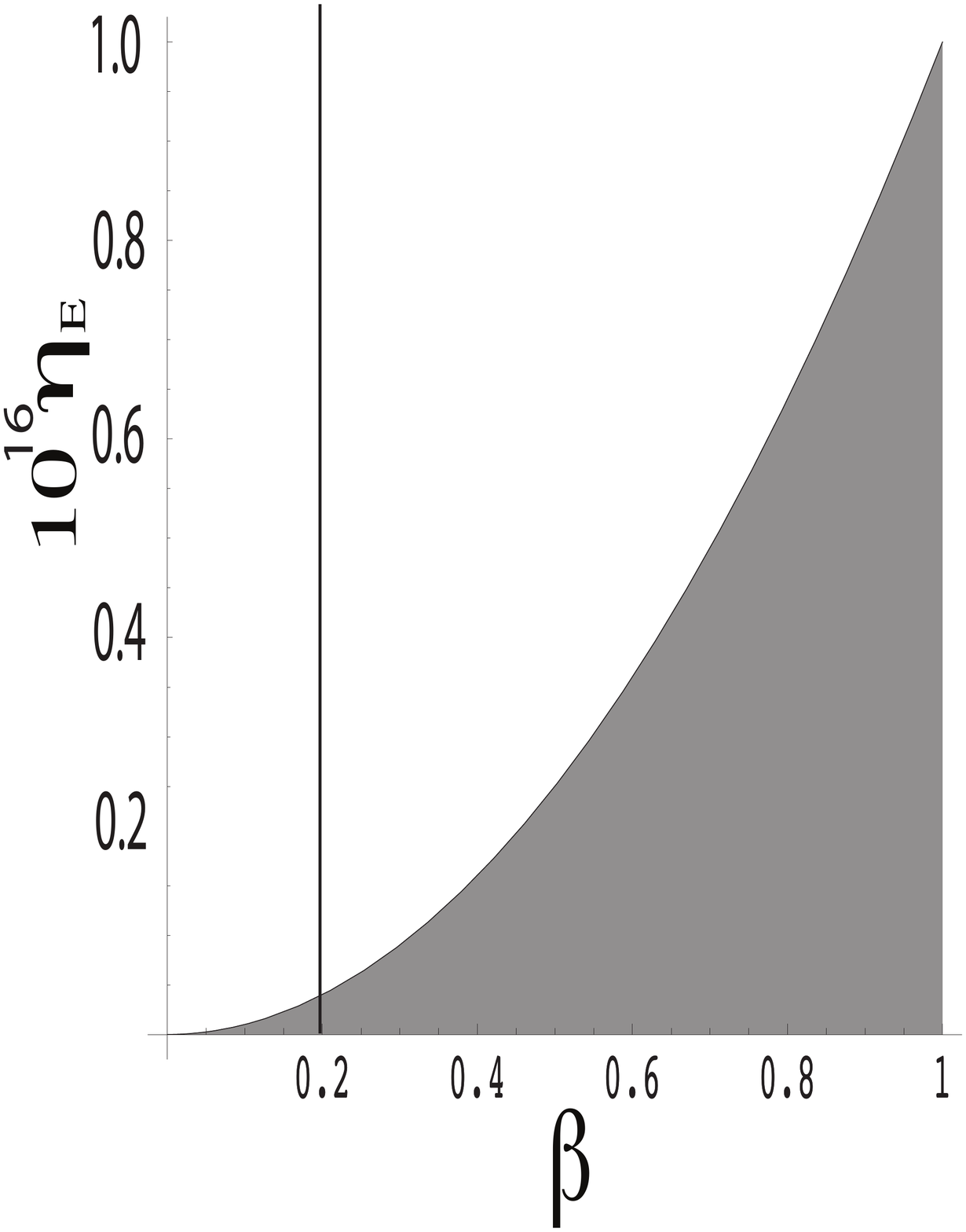}                            
\caption{An estimate of the allowed region in the $(\eta_{_{DM,E}},\beta)$ parameter space for minimal WIMP DM models. The curves in the figures give an estimate of  for $\eta_{_{DM,E}}$ for a given value of $\beta$ from the two loop diagrams in Fig.~\ref{gaugeloop}. The shaded region is unlikely for typical WIMP models. Using the observational constraint $\beta < 0.2$, the allowed region is further restricted to the left of the vertical line. The estimates in the above figures for $\eta_{_{DM,E}}$, for $\beta < 0.2$, are far below the current experimental bounds $\eta_{_{DM}} \lesssim 10^{-5}, \>\eta_{_{E}}\lesssim 10^{-13}$. An  improvement of about five orders of magnitude would be required in E\"otv\"os experiments to fully probe the allowed parameter space for $\beta = 0.2$ by measuring $\eta_{_E}$. This is within reach of the  MiniSTEP~\cite{Lockerbie:1998ar} proposal.}
\label{fermionicWIMPetabeta}
\end{figure}
Expressing $g_\chi{\hat\xi}_\chi$ in terms of $\beta$ we then obtain
\bea
\label{eq:etadmgauge}
\frac{\eta_{_{DM}}}{\beta^2} & \gtrsim & \left(\frac{7}{9}\right) \left(\frac{\alpha}{\pi}\right)^2 \left |  \> \big  (\frac{Z_1}{A_1}-\frac{Z_2}{A_2 } \big ) \big\{  \frac{m_e}{m_N} + \sum_q  \frac{m_q}{m_N} \> (x_{q,p}  -  \> x_{q,n})\big\}  \> \right |\\
\label{eq:etaearthgauge}
\frac{\eta_{_{E}}}{\beta^2}& \gtrsim &  \left(\frac{7}{9}\right) \left(\frac{\alpha}{\pi}\right)^4\left(\frac{v}{m_N}\right)\left |  \> \big  (\frac{Z_1}{A_1}-\frac{Z_2}{A_2 } \big ) \big\{  \frac{m_e}{m_N} + \sum_q  \frac{m_q}{m_N} \> (x_{q,p}  -  \> x_{q,n})\big\}  \> \right | \\
\nn
&& \times \left[ \frac{g_h(N_p+N_n)+ (m_e/v) N_e}{(N_p + N_n)+ (m_e/m_N) N_e}\right]\ \ \ .
\eea
Numerically the bounds in Eqs.~(\ref{eq:etadmgauge}) and (\ref{eq:etaearthgauge}) are (for Be and Ti laboratory samples with $|Z_1/A_1 -Z_2/A_2| \simeq 1/72 $)
\bea
\frac{\eta_{_{DM}}}{\beta^2} &&\gtrsim  10^{-10}, \nn \\
\label{eq:wimpbounds}
\frac{\eta_{_{E}}}{\beta^2} &&\gtrsim 10^{-16},
\eea
which are shown in Fig.~\ref{fermionicWIMPetabeta} as the  allowed regions for typical miminal WIMP DM models in the $(\eta_{_{DM,E}}, \beta)$ parameter space. The curve in the left and right figures gives as estimate of the minimum size for $\eta_{_{DM}}$ and  $\eta_{_E}$ respectively as a function of $\beta$. One can also estimate the ratio $\eta_{_E}/\eta_{_{DM}}$ from Eqs.~(\ref{eq:etadmgauge}) and (\ref{eq:etaearthgauge})  to be approximately 
\bea
\label{eta-E-DM-ratio}
\eta_{_E}/\eta_{_{DM}} \simeq 10^{-6}  \ \ \ .
\eea

For $\beta =0.2$, marked by the vertical lines in Fig.~\ref{fermionicWIMPetabeta}, the current upper bound from galactic dynamics~\cite{Kesden:2006vz}, we see that the lower bounds for typical WIMP DM models are $\eta_{_{DM}} > 4\times 10^{-12}$ and $\eta_{_{E}}> 4 \times 10^{-18}$. These lower bounds are far below the current experimental upper bounds shown in Eq.(\ref{bounds}). An improvement of about five to seven orders of magnitude in E\"otv\"os experiments would be required in order to probe these expectations of WIMP DM models. The MiniSTEP~\cite{Lockerbie:1998ar} experiment, which is currently under study, is expected to reach a sensitivity for $\eta_{_E}$ of about $10^{-18}$ and might be able to probe the lower bounds of these WIMP models. However, if $\beta < 0.05$ as indicated by a recent analysis~\cite{Bean:2008ac} of the CMB and large scale structure formation,  the lower bounds on $\eta_{_{DM,E}}$ are far beyond current and future planned experiments. If an effect is detected in $\eta_{_{DM,E}}$ far above the expectations in Fig.(\ref{fermionicWIMPetabeta}) it would suggest the possibility that the coupling of $\phi$ to the SM fermions is mostly via $h-\phi$ mixing corresponding to the last term appearing in Eqs.(\ref{tripletDM-fermions}) and (\ref{doubletDM-fermions}). One could extract a value for $\sin\theta$ and derive implications for various DM scenarios as discussed in the following section. Another possibility that might explain an effect above the expectation in Fig.~\ref{fermionicWIMPetabeta} would be a stronger induced coupling of $\phi$
to ordinary matter in non-minimal DM models; for example a one loop coupling of $\phi$ to ordinary matter (see Fig.~\ref{1-loop-squark}) in the presence of additional squark degrees of freedom.

\section{WEP Tests, Direct Detection, and Higgs Boson Decays}
\label{sectionVI}



As observed in Ref.~\cite{Bovy:2008gh}, the presence of a non-vanishing $\beta$ of astrophysically interesting magnitude, together with present limits on $\eta_{_{E, DM}}$ can imply upper bounds on the size of DM-nucleus cross sections relevant for direct detection experiments. Here we analyze these bounds in detail for the illustrative cases of scalar  DM scenarios and argue that upper bounds on the DM-nucleus cross sections are  less stringent than obtained in Ref.~\cite{Bovy:2008gh}. We further comment on the analysis of Ref.~\cite{Bovy:2008gh} at the end of section \ref{sectionVII}. We also consider the implications of a dark force for the DM relic density and derive corresponding constraints. Finally, using a light scalar triplet, as part of a multicomponent DM scenario,  we show how the presence of a non-vanishing $\beta$ -- together with experimental limits on $\eta_{_{E, DM}}$ -- can preclude observable shifts in the rate for the Higgs boson to decay to two photons as one might otherwise expect.


To include the full set of possible renormalizable interactions between the DM, SM fields, and ultralight scalar, we expand the scalar potential of Eq.~(\ref{Higgs-S}), imposing the $Z_2^\chi$ ($\chi\to -\chi$) symmetry need to prevent DM decays:
\bea
\label{eq:vhschi}
V(H,S,\chi)&=&V(H,S)+\frac{1}{2}\, M_0^2 \chi^2+\frac{\lambda_\chi}{4}\chi^4+a_2 H^\dag H \chi^2 + g_\chi \chi^2 S + \lambda_{\chi s} \chi^2 S^2 \ \ \ .
\eea
\begin{figure}
\includegraphics[width=2in, height=2in]{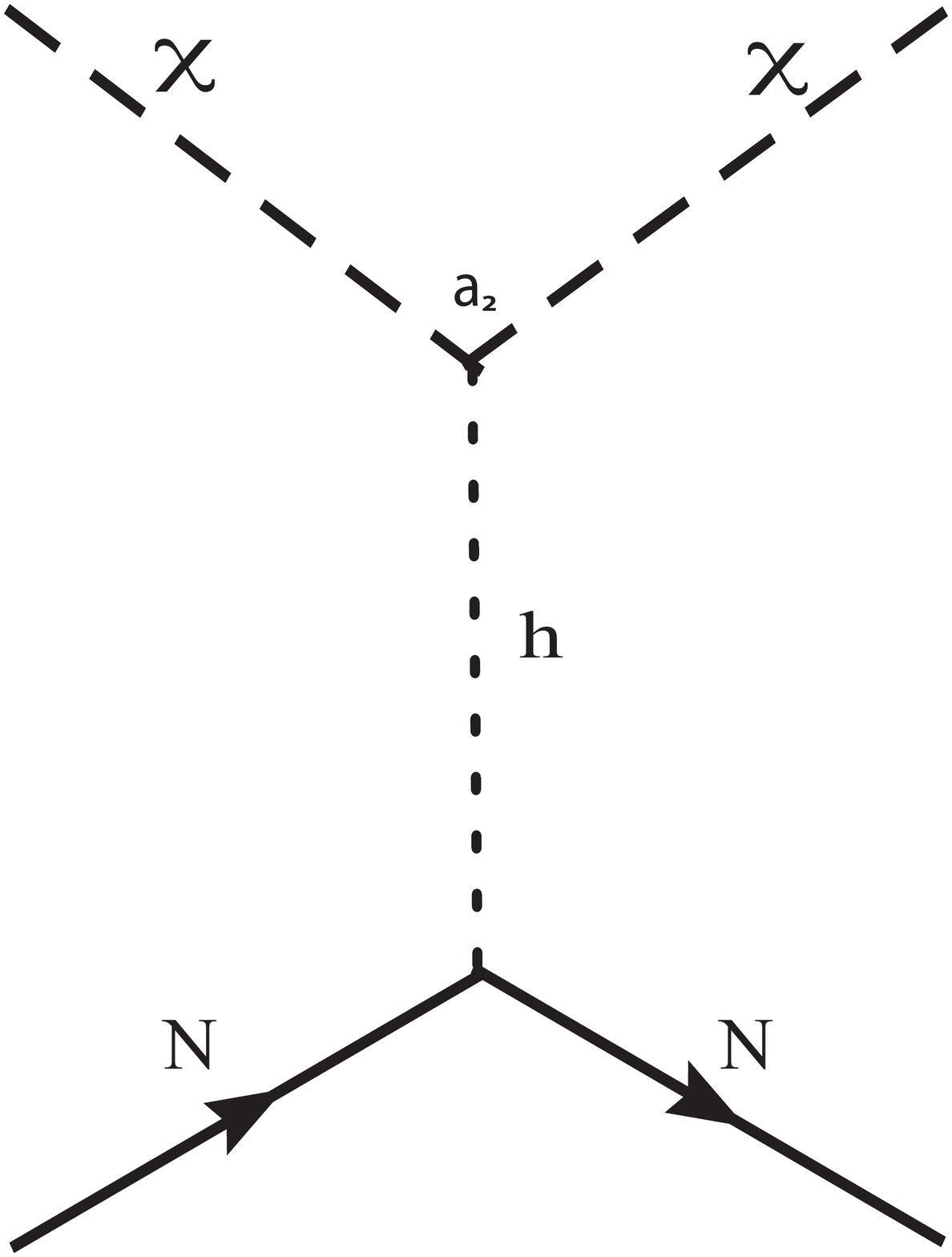}
\caption{Direct detection process for the scalar singlet and triplet $\chi$ via t-channel Higgs exchange with a nucleon. The magnitude of the detection rate is determined by the strength of the coupling $a_2$. If $\chi$ couples to the  ultralight scalar $\phi$, the size of $a_2$ and thus the detection rate is bound by WEP violation constraints. }
\label{singletDM}
\end{figure}
For the scalar singlet case, $\chi$ is a real field, while for the real triplet with components $\chi^0$ and $\chi^\pm$, one has \cite{FileviezPerez:2008bj}
\be
\chi^2=\left(\chi^0\right)^2 + 2\chi^+\chi^-\ \ \ .
\ee
We take $M_0^2$ and $a_2$ to be positive in order prevent a non-vanishing vev for $\chi$ and the occurrence of phenomenologically unacceptable cosmological domain walls. The experimental constraints on this DM model, for $g_\chi=0$,  were recently explored in \cite{Barger:2007im, He:2008qm}.  
\OMIT{The potential for a real scalar triplet $\chi$ has the same form as in Eq.(\ref{eq:vhschi}) with an additional quartic term of the form  $(\chi^\dagger T^a \chi) ( \chi^\dagger T^a \chi) $ which does not play a role in the analysis below.}

After electroweak symmetry breaking, the $H^\dag H \chi^2$ term generates a contribution to the DM mass:
\bea
\label{singlet-DM-mass}
M_\chi^2 = M_0^2 + a_2 v^2.
\eea
Henceforth, we will take $v=246$ GeV, $M_\chi^2$, $a_2$, and the mass of the SM-like Higgs boson ($m_h$)  as independent parameters. All of them govern the $\chi$-nucleus cross section, whose leading order amplitude is generated by $t$-channel Higgs exchange as in Fig.~\ref{singletDM} and is given by 

\bea
{\cal M} \simeq \frac{2 a_2 g_h v}{m_h^2} \bar{N}N\ \ \ ,
\eea
where we have neglected the $t$-dependence of the amplitude for simplicity. Note that since the real triplet has zero hypercharge, the elastic DM-nucleus scattering has no contribution form $Z$-boson exchange at tree level. The corresponding cross section is
\bea
\label{singlet-cross section}
\sigma_{\chi N} \simeq \frac{a_2^2 \>g_h^2 v^2 m_N^2}{\pi  (M_\chi + m_N)^2 m_h^4} \ \ \ ,
\eea
where, for simplicity, we have dropped the dependence on momentum transfer to the nucleus. 
[Recall that  $g_h \simeq 1.71 \times 10^{-3}$  is the coupling of the Higgs to the nucleon as defined in Eq.(\ref{higgs-nucleon})]. Note that the cross section decreases for increasing $M_\chi$ or decreasing $a_2$. Note also that the coupling $a_2$, together with the masses $M_\chi$ and  $m_h$, control the $\chi$ relic density through the annihilation diagrams of Fig.~\ref{annihilationDM}. For $M_\chi\sim m_h/2$ for singlet DM or a light triplet in the multicomponent DM scenario, the Higgs exchange contribution becomes large, requiring a suppression of $a_2$ in order to maintain the observed CDM relic density. In what follows, we will generally avoid this regime. 
\begin{figure}
\includegraphics[height=1.5in, width=6in]{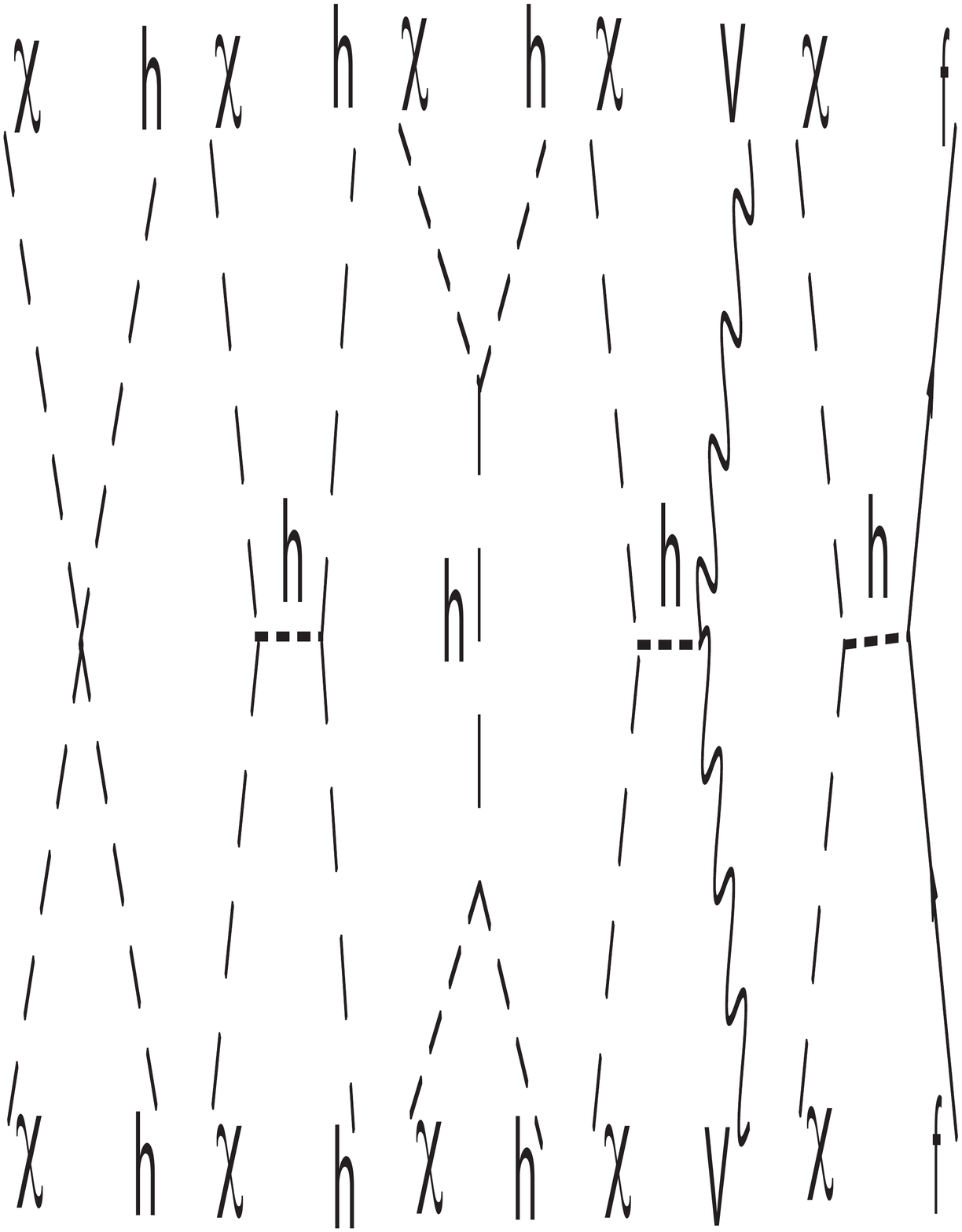}
\caption{Annihilation diagrams for the scalar singlet DM. For the scalar  triplet DM one has in addition the usual annihilation diagrams mediated by gauge interactions. If the gauge interactions of the triplet dominate the dynamics of annihilation, a DM mass of around 2 TeV  is needed to saturate the relic density.}
\label{annihilationDM}
\end{figure}

\subsection{WEP Tests and Ultralight-Scalar-Higgs Mixing}
\label{section6A}


A relation between the Higgs-exchange contribution to $\sigma_{\chi N}$ and $\eta_{_{E, DM}}$ arises for non-vanishing $\beta$ because the parameters $a_2$, $M_\chi$, and $m_h$ that enter the cross section also control the strength of the DM-loop induced mixing between the Higgs boson and the ultralight scalar. After electroweak symmetry-breaking, these loop effects generate contributions to the mass-squared parameters $\mu_{hs}^2$ and $\mu_S^2$. The parameter $\mu_S^2$ contributes only to the mass $m_\phi^2$ and $\mu_{hS}^2$ contributes to
$\sin \theta$ and $m_\phi^2$.  As with the contributions from Higgs loops to $\mu_S^2$ discussed earlier in Section \ref{sec:finetune}, the DM loop contributions to this mass-squared parameter will also require the introduction of fine tuning to maintain a sufficiently small mass for the ultralight scalar. Furthermore, as already mentioned and discussed in more detail in section \ref{sectionVII}, the finite renormalized parameter $\mu_S^2$ must be further restricted in parameter space in order to maintain $m_\phi < 10^{-25}$ eV along with a dark force large enough to be observed. We implicitly assume that we are in this region of parameter space, conducive to the observation of a long range dark force. We will discuss the implications of other regions in parameter space in section \ref{sectionVII}.

We begin by observing that  in addition to the direct coupling $\delta_1$ of $S$ to the Higgs via the operator $H^\dagger HS$, a  DM-induced $\phi$-matter coupling arises from the one-loop contribution to this operator  through the second diagram of Fig. \ref{fig:singletmixing}. After renormalization in the $\overline{\mathrm{MS}}$ scheme, the resulting finite coefficient of his operator is
\be
\label{eq:deltaonechi}
\delta_1^\mathrm{ren} = \delta_1(\mu)+ \kappa \frac{ g_\chi a_2}{ \> 4\pi^2} \, \ln\frac{M_0^2}{\mu^2}\ \ \ ,
\ee
where
\bea
\kappa = \begin{cases} 1, \>\> \text{singlet} \>\chi, \\ 3, \>\> \text{triplet} \>\chi , \end{cases}
\eea
 The factor of $\kappa =3$ appears in the case of the triplet $\chi$ due to the three components of the triplet traversing the loop in the second diagram of Fig.~\ref{fig:singletmixing}. Here $\delta_1(\mu)$ is the finite, scale-dependent coupling counterterm  whose numerical value is {\em a priori} unknown and whose presence is required to ensure renormalization group (RG) invariance of the physical properties of the $\phi$ and $h$. Note that the mass parameter $M_0^2$ (taken here to be positive) rather than $M_\chi^2$ appears in the argument of the logarithm since we are working in the theory before electroweak symmetry-breaking. 

We also observe that the $ \chi^2 S$ interaction will generate a contribution to the mass parameter $\mu_S^2$ as it yields the non-vanishing contribution to the $S$ self energy:
\be
\label{eq:selfchi}
\Sigma(p^2)_{\chi^2 S}  =  -\frac{g_\chi^2}{16\pi^2} \left[\frac{1}{\varepsilon}-\gamma+\ln 4\pi +\ln\mu^2 -F_0(M_0^2, M_0^2, p^2)\right] \ \ \ .
\ee
As with the case of the logarithmically divergent Higgs contribution $\Sigma(p^2)_{H^\dag H S}$ of Eq.~(\ref{eq:selfa}), the DM-loop contribution to the self energy requires a corresponding $\mu$-dependence in $\delta \mu_S^2(\mu)$ to maintain RG invariance of the pole mass that governs the range of the dark force. Large DM-loop contributions to $\mu_S^2$ can be minimized at all scales by taking $g_\chi$ to be sufficiently small: $g_\chi\lesssim 4\pi m_S^\mathrm{pole}$. Doing so, however, would preclude a value of $\beta$ of astrophysically relevant strength. Alternatively, one may allow for a much larger, phenomenologically interesting magntiude for $g_\chi$ and maintain a small $\mu_S^2$ by invoking fine-tuning between the one-loop contribution of Eq.~(\ref{eq:selfchi}) and $\delta m^2(\mu)$. 

A similar set of alternatives applies to the renormalized coupling $\delta_1^\mathrm{ren}$. One could require that the product $g_\chi a_2$ be sufficiently small in magnitude, with a correspondingly tiny $\delta_1(\mu)$, so that the induced $H$-$S$ mixing is consistent with the present bounds on $\eta_\mathrm{E,\, DM}$. To obtain a large $\beta$, one must then take $a_2$ to be sufficiently small, implying an upper bound on the Higgs exchange contribution to the DM-nucleus cross section via Eq.~(\ref{singlet-cross section}). This choice is essentially the strategy followed in Ref.~\cite{Bovy:2008gh} to obtain upper bounds on $\sigma_{\chi N}$.  However, as seen in Eq.(\ref{mu-kappa-delta})$,\delta_1$ contributes to $\mu_{hS}$ and thus to the mass $m_\phi$ via Eq.(\ref{mphi-approx}). The condition of $m_\phi < 10^{-25}$ eV gives a much stronger naturalness constraint on $a_2$ forcing it to be essentially zero for a non-zero dark force. The constraints from $\eta_{_{E,DM}}$ are thus not relevant in such a naturalness analysis. We will also discuss this in more detail in section \ref{sectionVII}.

\OMIT{
We show that in the presence of a fifth force in the dark sector, mediated by an ultralight scalar, this direct detection cross section can be constrained from existing bounds on WEP violation. In particular one can tie together any WEP violation 
observed in galactic dynamics and laboratory E\"otv\"os experiments to the direct detection rate for DM.}

\OMIT{If the DM $\chi$ couples significantly to an ultralight long range scalar, one would measure a non-zero value for $\beta = \frac{M_P}{\sqrt{4\pi}} \frac{g_\chi}{2 M_\chi^2}$ where $g_\chi$ is the coupling of the gauge singlet light scalar $S$ to the DM
\bea
\delta{\cal L} = g_\chi \> \chi^2 S.
\eea
 The above coupling will in general contribute to the 
super-renormalizable operator
\bea
\label{dim3mixing}
\frac{\delta_1}{2}\> H^\dagger H S,
\eea
which first appeared in Eq.(\ref{tripletmixing}). This contribution would be in addition to a tree level value for $\delta_1$ if a direct coupling between the Higgs and $S$ is allowed. As already discussed, after 
electroweak symmetry breaking this operator will generate 
$h-S$ mixing so that ordinary matter will experience a long range fifth force  indirectly via its interaction with the Higgs and E\"otv\"os experiments can constrain the value of $\delta_1$. The one loop contribution to $\delta_1$ is shown by the first diagram in Fig.~\ref{tripletmixing}. This diagram is divergent and the value of $\delta_1$ will be determined by a counterterm subtraction. The analysis of \cite{Bovy:2008gh} estimated the size of $\delta_1$ using the assumption of naturalness. In this framework  one assumes that there is no fine tuning between the bare and counterterms and  the size of $\delta_1$ is estimated by computing the first diagram in Fig.~\ref{tripletmixing} with a UV cutoff.  The size of $\delta_1$ as determined by the first diagram in Fig.~\ref{tripletmixing} then depends on a UV cutoff $\Lambda$, the coupling $a_2$ which determines the direct detection cross-section, and $g_\chi$ which is constrained by $\beta$. The constraint on $\delta_1$ from $\eta_{_{DM,E}}$ then allows one to constrain the direct detection cross-section in terms of WEP violation constraints on $\beta$ and $\eta_{_{DM,E}}$. This was the approach used by \cite{Bovy:2008gh} and their final constraints were dependent on the UV cutoff $\Lambda$ and  implicitly assumed that this DM loop contribution to $\delta_1$ is at least comparable to any non-zero tree level value for $\delta_1$ from a direct coupling of the Higgs with $S$.    }
\begin{figure}
\includegraphics[width= 6 in, height =2in]{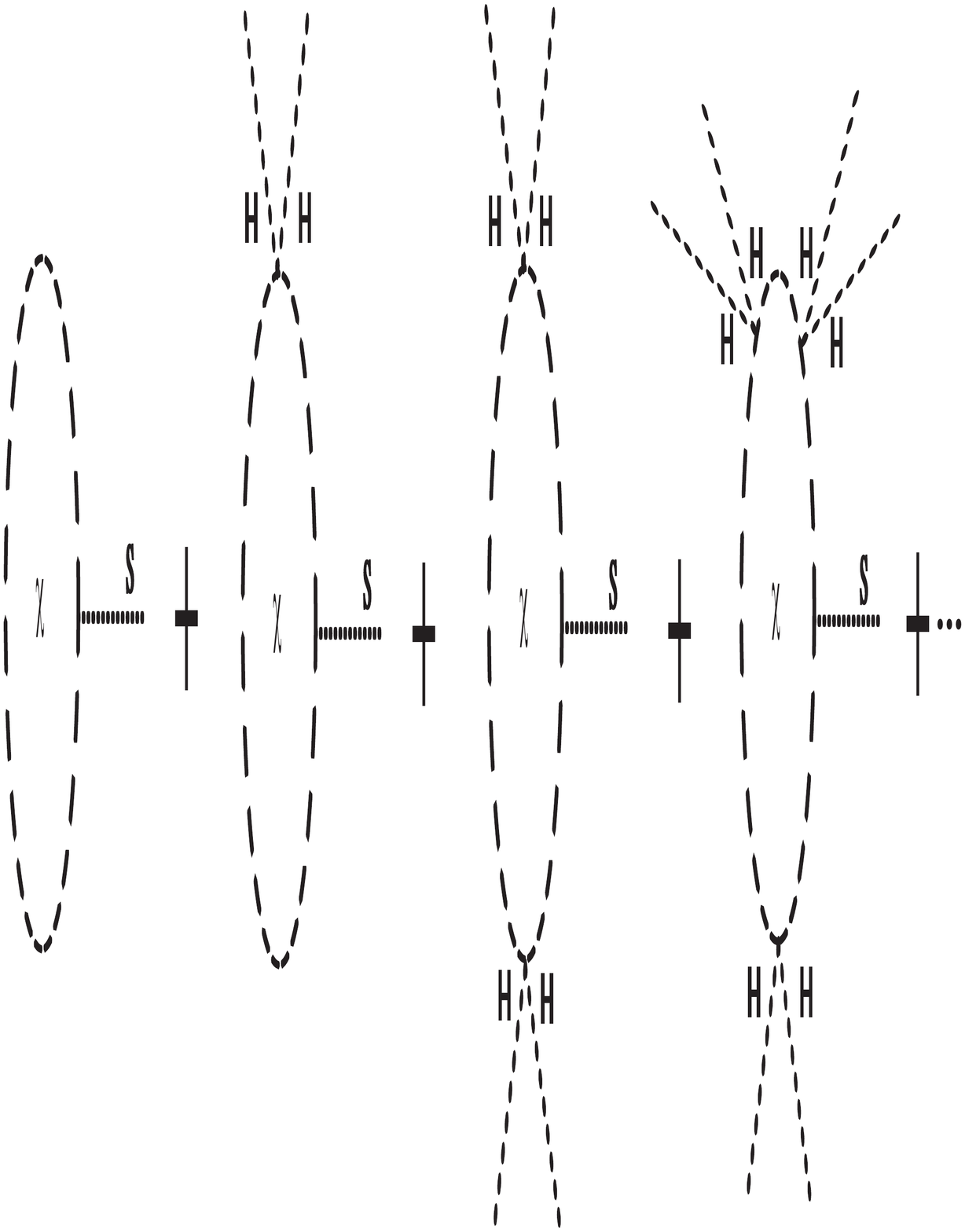}
\caption{One loop diagrams which which contribute to the effective potential $V(H,S)$ with one external $S$ field. After
electroweak symmetry breaking the effective potential contributes to Higgs-ultralight-scalar mixing. The first two diagrams are UV divergent and contribute to the renormalization of the $S$-tadpole and the coupling $\delta_1$ respectively. The remaining diagrams mix into higher dimensional operators and give a finite contribution to Higgs-ultralight-scalar mixing as explained in appendix \ref{appexB}. }
\label{fig:singletmixing}
\end{figure}

In what follows, we will instead allow for fine tuning in both $\delta_1$ since we have already allowed fine tuning for $\mu_S$. We show that assuming $\mu_S^2$ is restricted in parameter space to satisfy $m_\phi < 10^{-25}$eV for any value of
$\mu_{hS}^2$ in Eq.(\ref{mphi-approx}), we can obtain  upper bounds on $\sigma_{\chi N}$ by analyzing finite, one-loop contributions to $\mu_{hS}^2$, from higher dimensional operators after EWSB, and their implications for WEP tests. The other regions in parameter space and their implications will be discussed in section \ref{sectionVII}. To that end, consider the third diagram of  Fig.~\ref{fig:singletmixing}, which generates a contribution to the dimension five operator
\bea
C_2\> (H^\dagger H )  (H^\dagger H) S,
\eea
where in  the coefficient $C_2$  is finite and given by
\OMIT{\footnote{There is room in parameter space where $M_0^2$ can be negative \cite{Burgess:2000yq} while maintaining the requirements for a
positive $M_\chi^2$ in Eq.(\ref{singlet-DM-mass}), a potential bounded from below, and a global minimum at $\langle H\rangle =v/\sqrt{2}, \langle \chi\rangle=0$. However, in the computation of $C_2$ we have assumed that $M_0^2$ is positive for simplicity. For a negative $M_0^2$ one would have to quantize the theory with a shifted $\chi$ field so that one has a well defined propagator for the $\chi$ loop in the computation of $C_2$. We assume a positive $M_0^2$ throughout.}}
\bea
\label{dim5mixing}
C_2 = \kappa \frac{ \>a_2^2}{8 \pi^2} \frac{g_\chi}{ M_0^2}.
\eea
Since this contribution is finite there is no counterterm involved in determing the value of $C_2$. After electroweak symmetry breaking this term will generate a contribution to the off-diagonal elements in the $h$-$S$ mass-squared matrix
\bea
 \mu_{hS}^2=  2 C_2 v^3 + \delta_1 v,
\eea
leading to an $h$-$S$  mixing angle $\theta$
\bea
\tan \theta = \frac{x}{1+ \sqrt{1+x^2}}, \qquad x=\frac{\mu_{hS}^2}{\mu_h^2}=\frac{  2 C_2 v^3 + \delta_1 v}{m_h^2},
\eea
which was defined in Eqs.~(\ref{mixing}) and  (\ref{notation}). Since this mixing implies a coupling of $\phi\approx S$ to matter, the loop-induced coefficient $C_2$ will contribute to the  E\"otv\"os parameters $\eta_{_{DM,E}}$.  Given the dependence of $C_2$ on $a_2$ and the absence of any fine-tuning in this parameter, we obtain an upper bound on $\sigma_{\chi N}$ for non-vanishing $\beta$ as described below.

Before doing so, we observe the contribution to $\mu_{hS}^2$ from  full series of diagrams appearing in Fig.~\ref{fig:singletmixing} (plus the tadpole graph generated by the $\chi^2 S$ interaction) can be  evaluated in a straightforward way as outlined in the appendix ~\ref{appexB}. After renormalization, the result is
\be
\label{eq:resum}
\mu_{hS}^2= v\left[\delta_1(\mu)+ \kappa \frac{g_\chi a_2}{4\pi^2}\left(\ln\frac{M_\chi^2}{\mu^2}-1\right)\right]+ \kappa \frac{g_\chi a_2^2}{4\pi^2}\, \frac{v^3}{M_\chi^2}\ \ \ .
\ee
Apart from an overall constant in the first term and the replacement $M_0\to M_\chi$, this expression is the same as we obtained using the contributions to the $H^\dag H S$ and $(H^\dag H)^2 S$ operators from the second and third diagrams of Fig.~~\ref{fig:singletmixing}. The expression in Eq.~(\ref{eq:resum}) has the advantage that it depends on the tree-level $\chi$ mass after electroweak symmetry breaking rather than on the parameter $M_0$ as in the effective operator analysis. We will henceforth use the finite, second term in Eq.~(\ref{eq:resum}) to derive an upper bound on Higgs exchange contributions to $\sigma_{\chi N}$ .

\OMIT{Following the notation of Eq.~(\ref{Vmass}) and  (\ref{mixing}), the quadratic terms are
\bea
\label{Vmass2}
V_{mass} = \frac{1}{2}  (\mu_h^2 h^2 + \mu_S^2 S^2 +\mu_{hS}^2 h S),
\eea
where 
 Again the lighter mass eigenstate $h_-$ is idenified with the physical long range scalar $\phi$ with mass $m_\phi= m_-$, as defined in Eq.~(\ref{mpm}). }
\OMIT{ 
\begin{figure}
\includegraphics[width= 3 in, height = 3 in]{tripleteta}
\caption{Contribution to $\eta_{DM}$ due to mixing between the light scalar $\phi$ and the Higgs.}
\label{tripleteta}
\end{figure}}
To that end, we  write the mixing angle as
\bea
\label{theta-sigma-beta-1}
\sin \theta  &\approx& \tan \theta \approx x \approx  \kappa \frac{a_2 ^2 }{4 \pi^2} \frac{ g_\chi v^3}{M_\chi ^2 m_h^2} + \frac{\delta_1^\mathrm{ren} v}{m_h^2} =  \kappa \frac{a_2 ^2 }{\pi^{3/2}} \frac{ v^3}{M_P m_h^2} \beta + \frac{\delta_1^\mathrm{ren} v}{m_h^2}\ \ \ ,
\eea
where $\delta_1^\mathrm{ren}$ denotes the quantity in square brackets in Eq.~(\ref{eq:resum}). The mixing angle $\sin \theta$ also characterizes the universal $H$-$S$ mixing contribution to the E\"otv\"os parameters $\eta_{_{E,DM}}$. We now require that the contribution from each term on the RHS of Eq.~(\ref{theta-sigma-beta-1}) to $\eta_{_{E,DM}}$ be no larger than the experimental limits on these parameters. As discussed previously, the different parametric dependence of each term and avoiding slices of parameter space with unnatural cancellations between the two allows us to treat each one separately. Considering only the first term proportional to $a_2^2$, using Eqs.~(\ref{eq:univeta}), and (\ref{eq:earthuniv}) with 
\begin{figure}
\includegraphics[height=2.5in,width=3in]{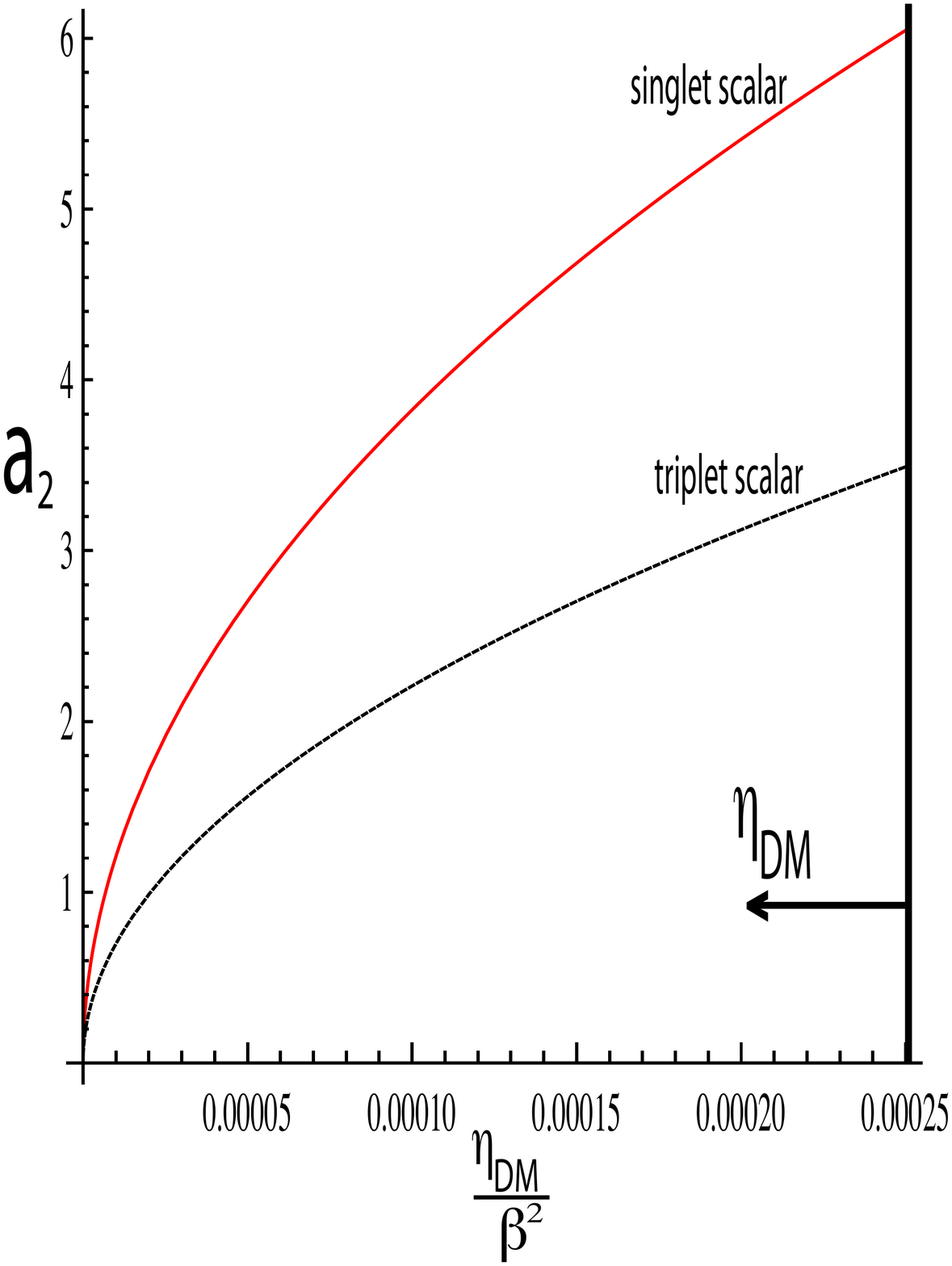} 
\hfill
\includegraphics[height=2.5in,width=3in]{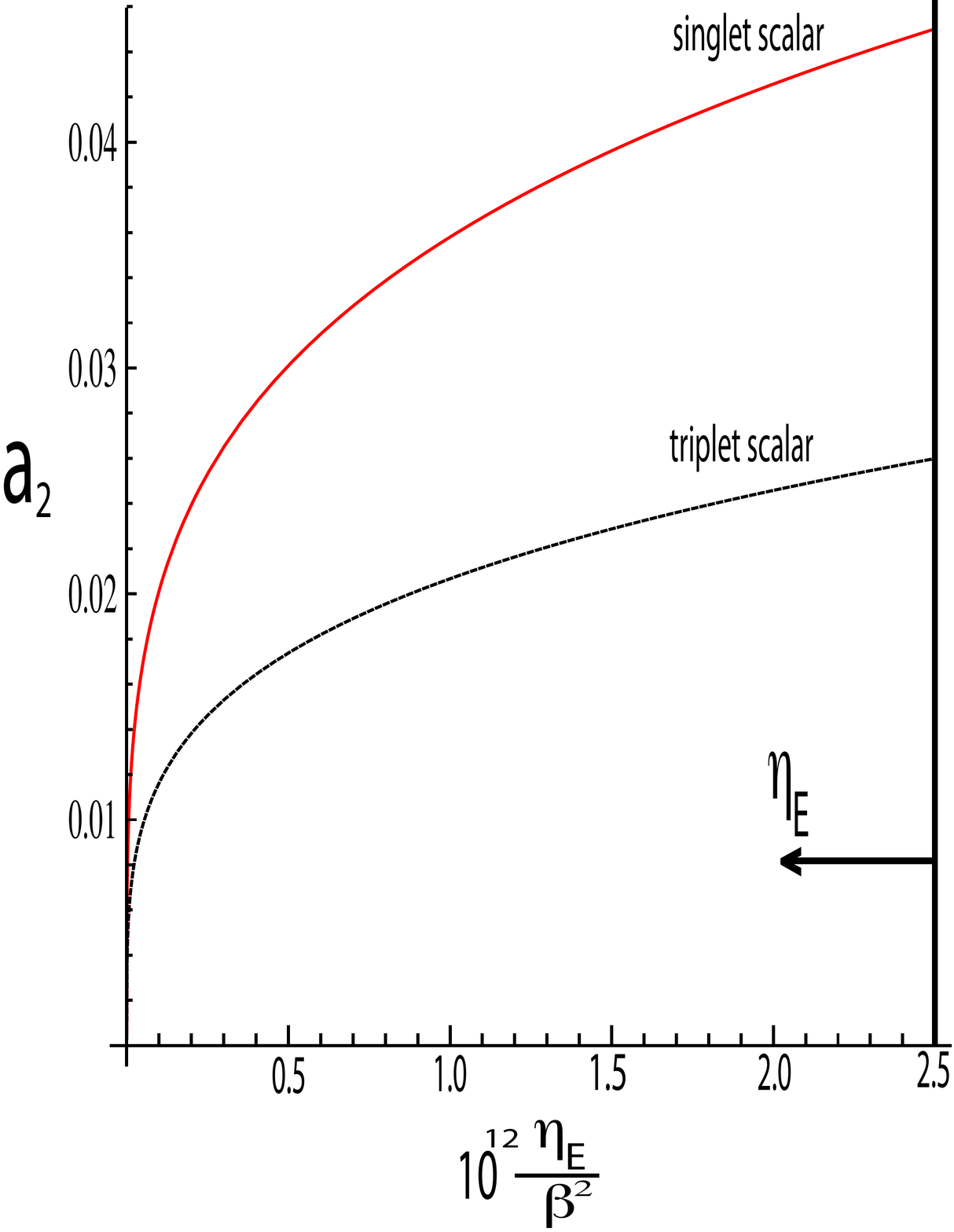}
\includegraphics[height=2.5in,width=3in]{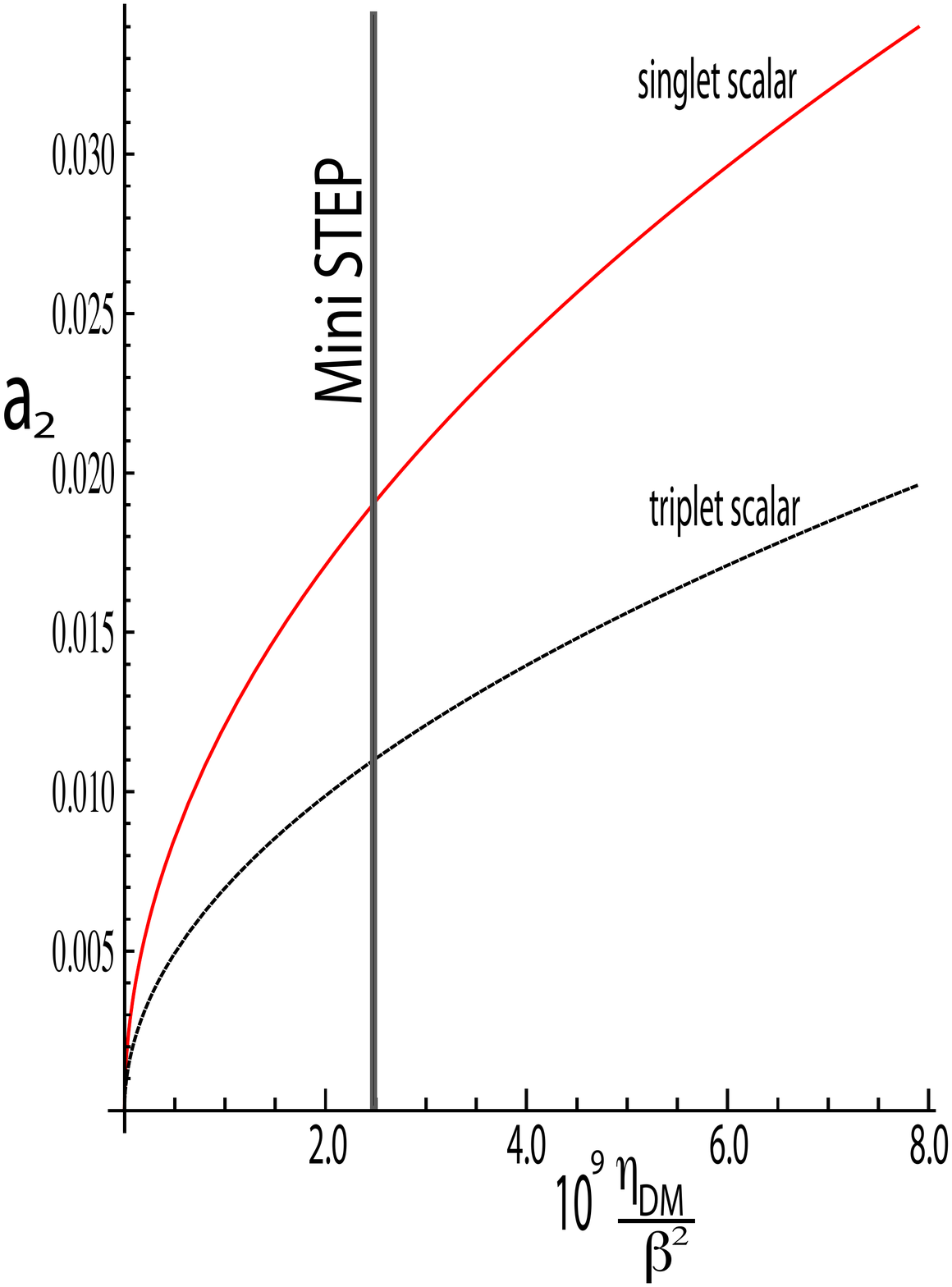}
\hfill
\includegraphics[height=2.5in,width=3in]{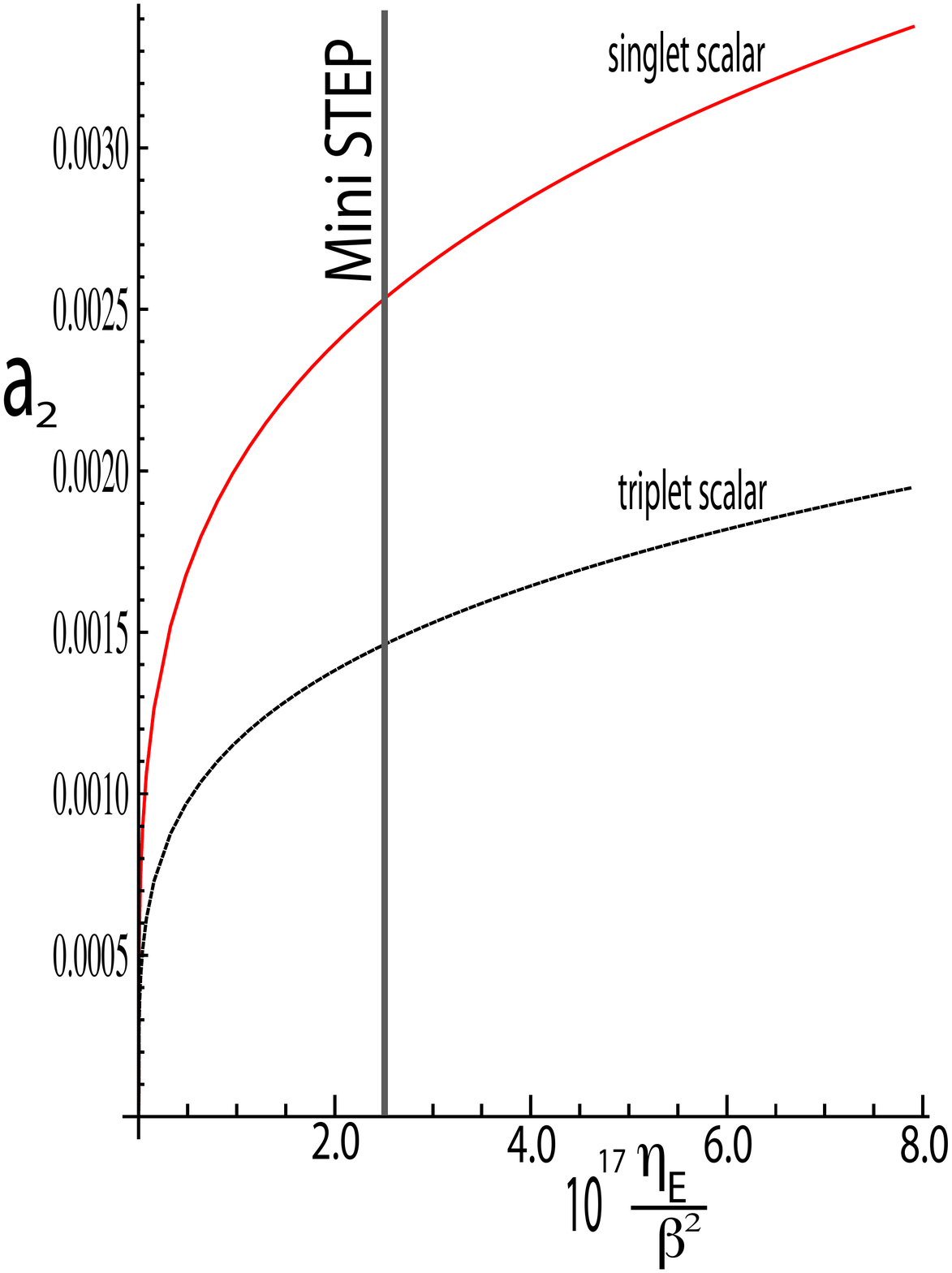}
\caption{Upper bound on $a_2$ in the singlet (red-solid) and real triplet (black-dotted) scalar DM models as a function of $\eta_{_{DM}}/\beta^2$ (left panel) and $\eta_{_E}/\beta^2$ (right panel). We have used $m_h=120$ GeV and assume $\beta=0.2$ to discuss the resulting bounds on $a_2$ from E\"otv\"os experiments. In the top left and right plots, the vertical black lines on the right correspond to the upper bounds $\eta_{_{DM}} < 10^{-5}$ and $\eta_{_E} < 10^{-13}$ respectively. These vertical black lines will move to the left with further improvements in E\"otv\"os experiments as indicated by the left-pointing arrow in each plot.
The bottom left and right plots show the region closer to the expected future bounds, from the MiniSTEP experiment, of $\eta_{_{DM}}<10^{-10}$ and $\eta_{_E}<10^{-18}$ respectively as indicated by the vertical black line in each plot.  We explore the implications of these bounds on $a_2$ from E\"otv\"os experiments for specific observables in sections \ref{section6} and \ref{section7}. }
\label{a2bound}
\end{figure}
\be
{\bar g}\rightarrow -\sin\theta\frac{m_N}{v}
\ee
and expressing $g_\chi$ in terms of $\beta$, we obtain
\bea
\label{eq:etadmmix}
\frac{\eta_{_{DM}}}{\beta^2} & \simeq & a_2^2\, \left(\frac{7\kappa}{18\pi}\right)\, \left(\frac{v}{m_h}\right)^2\, \left |  \> \big  (\frac{Z_1}{A_1}-\frac{Z_2}{A_2 } \big ) \big\{  \frac{m_e}{m_N} + \sum_q  \frac{m_q}{m_N} \> (x_{q,p}  -  \> x_{q,n})\big\}  \> \right |\\
\label{eq:etaemix}
\frac{\eta_{_{E}}}{\beta^2} & \simeq & a_2^4\, \left(\frac{7\kappa}{36\pi^4}\right)\, \left(\frac{v}{m_h}\right)^4\, \left(\frac{v}{m_N}\right)\, \left |  \> \big  (\frac{Z_1}{A_1}-\frac{Z_2}{A_2 } \big ) \big\{  \frac{m_e}{m_N} + \sum_q  \frac{m_q}{m_N} \> (x_{q,p}  -  \> x_{q,n})\big\}  \> \right |\\
\nn
&& \times \left[ \frac{g_h(N_p+N_n)+ (m_e/v) N_e}{(N_p + N_n)+ (m_e/m_N) N_e}\right]\ \ \ .
\eea

Eqs.~(\ref{eq:etadmmix}) and (\ref{eq:etaemix}) imply that for fixed $\beta$ and $m_h$, the experimental bounds on $\eta_{_{E,DM}}$ translate into bounds on $a_2$ as shown in Fig. \ref{a2bound}. The  solid red curves and the dashed black curves correspond to the bounds on $a_2$ in the singlet $\chi$ and real triplet $\chi$ models respectively. For $\beta =0.2$ the current bounds in E\"otv\"os experiments imply $\eta_{_{DM}}/\beta^2 < 2.5 \times 10^{-4}$ and $\eta_{_E}/\beta^2 < 2.5 \times 10^{-12}$ respectively.  The tighter bound from $\eta_{_E}$ implies $a_2 < 0.042 $ and $a_2 < 0.026$ for the singlet and triplet $\chi$ models respectively.  The possible future experiment like MiniSTEP  is expected to improve the the sensitivity of $\eta_{_{DM,E}}$ by five orders of magnitude. From  Eq.(\ref{eq:etaemix}) this would tighten the bound on $a_2$
by an additional  factor of $10^{5/4}$ for a non-zero $\beta$. 

\subsection{WEP Tests and Scalar Singlet DM Relic Density}
For the scalar singlet DM, the DM relic density is determined entirely by the parameter $a_2$ for fixed DM and Higgs masses. This feature can be seen from Fig.~\ref{annihilationDM}, where $a_2$ enters the amplitude for each annihilation diagram and, thus,  determines the DM annihilation rate. In particular, the value of $a_2$ must be sufficiently large so that the DM relic density does not over saturate the observed value. Thus, the requirement that the singlet DM relic density smaller than the total DM relic density, $\Omega_{DM}^S \leq \Omega_{DM}$, leads to a minimum value $a_2^{min}$ for fixed values of  $M_\chi$ and $m_h$. If the WEP bounds on $a_2$ imply that $a_2 < a_2^{min}$, then a dark force of the corresponding strength will be ruled out.

In order to illustrate this interplay, we refer to Fig.~3 of \cite{He:2008qm}. The parameters $a_2$ and $M_\chi$ are the same as $\lambda$ and $m_D$ respectively in the notation of \cite{He:2008qm}. From Fig.~3 of that work, we see that for DM masses in
the  0-50 GeV range, the required value of $a_2$ ranges from $\sim 0.16 - 0.05$ respectively for $m_h=120$ GeV. On the other hand, from the upper curve in the top right graph of Fig.~\ref{a2bound}, we see that $a_2 \lesssim 0.045 $ for $\beta=0.2$ from the current bound of $\eta_{_E} < 10^{-13}$ (vertical black line at right). This WEP constraint $a_2 < 0.045$ implies an over-density of DM in the range $0 <M_{\chi} < 50 $ GeV thus ruling out the possibility of a dark force with $\beta > 0.2$ in this range of parameter space. We give sample points in the parameter space of singlet DM models in Table \ref{relic-table}.

A more detailed analysis can be performed to rule out even smaller values of $\beta$ depending on the DM mass in the scalar singlet model. Future E\"otv\"os experiments with the sensitivity of MiniSTEP \cite{Lockerbie:1998ar} which are expected to reach a sensitivity  of $\eta_{_E} < 10^{-18}$, could require bound of $a_2 \lesssim 0.0025$ for $\beta =0.2$ as seen in the bottom right graph of Fig.~\ref{a2bound}. In this case, one can rule out $\beta < 0.2$ even for DM masses above 60 GeV which require smaller values of $a_2$ in order to get the right relic density. As seen in Fig.~3 of \cite{He:2008qm}, larger values of the Higgs mass typically imply 
much larger values of $a_2$. For example, a Higgs mass of 200 GeV implies a range of $a_2$  of $\sim 0.42 - 0.05$ for the DM mass range of $0-80$ GeV thus ruling out the possibility of $\beta >0.2$ in order to prevent an over-density of DM. Thus, the bound on $a_2$ from WEP constraints is  a powerful probe of a dark force in the scalar singlet DM model. 

\begin{table}
\begin{tabular}{|c | c  | c | c| }
\hline \hline
$a_{2\text{relic}}$ \qquad \qquad & $M_\chi  $(GeV) \qquad \qquad &  Expectation for$\frac{\eta_{_E}}{\beta^2}$ &$\beta =0.2$\\
\hline \hline 
0.15 & 20 & $4 \times 10^{-10}$  & Excluded \\
0.10 & 40 & $7 \times 10^{-11}$  & Excluded\\
0.02 & 100 & $1 \times 10^{-13}$ & Allowed \\
\hline
\end{tabular}
\caption{The first two columns give sample points in the $(a_{2\text{relic}}, M_\chi)$ space of scalar singlet DM models with a Higgs mass of $m_h=120$ GeV. The third column gives an expectation for $\eta_{_E}/\beta^2 $ from Eq.(\ref{eq:etaemix}). The fourth column uses the current bound of $\eta_{_E} < 10^{-13}$ to determine whether a dark force  of $\beta=0.2$ is ruled out. One can equivalently compare the different values of $a_{2\text{relic}}$ with the WEP bound on $a_2$, at $\beta=0.2$, in top right graph of Fig.~\ref{a2bound} at the far right verticle line.     }
\label{relic-table}
\end{table}

For the scalar real triplet DM model, the DM relic density is determined by gauge interactions in addition to the parameter $a_2$. In this case, the WEP bound on $a_2$ shown in Fig.~\ref{a2bound} does not necessarily rule out a dark force since the correct relic density can still be obtained from annihilation diagrams that proceed via gauge interactions that are independent of $a_2$. For example, the bound of $a_2 <0.02$ implied by $\eta_{_E} < 10^{-13}$ for $\beta=0.2$, as shown in the top right graph of Fig.~\ref{a2bound}, implies that the annihilation rate will be dominated by gauge interactions.

\subsection{WEP Tests and DM-Nucleus Cross-Sections}
\label{section6}

The current  bounds on $a_2$ for a non-zero $\beta$ in the dark sector, will also lead to upper bounds on the Higgs exchange contributions to the direct detection cross-section.
 From Eq. (\ref{singlet-cross section}),  the parameter $a_2$  can be written in terms of the tree level cross-section $\sigma_{\chi N}$, which proceeds via a t-channel higgs exchange, as
\bea
\label{relations}
 a_2^2 = \left [\frac{\pi (M_\chi + m_N)^2 m_h^4}{g_h^2 v^2 m_N^2} \right ] \sigma_{\chi N}\Bigg |_{\text{Higgs exch}}.\eea
Substituting into Eqs.~(\ref{eq:etadmmix}) and (\ref{eq:etaemix}), defining the quantities
\bea
\label{eq;fdef}
F & \equiv &  \left |  \> \big  (\frac{Z_1}{A_1}-\frac{Z_2}{A_2 } \big ) \big\{  \frac{m_e}{m_N} + \sum_q  \frac{m_q}{m_N} \> (x_{q,p}  -  \> x_{q,n})\big\}  \> \right |,\\
\label{eq:edef}
E & \equiv & \left[ \frac{g_h(N_p+N_n)+ (m_e/v) N_e}{(N_p + N_n)+ (m_e/m_N) N_e}\right] \approx g_h,
\eea
we obtain the following relations between $(M_\chi+m_N)^2 \sigma_{\chi N}$ and the Higgs-exchange contributions to the DM-nucleus cross section:
\bea
\label{eq:sigmadm}
\left(M_\chi+ m_N\right)^2\, \sigma_{\chi N}\Big\vert_\mathrm{Higgs\, exch} & = & \left(\frac{18}{7\kappa}\right)\, g_h^2 \left(\frac{m_N}{m_h}\right)^2\, \frac{1}{F}\, \frac{\eta_{_{DM}}}{\beta^2}\\
\left(M_\chi+ m_N\right)^2\, \sigma_{\chi N}\Big\vert_\mathrm{Higgs\, exch} & = & \left(\frac{6\pi}{\sqrt{7\kappa}}\right)\, g_h^2 \left(\frac{m_N}{m_h}\right)^2\, \left[\left(\frac{m_N}{v}\right) \frac{1}{FE}\right]^{1/2}\, \frac{\eta_{_{E}}^{1/2}}{\beta}\ \ \ .
\eea

The experimental limits on the  E\"otv\"os parameters, together with the foregoing expressions, lead to bounds on the Higgs exchange contributions to the DM-nucleus cross sections. These bounds can be brought into the numerically convenient form as a function of  $\eta_{_{DM}}/\beta^2$ for the singlet $\chi$ as
\be
\label{num-DM-S}
 \left [\frac{M_\chi + m_N}{100 \>\text{GeV}}\right ]^2 \frac{\sigma_{\chi N}}{1\> \text{pb}}  \\
< (1.1 \times 10^{4}) g_h^2 \left [  \frac{100 \>\text{GeV}}{m_h} \right ]^2 \Big | \frac{Z_1}{A_1} - \frac{Z_2}{A_2}\Big |^{-1}\frac{\eta_{_{DM}}}{\beta^2 },
\ee
and for the triplet $\chi $ as
\be
\label{num-DM-T}
 \left [\frac{M_\chi + m_N}{2 \>\text{TeV}}\right ]^2 \frac{\sigma_{\chi N}}{1\> \text{pb}}  \\
< 9.2\> g_h^2 \left [  \frac{100 \>\text{GeV}}{m_h} \right ]^2 \Big | \frac{Z_1}{A_1} - \frac{Z_2}{A_2}\Big |^{-1}\frac{\eta_{_{DM}}}{\beta^2 }.
\ee
\begin{figure}
\includegraphics[width=3in, height=2.5in]{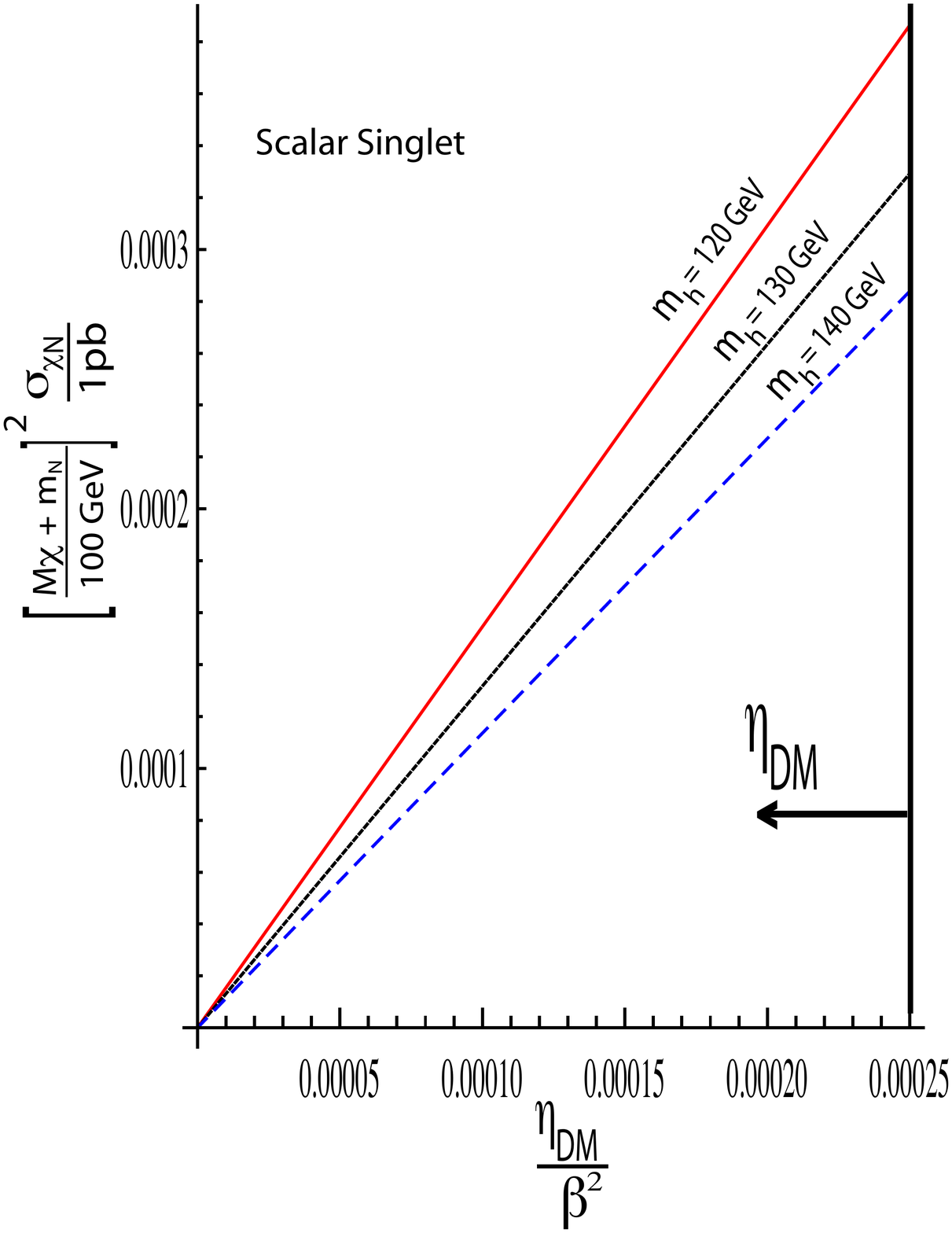}
\hfill
\includegraphics[width=3in, height=2.5in]{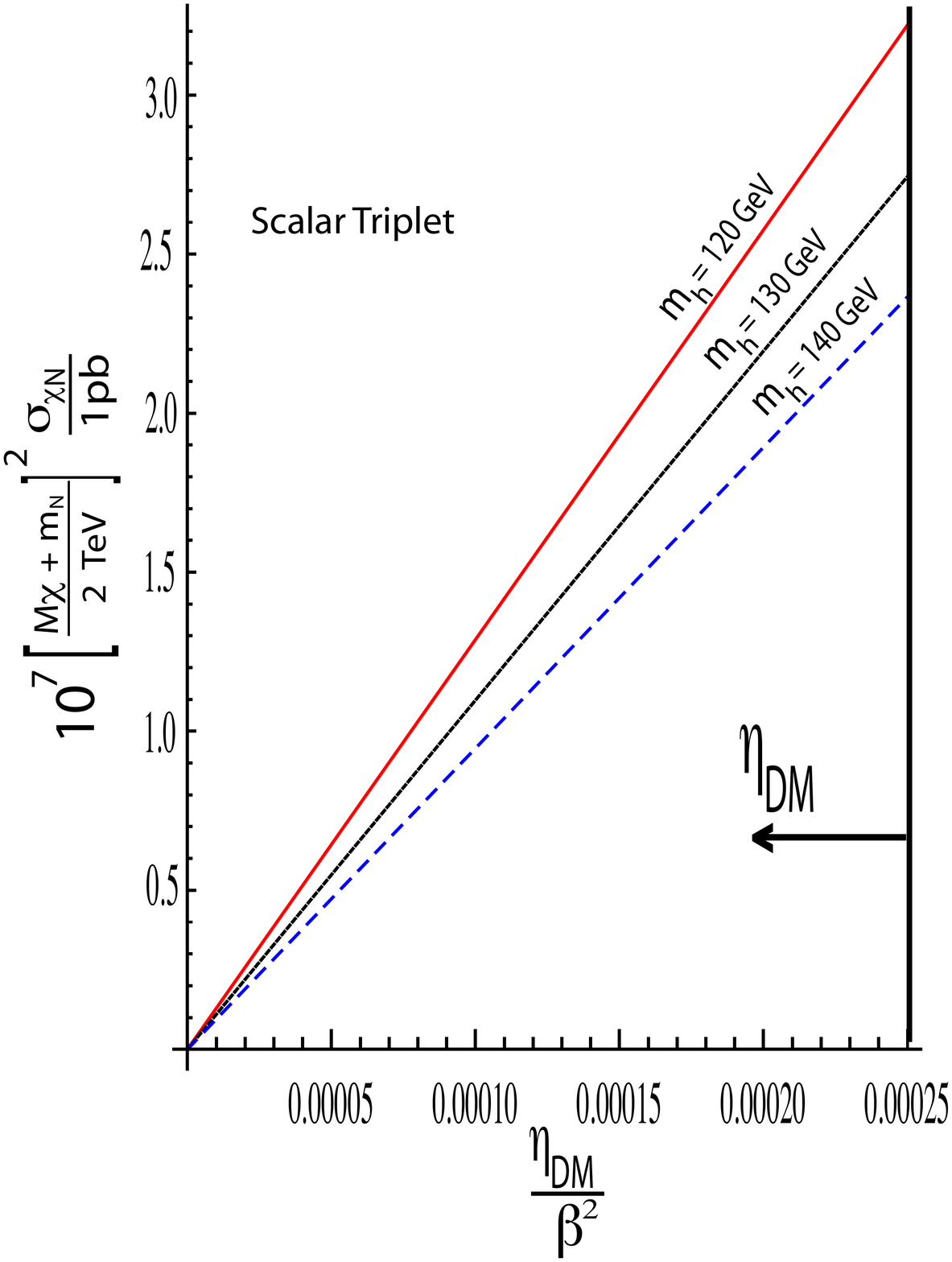}
\includegraphics[width=3in, height=2.5in]{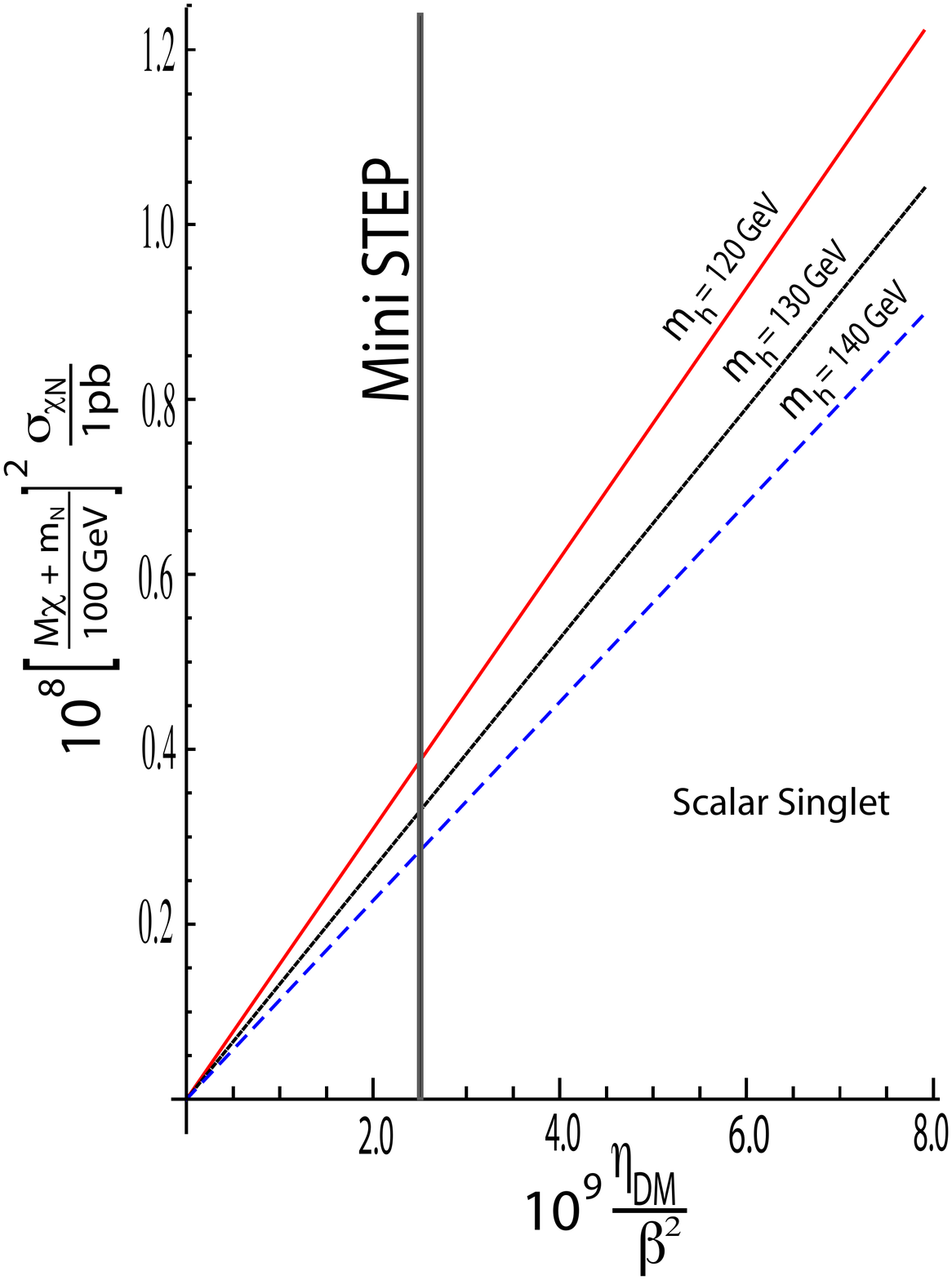}
\hfill
\includegraphics[width=3in, height=2.5in]{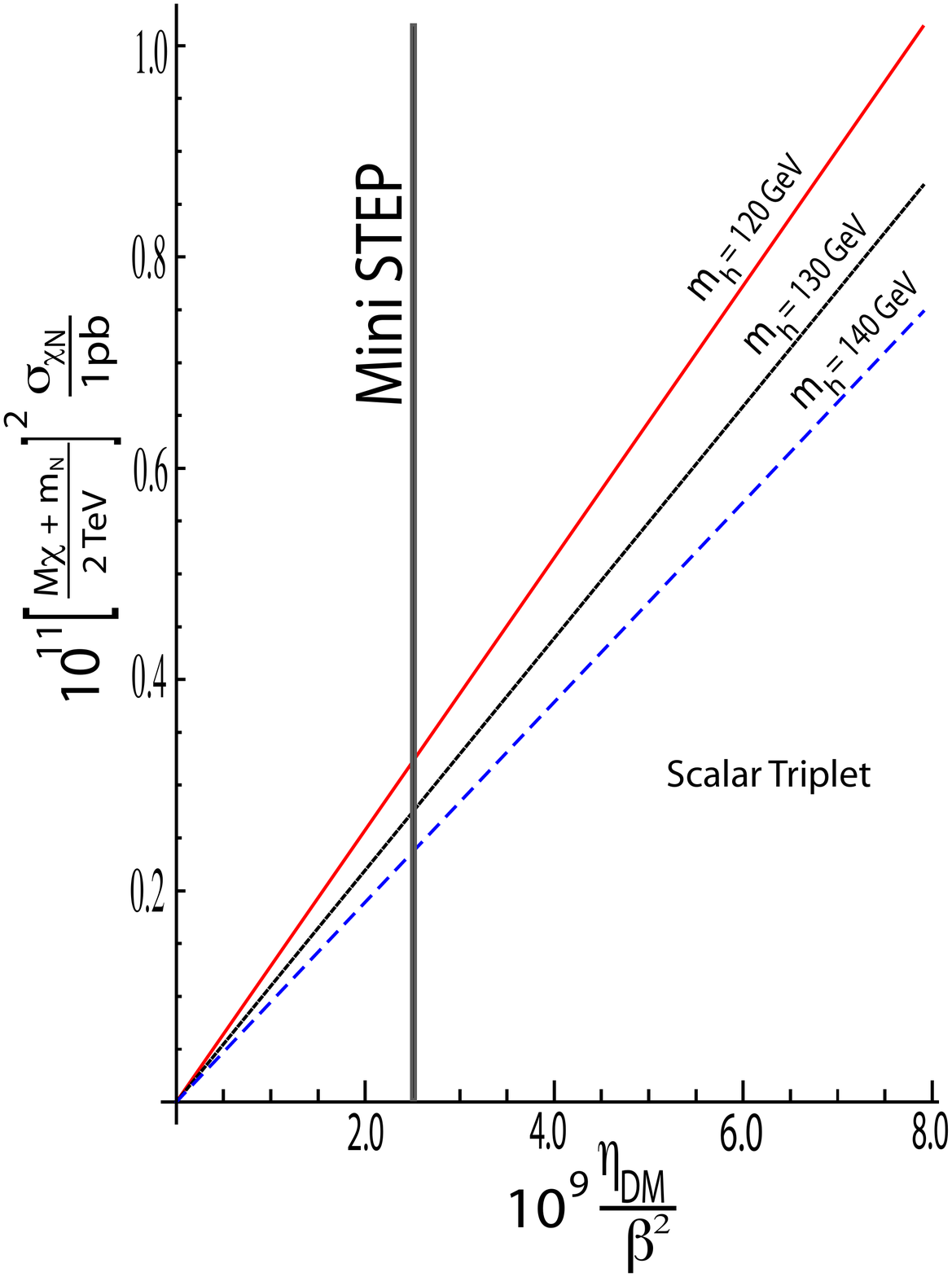}
\caption{Upper bounds on the Higgs exchange tree level direct detection cross section of scalar singlet(left panel) and real triplet(right panel) DM implied by a dark force as a function of $\eta_{_{DM}}/\beta^2$. To be specific, we assume $\beta=0.2$ and discuss the implied bounds. In the top left and top right plots the vertical black line on the right corresponds to $\eta_{_{DM}}<10^{-5}$. In the bottom left and bottom right plots, the vertical black line corresponds to the expected future sensitivity of $\eta_{_{DM}}<10^{-10}$. In all plots, the three lines from to bottom correspond to the bounds for the Higgs masses of 120, 130, and 140 GeV respectively. The size of these bounds compared to current and future sensitivities for direct detection experiments is discussed in the text.}
\label{singletWEP}
\end{figure}
Similarly, the bounds as a function of $\eta_{_E}/\beta^2$ can be brought into the numerically convenient form  for the singlet $\chi$ as
\bea
\label{num-E-S}
\left [\frac{M_\chi + m_N}{100 \>\text{GeV}}\right ]^2 \left[ \frac{\sigma_{\chi N}}{1 \>\text{pb}} \right] < 8.4\times 10^{2} \>  g_h^2\left [ \frac{100 \> \text{GeV}}{m_h}\right ]^2  \left | \frac{Z_1}{A_1} - \frac{Z_2}{A_2}\right |^{-1/2} \frac{\eta_{_{E}}^{1/2}}{\beta },
\eea
and for the real triplet $\chi$ as
\bea
\label{num-E-T}
\left [\frac{M_\chi + m_N}{2 \>\text{TeV}}\right ]^2 \left[ \frac{\sigma_{\chi N}}{1 \>\text{pb}} \right] < 0.7 \>  g_h^2\left [ \frac{100 \> \text{GeV}}{m_h}\right ]^2  \left | \frac{Z_1}{A_1} - \frac{Z_2}{A_2}\right |^{-1/2} \frac{\eta_{_{E}}^{1/2}}{\beta },
\eea
where  $ \left | \frac{Z_1}{A_1} - \frac{Z_2}{A_2}\right |\simeq1/72$ for Beryllium and Titanium samples in E\"otv\"os experiments.
The upper bounds for the Higgs exchange tcontribution to the direct detection cross section of the singlet (left panel) and real triplet (right panel)   DM in the presence of a dark force, as determined by  Eqs.(\ref{num-DM-S}), (\ref{num-DM-T}), (\ref{num-E-S}), and (\ref{num-E-T}), are shown in Figs.~\ref{singletWEP} and \ref{singletWIMPetaE} as a function of $\eta_{_{DM}}/\beta^2$ and 
$\eta_{_E}/\beta^2$ respectively. In each graph we show three sample curves  corresponding to Higgs mass choices of $m_h=120$, 130, and 140 GeV as indicated.
\begin{figure}
\includegraphics[width=3in, height=2.5in]{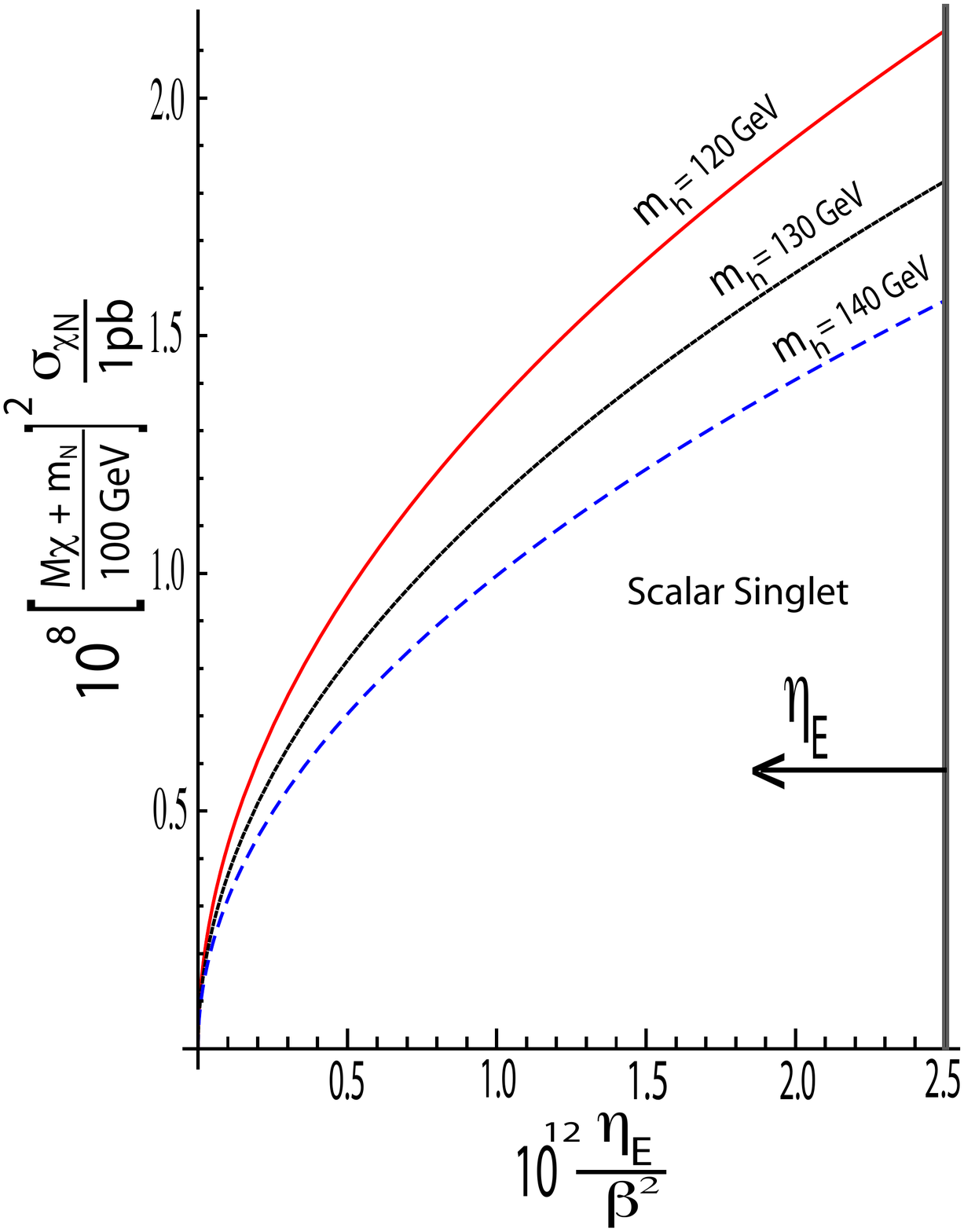}
\hfill
\includegraphics[width=3in, height=2.5in]{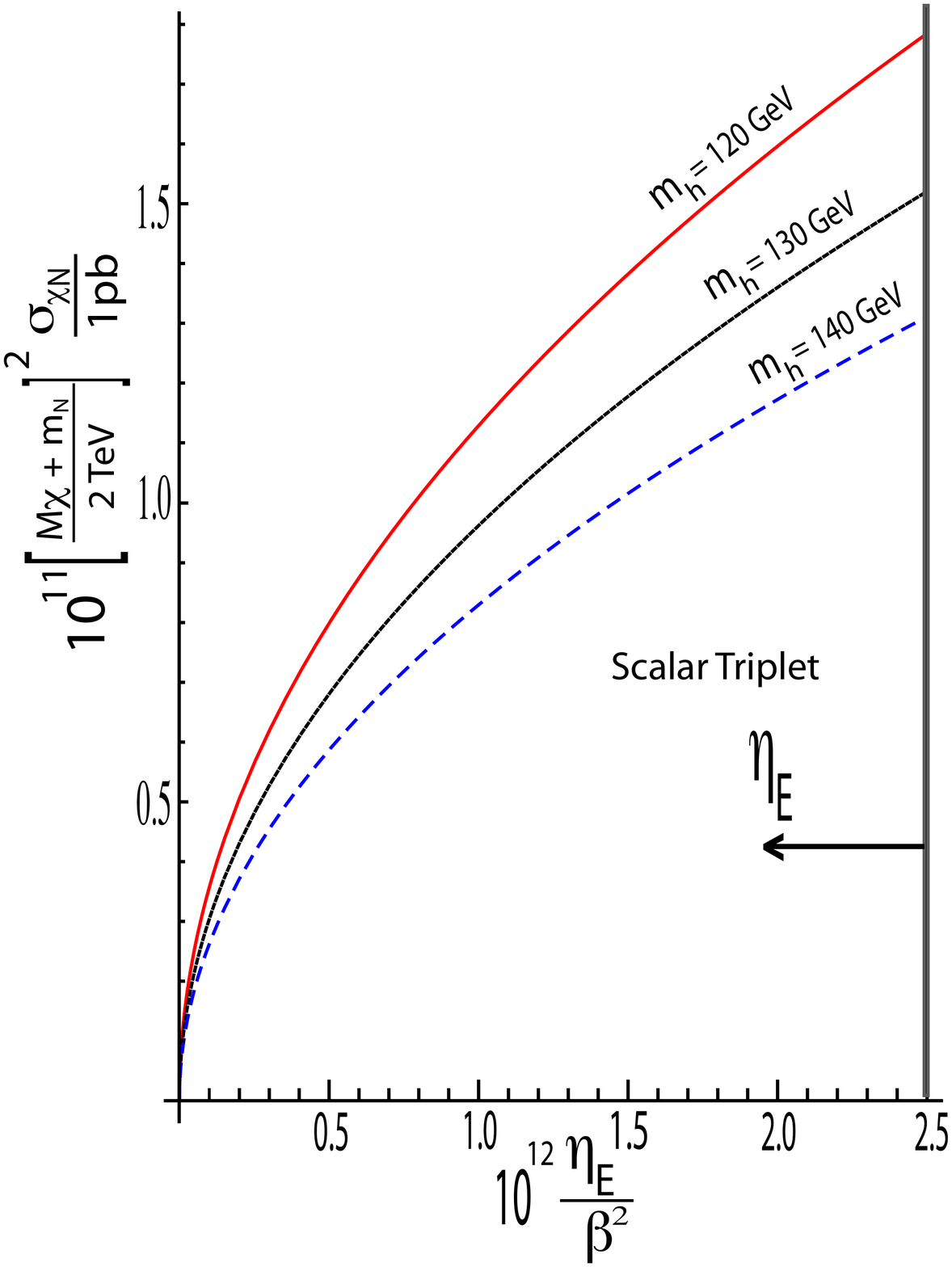}
\includegraphics[width=3in, height=2.5in]{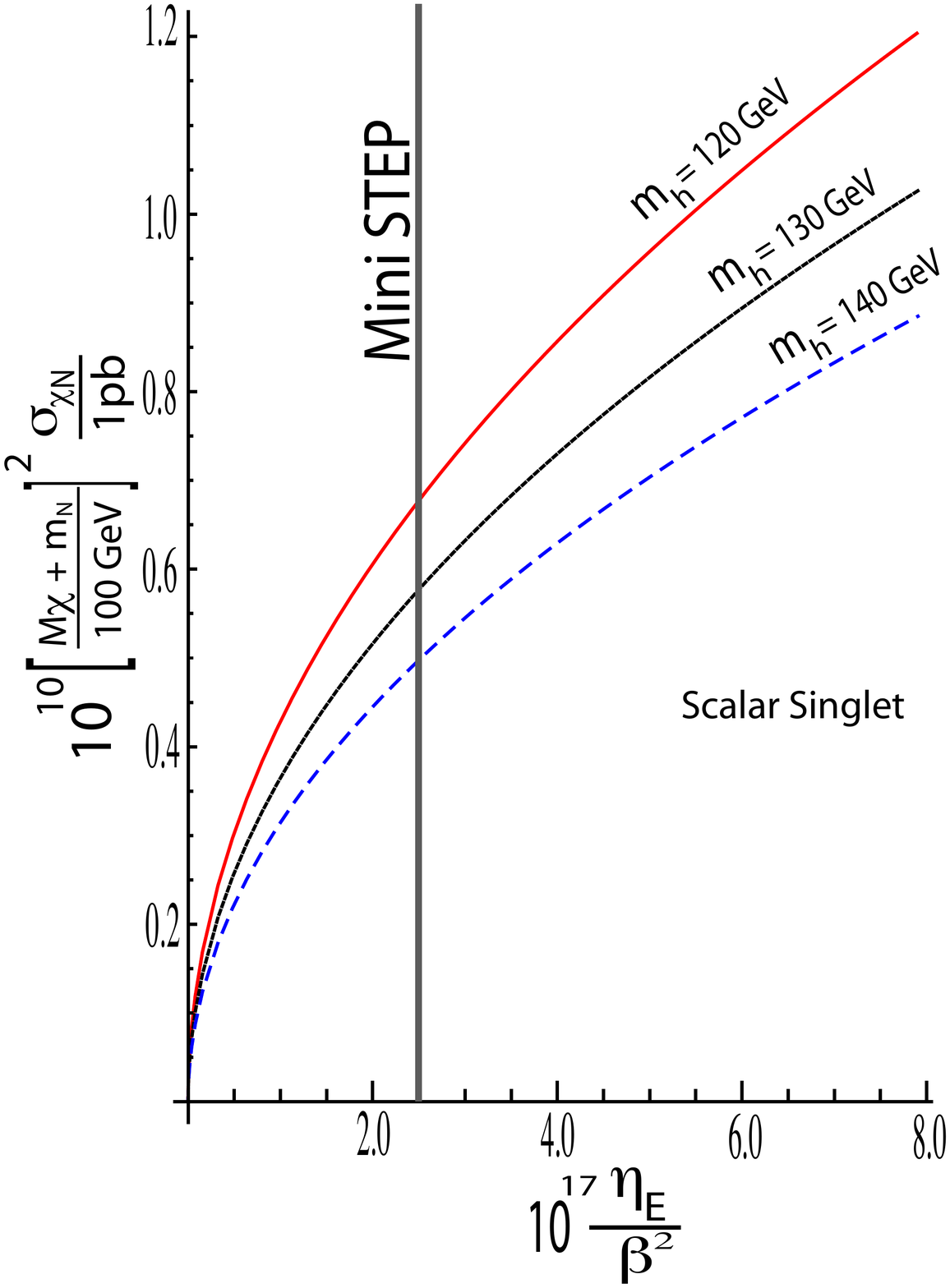}
\hfill
\includegraphics[width=3in, height=2.5in]{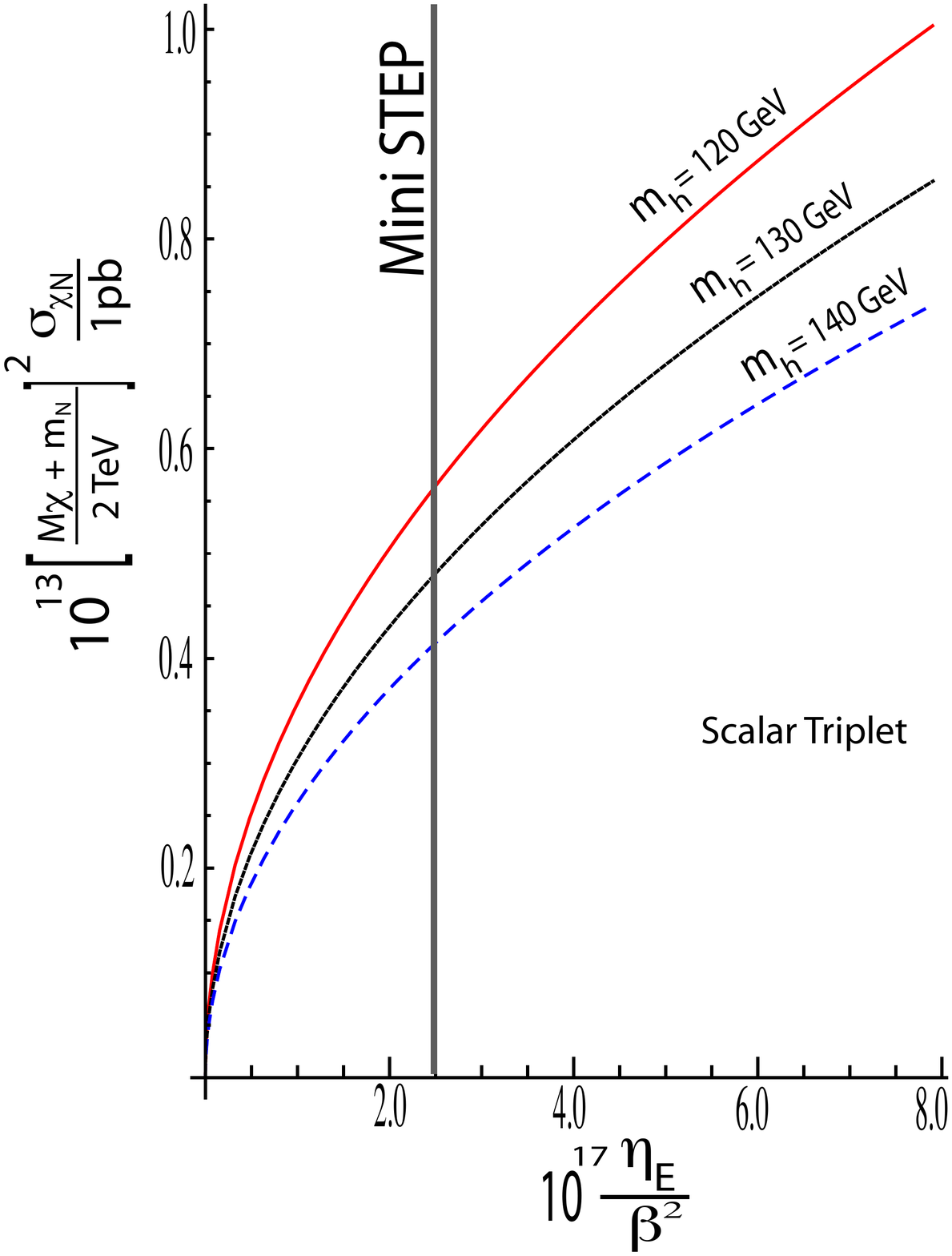}
\caption{Upper bounds on the Higgs exchange tree level direct detection cross-section of scalar singlet(left panel) and real triplet(right panel) DM implied by a dark force as a function of $\eta_{_E}/\beta^2$. To be specific, we assume $\beta=0.2$ and discuss the implied bounds. In the top left and top right plots, the vertical black lines on the right correspond to the current bound of $\eta_{_{E}}< 10^{-13}$. The the bottom left and bottom right plots, the vertical black line corresponds to the expected future sensitivity of $\eta_{_E} =10^{-18}$. In all plots, the three curves from top to bottom, correspond to the Higgs masses of 120, 130, and 140 GeV respectively. The size of these bounds with current and future sensitivities of direct detection experiments is discussed in the text.
}
\label{singletWIMPetaE}
\end{figure}
In the top row of Figs.~\ref{singletWEP} and \ref{singletWIMPetaE}, the vertical black lines on the right correspond to
the current WEP bounds of $\eta_{_{DM}}< 10^{-5}$ and $\eta_{_E}<10^{-13}$ for our benchmark value of $\beta=0.2$ which we use throughout this discussion.
In the bottom row of Figs.~\ref{singletWEP} and \ref{singletWIMPetaE}, the vertical black lines labeled `MiniSTEP' correspond to the expected  sensitivity of $\eta_{_{DM}}\sim10^{-10}$ and $\eta_{_E} \sim10^{-18}$ from a possible future experiment like MiniSTEP~\cite{Lockerbie:1998ar}.
We see that the bounds on the Higgs exchange contribution to the DM-nucleus cross-sections are typically much stronger from WEP violation constraints on $\eta_{_E}$ compared to those on $\eta_{_{DM}}$. However, since the DM-nucleus cross-section bounds depend linearly on $\eta_{_{DM}}$ and on the square root of $\eta_{_E}$, with enough improvement the bound from $\eta_{_{DM}}$ could become stronger. In the following discussion, we focus only on the direct detection bounds from $\eta_{_E}$ shown in Fig.~\ref{singletWIMPetaE}.

For the scalar singlet DM, the DM-nucleus cross-section bound from WEP tests does not yield any more information than the bound on $a_2$ which has already been discussed. This is due to the fact that $a_2$ determines the DM matter relic density entirely for fixed DM and Higgs masses. If the WEP violation bound on $a_2$  is too strong, the resulting DM relic density will be too large over-closing the universe and thus ruling out the dark force. The bound on the DM-nucleus cross-section resulting from the corresponding WEP violation bound on $a_2$, is thus not useful since it is already ruled out.  

However, the bound on the scalar singlet DM-nucleus cross-section can be useful in constraining the size of $\beta $ in a multicomponent DM scenario where the scalar singlet is only a fraction of the DM. For larger values of $\beta$, as already discussed, the bounds on $a_2$ from WEP tests are too strong leading to an over-closed universe. For smaller values of $\beta$ the bound on $a_2$ becomes weaker as seen from Eqs. (\ref{eq:etadmmix}) and (\ref{eq:etaemix}). For small enough values of $\beta$, the upper bound on $a_2$ would be consistent with an under-relic-density of the singlet scalar. A multicomponent DM scenario can also have $a_2$ consistent with an under-relic-density for the scalar singlet and in this case the WEP violation constraints on $a_2$ can lead to interesting bounds on the DM-nucleus cross-section. 

For the scalar real triplet DM, the DM relic density is determined by $a_2$ and gauge interactions in general. However, the tree level DM-nucleus cross-section proceeds only via a t-channel Higgs exchange and its size is determined by $a_2$. We point out that
the bound in Fig.~\ref{singletWIMPetaE} constrains the tree level Higgs exchange diagram but not the  one loop diagrams, which proceed via gauge interactions and the Higgs coupling to the nucleus and is independent of $a_2$. Thus, if the observed DM-nucleus cross-section is of the size explained by this one loop diagram a dark force cannot be ruled out.

In Table~\ref{table} we show the sensitivities of current and future DM detection experiments, taken from Table I of \cite{Barger:2007im}. 
\begin{table}
\begin{tabular}{|c||c|c|c|}
\hline
$\text{M}_\chi = 50\>\text{GeV}   \>\> $ & $\>\>\text{Experiment} \>\>$ & $\>\>\text{Sensitivity}\>\>$ & $\>\>\text{Sensitivity}\>\>$ \\
 && $\>\> \sigma_{\chi N} \> (\text{pb})\>\>$ & $\>\> \left [ \frac{M_\chi + m_N}{100 \>\text{GeV}}\right]^2 \left [\frac{\sigma_{\chi N}}{1 \>\text{pb}}\right ]\>\>$ \\ 
 &&&\\
\hline
\hline
 & CDMS~\cite{Akerib:2005za} & $\>\>1.6 \times 10^{-7}\>\>$ & $\>\>4.1 \times 10^{-8} \>\> $ \\
  & XENON10~\cite{Angle:2007uj} & $\>\>4.5 \times 10^{-8}\>\>$ & $\>\>1.2 \times 10^{-8} \>\> $ \\
& CDMS (2007~\cite{Ni:2006fh}) & $\>\>1 \times 10^{-8}\>\>$ & $\>\>3 \times 10^{-9} \>\> $ \\
& WARP (140 \text{kg})~\cite{Brunetti:2004cf}& $\>\>3 \times 10^{-8}\>\>$ & $\>\>8\times 10^{-9} \>\> $ \\
& SuperCDMS (Phase A)~\cite{Akerib:2006rr} & $\>\>1 \times 10^{-9}\>\>$ & $\>\>3 \times 10^{-10} \>\> $ \\
& WARP (1 ton)~\cite{Aprile:2002ef} & $\>\>2 \times 10^{-10}\>\>$ & $\>\>5  \times 10^{-10} \>\> $ \\
\hline
\end{tabular}
\caption{Sensitivities for DM direct detection cross sections in different experiments. These sensitivities are for 50 GeV DM corresponding to the most sensitive mass window. We see that the XENON10, CDMS (2007), WARP (140 kg), SuperCDMS, and WARP (1 ton) experiments have enough sensitivity to probe the bounds on the direct detection cross sections in Fig.~\ref{singletWIMPetaE} for singlet DM coupled to a WEP violating force. }
\label{table}
\end{table}
We see from Fig.~\ref{singletWIMPetaE} that it will be difficult for current and future direct detection experiments to probe the upper bound on the DM-nucleus-Higgs-exchange  cross-sections, for scalar triplet DM, for values of $\beta$ that are astrophysically interesting allowing one to rule out this possibility. One would need a significant deviation from the expected cross-section from one loop gauge diagram, indicating a large value of $a_2$, to rule out a significant dark force.
For smaller enough values of $\beta$, the DM-nucleus cross-section bounds should be within reach of current or future experiments. The bounds we have derived on the DM-nucleus cross-sections are much weaker than those in \cite{Bovy:2008gh} since  our analysis constrains higher dimension operators with finite coefficients while the work of \cite{Bovy:2008gh} had to rely on naturalness arguments to constrain renormalizable couplings. As we have discussed earlier, since the ultralight scalar mass is itself fine tuned we have avoided using naturalness arguments.

\subsection{WEP Tests and Higgs Decays}
\label{section7}

We have shown in the last section that WEP constraints lead to upper bounds on the tree level DM-nucleus cross-sections for the scalar singlet and real triplet $\chi$ models. However, if the dark sector is made up of a rich spectrum of DM particles of different species, direct detection of any species that makes up only a tiny fraction of the relic density becomes difficult.  One example of such a DM species is the neutral component of the real triplet scalar $\chi$ with a mass far below a TeV. For masses below 500 GeV, the triplet DM will make up less than 10\% of the relic density~\cite{Cirelli:2005uq}. The astrophysical effects of a dark force experienced by such a species would be too small to be detected. In this section, we show that when direct detection experiments or astrophysical observations fail to constrain dark forces, collider signals might still harbor information on dark forces.
\begin{figure}
\includegraphics[width=4.3in, height=2 in]{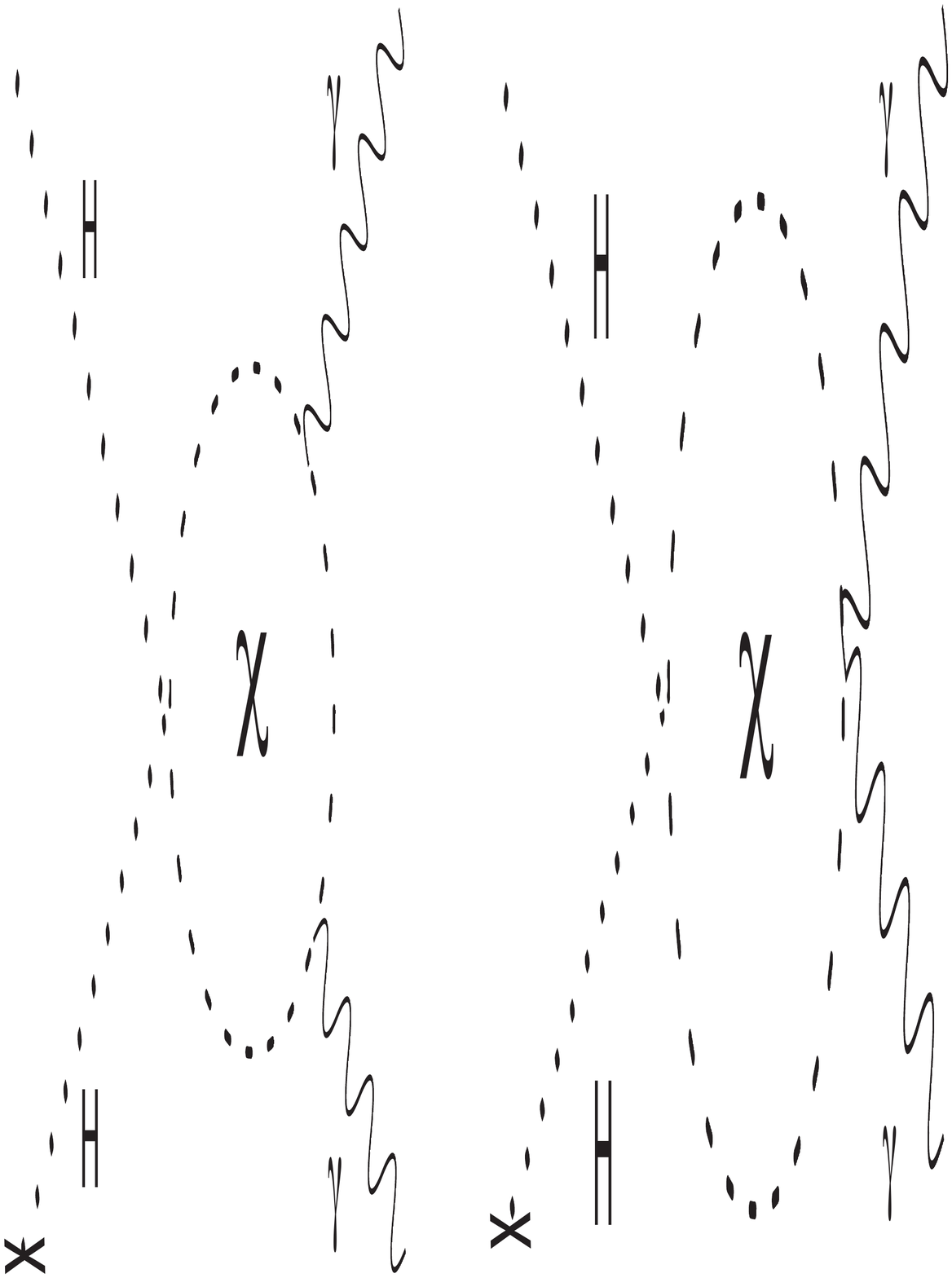}
\caption{Contributions to the $h\to \gamma \gamma $ rate from virtual $\chi^{\pm}$ loops. }
\label{tripletphotons}
\end{figure}
Fig.~2 of \cite{FileviezPerez:2008bj} shows the size of the shift in the $h\to \gamma \gamma $ rate for typical values of the parameter $a_2$. We have reproduced this figure as shown on the left in Fig.~\ref{a2shiftrate}. We plot the quantity
\bea
\delta(\%) \equiv 100\times \frac{\Gamma (h\to \gamma \gamma) - \Gamma^{SM} (h\to \gamma \gamma)}{\Gamma^{SM}  (h\to \gamma \gamma)},
\eea

For specificity we focus on the real scalar triplet $\chi$ discussed in the last section, but with a mass less than 200 GeV, and examine the implications of a dark force on collider signals. The analysis of
\cite{FileviezPerez:2008bj} showed that one potential signature of the scalar triplet would be a  modification of the $h\to \gamma \gamma$ decay rate due to the virtual charged components of the $\chi$ triplet traversing the loop  shown in Fig.~\ref{tripletphotons}. In the rest of this section we focus on this channel. For a heavy Higgs, a similar analysis can be done for $h\to \gamma Z, ZZ, W^+ W^-$.  As already discussed, WEP constraints imply an upper bound on the parameter $a_2$ which determines the size of the contribution of Fig.~\ref{tripletphotons} to $h\to \gamma \gamma$. The WEP bound on $a_2$ translates into a bound on $\delta(\%)$ which is shown in the right panel of Fig.~\ref{a2shiftrate} for different values of  $M_\chi$. 
Comparing the left plot of Fig.~\ref{a2shiftrate} with the top right graph in Fig.~\ref{a2bound}, we see that the current bounds on $a_2$ from $\eta_{_E}$ for a non-zero $\beta$ can give non-trivial bounds on $\delta(\%)$ that can be tested in colliders. The right plot in Fig.~\ref{a2shiftrate} gives the upper bound on $\left | \delta (\%) \right |$ as a function of $\eta_{E}/\beta^2$. For $\beta =0.2$, we have the  bound  $\eta_{_E}/\beta^2 < 2.5\times 10^{12}$ coming from the current bound of $\eta_{_E} < 10^{-13}$. We see that the bound on $\delta (\%)$ for a dark force of $\beta =0.2$ is well below one percent. Thus, any observed shift  in $h\to \gamma \gamma$, that cannot be explained by  physics observed at colliders and unrelated to $\chi$, requires a significant contribution from the $\chi $ loop  implying a value for  $\beta$ much smaller than 0.2. If a non-zero value of $a_2$ is extracted from a study of $h\to \gamma \gamma$ decays, one can estimate
the size of $\eta_{_{DM,E}}/\beta^2$ from Eqs. (\ref{eq:etadmmix}) and (\ref{eq:etaemix}) respectively and use the current bounds on $\eta_{_{DM,E}}$ to constrain the size of $\beta$. For example,  using $m_h=120$ GeV and the current bound of $\eta_{_E} < 10^{-13}$, the non-zero values of $a_2 =\sqrt{\pi},1.0, 0.5$ would imply that $\beta < 7\times 10^{-5}, 2\times 10^{-4}, 9\times 10^{-4}$ respectively.

\begin{figure}
\includegraphics[height=3in, width=3in]{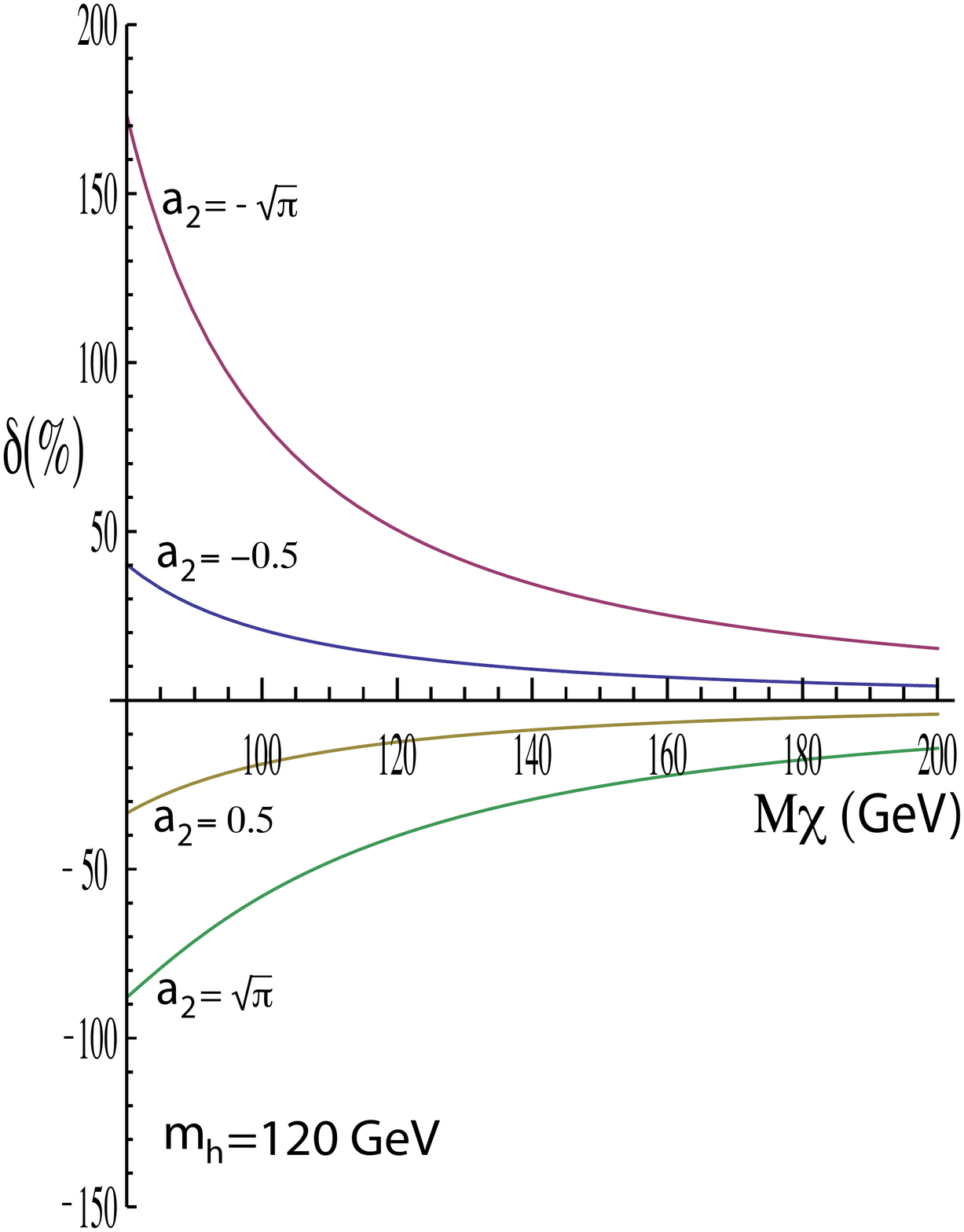}
\hfill
\includegraphics[height=3in, width=3in]{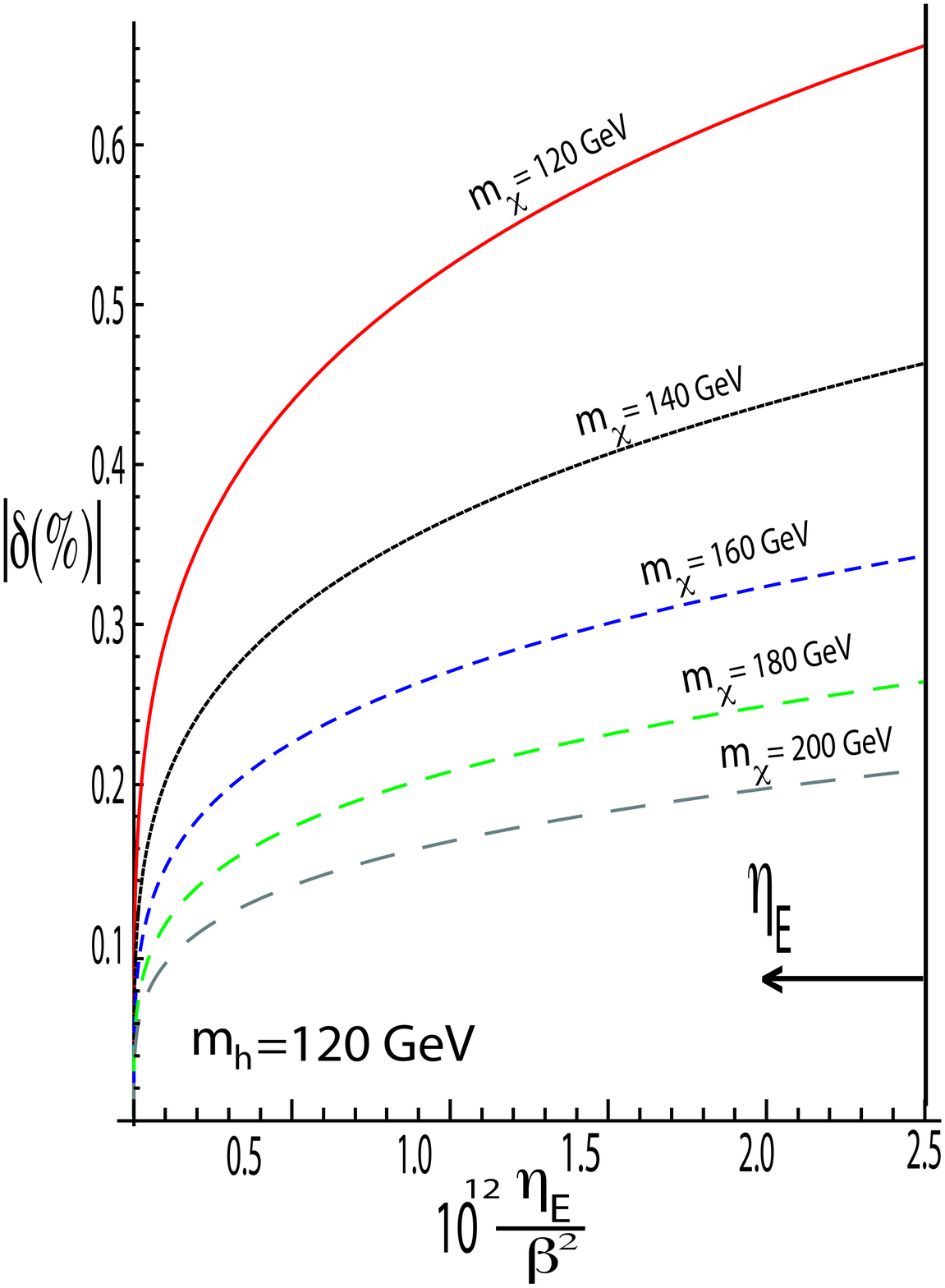}
\caption{The left plot shows the deviation of the $h\to \gamma \gamma $ rate compared to the SM prediction for typical values of the parameter $a_2$ as a function of the triplet mass $M_\chi$. The right plot shows the magnitude of the allowed shift in the $h\to \gamma \gamma$ rate in the presence of a dark force. The bound on this allowed shift arises due to the bound on $a_2$ from WEP violation constraints on $\eta_E$ as seen in the top right plot of Fig.~\ref{a2bound}. The typical values of $a_2$ in the left plot above, which lead to sizable deviations in the $h\to \gamma \gamma$ rate, are too big to be compatible with WEP violation constraints. }
\label{a2shiftrate}
\end{figure}

\OMIT{One may also look at the implications of a dark force for other collider processes, sensitive to $a_2$, such as $gg\to h\to \chi^+ \chi^-$ depending on the feasibility of observing this channel for order one values of $a_2$. If such processes are observed they would imply a value of $a_2$ too large to be compatible with WEP tests. Similarly, one can generalize such an analysis for other types of DM models to see if one can probe dark forces at colliders. }


\section{Dark Force Parameter Space}
\label{sectionVII}

Before concluding, we generally discuss the regions in parameter space of $\text{SM} + \chi + \phi$ type models 
that are likely to give rise to an observable dark force. In particular, we discuss how the requirement $m_\phi < 10^{-25}$eV, neccessary to allow a dark force of intergalactic range, restricts the allowed parameter space. Recall that after EWSB and diagonalizing the mass matrix, the ultralight scalar mass is given by
\bea
\label{mphi-approx-1}
m_\phi^2 \simeq \mu_S^2 -\frac{\mu_{hS}^4}{4m_h^2}.
\eea
We showed in section \ref{section6A} that $\mu_{hS}^2$, which determines the mixing between the ultralight scalar and the Higgs, receives  finite contributions from higher dimension operators whose size we  constrained from WEP tests. In  Eq.(\ref{theta-sigma-beta-1}), the second term is the contribution to the mixing angle from the operator $H^\dagger HS$ after EWSB and the first term is the finite contribution from the sum of higher dimension operators induced via DM loops(see Fig.~\ref{fig:singletmixing}).  The parameter $\mu_S^2$  similarly receives finite contributions
from higher dimension operators. For example, attaching one extra external $S$ field to the DM loops in Fig.~\ref{fig:singletmixing} will generate a tower of operators that contribute to $\mu_S^2$ after EWSB. The lowest dimension non-renormalizable operator that contributes to $\mu_S^2$ will be
\bea
D_2 \> H^\dagger H H^\dagger H S^2,
\eea
with finite coefficient $D_2$ which can be estimated from NDA as
\bea
D_2 \sim \frac{a_2^2}{\pi M_P^2 } \beta^2.
\eea
We could sum the contribution of the entire tower of operators to $\mu_S^2$ as we did for the case
of $\mu_{hS}^2$. However, the explicit sum is not needed for the following discussion.

The requirement
that $m_\phi < 10^{-25}$ eV now imply three types of possible regions in parameter space: I,II, and III.  We discuss each of these regions in turn below and relate them to the analysis of previous sections. 

\textit{Region I}: In the first region of parameter space,  there are no intricate cancellations of any kind among the terms in Eq.(\ref{mphi-approx-1}).  Each term that goes into determining $\mu_S^2$ and $\mu_{hS}^2$ is required to be of the order of $m_\phi^2$. In this case, we can obtain an an approximate bound on $a_2$ from the $D_2$  which contributes to  $\mu_S^2 \sim m_\phi^2$ as
\bea
\label{strongest-a2-bound}
a_2^2 < \frac{4\pi}{\beta^2}\frac{M_P^2}{v^2} \frac{m_\phi^2}{v^2} < \frac{3\times 10^{-39}}{\beta^2 }.
\eea
For any observable non-zero value of $\beta$, the above bound essentially forces $a_2$ to be zero. As already discussed, such a small value of $a_2$ will lead to an over-relic-density of scalar singlet DM over-closing the universe and is ruled out. Similar arguments can be made for WIMP DM in which case $D_2$ will receive contributions that depend only on the gauge couplings and $\beta$, thus ruling out any observable value for $\beta$. In short,
this region of parameter space is incompatible with the existence of a phenomenologically interesting  dark force and thus not considered in this paper.

\textit{Region II}: In this region, $\mu_S^2$ of Eq.(\ref{mphi-approx-1}) is chosen such that the condition $m_\phi < 10^{-25}$ eV
is always satisfied regardless of the size of $\mu_{hS}^2$ and any finite contributions to it from higher dimension operators after EWSB. Furthermore, in this region there are no  intricate cancellations between terms that determine $\mu_{hS}^2$ so that one can put bounds on these from WEP tests.     This is phenomenologically the most interesting region and was the focus of this paper.  

\textit{Region III}: Finally, the third region corresponds to the case where there are intricate cancellations among various terms in $\mu_S^2$ and $\mu_{hS}^2$ individually. If there are intricate cancellations between large terms in $\mu_{hS}^2$, then we cannot extract meaningful bounds on $a_2$ or the Higgs exchange direct detection cross-section from WEP tests. In this special region of parameter space the  bounds on $a_2$ and the Higgs exchange contributions to the direct detection cross-sections derived in this paper do  not apply.

The analysis of Ref.~\cite{Bovy:2008gh} assumed no fine tuning of  $\delta_1$ or $\mu_{hS}^2$ against radiative corrections sensitive to the cutoff. As seen in Eq.(\ref{mphi-approx-1}), $\mu_{hS}$ contributes to the light scalar mass $m_\phi$ and in the absence of  fine tuning the strongest  bound on $a_2$ comes from $m_\phi < 10^{-25}$ eV forcing $a_2$ to be essentially zero. The bounds from WEP tests are relatively far weaker and not relevant for $a_2$ or equivalently for the DM-nucleus cross-section via Higgs exchange.  Our analysis differs in that we allow for fine tunning in all renormalizable parameters, and then examine the different regions in the space of these renormalized parameters and the corresponding implications for terrestrial experiments.

\section{Conclusion}
\label{section8}

The existence of a new long-range WEP-violating  attractive force in the dark sector, a dark force comparable to gravity, can have interesting cosmological consequences, including an accelerated rate of structure formation and an explanation of certain features of the DM distribution and other astronomical observations~\cite{Einasto:2006si, Farrar:2003uw, Nusser:2004qu}. Strong  constraints for such a dark force comes from a study~\cite{Kesden:2006vz,Kesden:2006zb} of the dynamics of satellite galaxies and the evolution of density perturbations constrained~\cite{Bean:2008ac} by the CMB spectrum. The current bounds indicate the strength of a possible dark force to be less than $\lesssim 20$\%  of  gravity from galactic dynamics and less than 5\% of gravity from the CMB spectrum. 

We considered the consequences of such a dark force for terrestrial experiments. Ordinary matter will feel the effects of a dark force via virtual DM as long as the DM candidate is not sterile. Depending on the DM model, a dark force can lead to constraints on E\"otv\"os experiments, DM-direct-detection experiments, and Higgs decay properties to be studied at future colliders. We studied several minimal models of DM to illustrate the prospective implications of an astrophysically relevant dark force for terrestrial experiments. 
\begin{itemize}
\item We derived lower bounds on the size of the E\"otv\"os parameters $\eta_{_{DM,E}}$ for a non-zero dark force for minimal DM models. These E\"otv\"os parameters measures the effect of WEP violation on ordinary matter which can arise through virtual DM that communicates the dark force to ordinary matter.  We find that for light scalar singlet DM, relic density considerations and the experimental limits on $\eta_{_{E}}$ rule out a dark force  having  strength of 20\% of gravity, in large regions of parameter space. Future experiments with improved sensitivity could probe a dark force of this magnitude for heavier singlet DM. For minimal WIMP DM, the expected magnitudes of $\eta_{_{DM,E}}$ lie well below current and prospective sensitivities of terrestrial E\"otv\"os experiments, but could be probed in a satellite-based experiment having the sensitivity of the MiniSTEP proposal. In non-mimimal WIMP DM models, it is possible to generate larger effects for E\"otv\"os experiments that could be detected by the Microscope experiment.

\item WEP tests imply constraints on Higgs-exchange contributions to the DM-nucleus cross-sections.
 For scalar singlet DM, these bounds apply to the entire cross section. If the scalar singlet DM saturates the DM relic density, these bounds on the DM-nucleus cross-section do not give any  information beyond the implications of WEP bounds on the DM relic density. If the scalar singlet does not saturate the DM relic density, allowing for stronger interactions with the Higgs, WEP tests provide useful bounds on the DM-nucleus cross-section.
 For WIMP DM, the WEP constraints on the Higgs-DM interactions give upper bounds on the contribution of DM-nucleus scattering via Higgs exchange, to the total cross-section.
 
 \item For singlet DM, the current WEP bounds on the DM-nucleus cross-section in the presence of a dark force 
 that is 20\% of gravity, are typically within reach of current and future direct detection experiments. For scalar WIMP
 DM with a mass in the TeV range, the corresponding bounds on the DM-nucleus cross-section are typically beyond the reach of current and future direct detection experiments. If these scalar WIMPs are detected, it will rule out a dark force greater than 20\% of gravity, implying a tighter upper bound on the dark force. 
 
 \item The WEP violation constraints on the DM-Higgs interactions can lead to constraints for collider physics. As a specific example, we derived testable bounds on the allowed shift in the $h\to \gamma \gamma$ rate when the Higgs couples to WIMP DM  in the real triplet scalar representation with a mass less than 200 GeV. Such a light triplet will only contribute a tiny fraction of the DM relic density and could be part of a multicomponent DM scenario. The implied bounds on the Higgs to two photon rate or a dark force comparable to gravity, is far below the sensitivity of the LHC or the ILC. An observed shift in $h\to \gamma \gamma$ attributed to the charged components of the triplet would rule out the dark force.
 
\item An observable scalar dark force with intergalactic range implies restrictions in the space of the renormalized parameters of the theory. These parameter space restrictions apply after the usual fine tuning of parameters against radiative corrections sensitive to the cutoff. 
\end{itemize}

Apart from these experimental implications, a notable theoretical consequence of an astrophysically interesting dark force mediated by an ultralight scalar $\phi$ is the need for substantial fine-tuning to preserve its tiny mass ($m_\phi < 10^{-25}$ eV). In the DM scenarios considered here, divergent loop contributions associated with the DM or SM particles that interact with $\phi$ would generate large contributions to $m_\phi$ that must be removed by fine-tuning unless the strength of the dark force is imperceptibly small. The discovery of such a dark force would introduce yet another mass hierarchy problem in particle physics. In our analysis, we have taken this need for fine-tuning at face value and have attempted to apply it consistently to the derivation of implications for terrestrial experiments. These consequences imply that direct detection experiments -- together with E\"otv\"os experiments and astrophysical observations of satellite galaxies and structure formation -- can be employed as part of a multifaceted probe of a long range force in the dark sector. 



\section*{Acknowledgments}

We would like to thank  Daniel Chung, Glennys Farrar, Jens Gundlach, Marc Kamionkowski, Thomas McElmurry, Frank Petriello, and Lian-Tao Wang for useful discussions. This work was supported in part by Department of Energy contracts
DE-FG03-92-ER40701 and DE-FG02-08ER41531, the Wisconsin Alumni Research Foundation, the Alfred P. Sloan Foundation, and the Gordon and Betty Moore Foundation.

\appendix
\section{Atomic Charge to Mass Ratio Under a Coupling to Ultralight Scalars}
\label{appexA}

Here we give details for obtaining the atomic charge to mass ratio for a coupling to ultralight scalars. As discussed in the text and shown in Eq.(\ref{atomic-eff}), we can obtain the atomic coupling to $\phi$ from the couplings of $\phi$ to SM particles by a matching calculation. From Eqs.(\ref{atomic-coup}) and (\ref{atomic-eff}), the coupling atomic coupling $g_A$ is given by
\bea
\label{gA}
\xi_A g_{_A} = \frac{1}{m_p} \langle A | \left(\sum_{q} g_q m_q\bar{q}q + \sum_{\ell }g_\ell m_\ell \bar{\ell} \ell \right)|A \rangle + c_g \langle A |   G^{\mu \nu}_a G_{\mu \nu}^a | A \rangle +  c_\gamma \langle A | F^{\mu \nu}F_{\mu \nu} | A \rangle . \nn \\
\eea
Note that with the normalization factor $\xi_A$ the above equation is dimensionally consistent, since for fermionic atoms $g_A$ is dimensionless and for scalar atoms it has dimension one. Next we exploit the properties of the energy-momentum tensor in the low energy effective theory. The trace of the energy momentum tensor is given by
\bea
\theta ^\mu _\mu =\frac{ \beta_3 }{2 g_3} G_{\mu \nu}^a G^{\mu \nu}_a + \frac{ \beta_e}{2e} F_{\mu \nu} F^{\mu \nu} + \sum_q m_q \bar{q}q+ \sum_\ell m_\ell \bar{\ell}\ell ,
\eea
where in this convention the QCD and  QED beta functions at one loop are 
\bea
\beta_3 = -\left[\frac{11}{3} C_2 (G) - \frac{n_f}{2} C_2 (N_c)\right] \frac{g_3^3}{16\pi^2}, \qquad \beta_e = \frac{e^3}{12\pi^2} n_f.
\eea
The mass of the atom $A$ is given by
\bea
\label{energy-mom-mass}
M_A &=& \langle A |\theta ^\mu _\mu | A \rangle\nn \\
&=& \langle A |\left(\frac{ \beta_3}{2 g_3} \text{Tr} [G_{\mu \nu} G^{\mu \nu}] + \frac{ \beta_e}{2 e} F_{\mu \nu} F^{\mu \nu}\right) | A \rangle + \langle A |\left( \sum_q m_q \bar{q}q + \sum_\ell m_\ell  \bar{\ell}\ell\right) | A \rangle .\nn \\
\eea
The atomic mass can also be expressed as
\bea
\label{binding-mass}
M_{A} = Z m_p +  (A-Z) m_n + Z m_e - {\cal E}_A,
\eea
where ${\cal E}_A$ is the binding energy of the atom $A$. From here one can write the derivative with respect to the quark and electron masses as
\bea
\label{change-binding-1}
m_q\frac{d M_A}{d m_q}  &=& Z m_q  \frac{d m_p}{d m_q} +  (A-Z) m_q  \frac{d m_n}{d m_q}  - m_q \frac{d {\cal E}}{d m_q},\nn \\
 m_\ell\frac{d M_A}{d m_\ell}  &=& Z m_e \delta_{e\ell}  -  m_\ell \frac{d {\cal E}}{d m_\ell}.
\eea
\OMIT{
\bea
\label{change-binding-1}
\sum_k m_k\frac{d M_A}{d m_k}  &=& Z \Big  ( m_e + \sum_q m_q  \frac{d m_p}{d m_q} \Big )+  (A-Z)\sum_q m_q  \frac{d m_n}{d m_q}  - \sum_k m_k \frac{d {\cal E}}{d m_k},
\eea}
We now use Eq.~(\ref{energy-mom-mass}) 
for  $M_A$ and the Feynman-Hellman theorem to calculate $m_{q,\ell}\frac{d M_A}{d m_{q,\ell}}$ and equate with Eq.~(\ref{change-binding-1}), we  obtain
\bea
\label{quark-electron-melt}
 m_q\frac{d M_A}{d m_q} &=& \langle A |  m_q \bar{q}q | A \rangle = Z  m_q  \frac{d m_p}{d m_q} +  (A-Z)  m_q  \frac{d m_n}{d m_q}  - m_q \frac{d {\cal E}_A}{d m_q},\nn \\
m_\ell\frac{d M_A}{d m_\ell} &=&  \langle A |  m_\ell \bar{\ell}\ell | A \rangle = Z m_e \delta_{e\ell} - m_\ell \frac{d {\cal E}_A}{d m_\ell}.
\eea
\OMIT{
\bea
\label{quark-electron-melt}
\sum_k m_k\frac{d M_A}{d m_k} &=& \langle A | \sum_q m_q \bar{q}q + \sum_\ell m_\ell  \bar{\ell}\ell | A \rangle \nn \\
&=& Z \Big  ( m_e + \sum_q m_q  \frac{d m_p}{d m_q} \Big ) +  (A-Z) \sum_q m_q  \frac{d m_n}{d m_q}  - \sum_k m_k \frac{d {\cal E}_A}{d m_k}.
\eea}
Using the relations of Eq.(\ref{quark-electron-melt}) in Eq.~(\ref{energy-mom-mass}) we obtain an expression for the atomic matrix element of the gluon operator
as
\bea
\label{glue-melt}
 \langle A | \frac{ \beta_3}{2 g_3} \text{Tr} [G_{\mu \nu} G^{\mu \nu}] | A \rangle  &=& M_A - Z \Big  ( m_e + \sum_q m_q  \frac{d m_p}{d m_q} \Big ) -  (A-Z)\sum_q m_q  \frac{d m_n}{d m_q}  \nn \\
 &+& \sum_k m_k \frac{d {\cal E}}{d m_k}  - \langle A | \frac{ \beta_e}{2 e} F_{\mu \nu} F^{\mu \nu} | A \rangle .
\eea
Using Eqs.(\ref{quark-electron-melt}) and (\ref{glue-melt}) in Eq.~(\ref{gA}) we finally arrive at the expression for $\xi_A g_A$:
\bea
\xi_A g_A &=& \frac{2 c_g g_3}{\beta_3}M_A +  \Big [ Z   (\zeta_e m_e + \sum_q \zeta_q m_q\frac{dm_p}{dm_q}) +  (A-Z) \sum_q \zeta_q m_q\frac{dm_n}{dm_q} -  \sum_k \zeta_k m_k \frac{d{\cal E}_A}{dm_k} \Big ] \nn \\
&+& \kappa \langle A | F^{\mu \nu}F_{\mu \nu} | A \rangle,
\eea
where we have introduced the index $k$ which runs over the light quarks $q$ and the charged leptons $\ell$ and the parameters $\zeta_k$ and $\kappa$ are given by
\bea
\label{zetakappa-appex}
\zeta_k = \frac{g_k}{m_p} - \frac{2 g_3}{\beta_3}c_g , \qquad \kappa = c_\gamma -   \frac{g_3 \beta_e}{e \beta_3}c_g.
\eea

We now utilize the expression for the nucleon mass in terms of the nucleon matrix element of the trace of the three flavor QCD energy momentum tensor
\bea
m_N = \langle N |\theta^\mu_\mu | N \rangle=  \langle N |\>\frac{ \beta_3 }{2 g_3} G_{\mu \nu}^a G^{\mu \nu}_a +  \sum_q m_q \bar{q}q \> | N \rangle,
\eea
the variation of the nucleon mass with respect to the mass of of a quark of flavor $q$  is given by
\bea
m_q \frac{dm_{N}}{dm_q} =  \langle N |  m_q \bar{q}q | N \rangle.
\eea
Once again we have used the non-relativistic normalization of nucleon states. The nucleon matrix elements on the RHS are extracted from pion-nucleon scattering data using chiral perturbation theory.
Experimentally their values are determined to be \cite{Belanger:2008sj,Jungman:1995df}
\bea
\label{nuclearmelt}
& x_{u,p} \equiv \frac{dm_p}{dm_u}  = \langle p|\bar{u}u|p\rangle \sim  0.019\, m_p/m_u, \nn\\ 
&x_{d,p} \equiv \frac{dm_p}{dm_d}  =  \langle p|\bar{d}d|p\rangle \sim  0.041\,  m_p/m_d, \nn \\
& x_{s,p} \equiv \frac{dm_p}{dm_s}  = \langle p|\bar{s}s|p\rangle \sim  0.14\, m_p/m_s, \nn\\
&x_{u,n} \equiv \frac{dm_n}{dm_u}  =  \langle n|\bar{u}u|n\rangle \sim  0.023\,  m_n/m_u, \nn \\ 
&x_{d,n} \equiv \frac{dm_n}{dm_d}  =  \langle n|\bar{d}d|n\rangle \sim 0.034\,  m_n/m_d, \nn \\ 
&x_{s,n} \equiv \frac{dm_n}{dm_s}  =  \langle n|\bar{s}s|n\rangle \sim  0.14\,  m_n/m_s
\,.
 \eea
These numbers are taken from table 6 of \cite{Jungman:1995df}. In general $m_k {d{\cal E}_A}/{dm_k}$ and $ \langle A | F^{\mu \nu}F_{\mu \nu} | A \rangle$  are not analytically calculable, at least for large atoms, and will contribute to the uncertainty in the atomic charge to mass ratio.  The atomic charge to mass ratio can be finally written as
\bea
\label{cmr-atomic-appex}
\hat{\xi}_A \Bigg  ( \frac{q}{\mu} \Bigg )_A = \frac{g_A \xi_A}{M_A} = \frac{2 c_g g_3}{\beta_3}+\frac{1}{M_A}\Big [ Z   (\zeta_e m_e + \sum_q \zeta_q m_q \> x_{q,p}) +  (A-Z) \sum_q \zeta_q m_q\>  x_{q,n} + \omega _A \Big ], \nn \\
\eea
where we have defined
\bea
\omega_A \equiv \kappa \langle A | F^{\mu \nu}F_{\mu \nu} | A \rangle  -\sum_k \zeta_k  m_k \frac{d{\cal E}_A}{dm_k}.
\eea

\section{Effective Potential for Higgs-Ultralight-Scalar Mixing}
\label{appexB}

As discussed in section \ref{sectionVI}, ordinary matter can couple to the ultralight scalar $\phi$, which mediates a long range WEP violating force, via its mixing with the Higgs.
Here we show the computation of the effective potential which generates this mixing after electroweak symmetry breaking.
This effective potential is generated at one loop via the sum of diagrams shown in Fig.~\ref{fig:singletmixing} for the scalar singlet $\chi$ and real scalar triplet $\chi$ models discussed in section ~\ref{sectionVI}. Working in unitary gauge where $H=h/\sqrt{2}$ and in d-dimensions, one can write the sum of all diagrams in Fig.~\ref{fig:singletmixing} as

\bea
-i V_{\text{eff}}^S(S,h) &=& -i \kappa g_\chi S \int_E \frac{d^d k}{(2\pi)^d} \sum_{n=0}^{\infty} \frac{(a_2 \> h^2)^n}{(k^2 + M_0^2)^{n+1}}  \nn \\
&=& -i \kappa g_\chi S \int_E \frac{d^d k}{(2\pi)^d} \frac{1}{(k^2 + M_0^2 + a_2 h^2)} \nn \\
&=& \kappa \frac{i g_\chi S }{16 \pi^2} (M_0^2 + a_2 h^2) \left [ \frac{1}{\epsilon} - \gamma_E + \ln 4\pi + 1 - \ln \left ( \frac{M_0^2 + a_2 h^2}{\mu^2}\right )\right ],
\eea
where the first line is obtained after performing a Wick rotation to Euclidean momentum space. The superscript in $V_{\text{eff}}^S(S,h)$ denotes that it is only the part of the effective potential linear in $S$. 
We see from the above result that the coefficient of the $S$ and $S h^2$ operators are UV divergent. These divergences are understood from the need to renormalize the tadpole graph of $S$ and the renormalizable coupling $\delta_1$ of Eq.(\ref{Higgs-S}), corresponding to the first two diagrapms in Fig.~\ref{fig:singletmixing}. The remaining diagrams mix into non-renormalizable operators and are finite. The counterterms needed to cancel the UV divergences are
\bea
i \delta  V_{\text{eff}}(S,h) = S   \left [ \kappa \frac{g_\chi M_0^2}{16\pi^2} \left ( \frac{1}{\epsilon} -\gamma_E + \ln 4\pi \right ) + \hat{b}_1(\mu) \right ] + S h^2   \left [ \kappa \frac{g_\chi a_2}{16 \pi^2} \left ( \frac{1}{\epsilon} - \gamma_E + \ln 4\pi \right ) + \frac{\hat{\delta}_1(\mu)}{4}\right ], \nn \\
\eea
where $\hat{b}_1(\mu )$ and $\hat{\delta}_1(\mu)$ are scheme dependent finite quantities. 

The quadratic terms in the  potential is given by
\bea
V_{\text{quad}} = \frac{1}{2} (\mu_h^2 h^2 + \mu_S^2 S^2 + \mu_{hS}^2 hS) ,
\eea
as first shown in Eq.(\ref{Vmass}). As seen from Eqs.(\ref{dim5mixing}) and (\ref{theta-sigma-beta-1}) the mixing angle for Higgs-ultralight-scalar mixing is given by
\bea
\sin \theta \simeq \frac{\mu_{hS}^2}{\mu_h^2} \simeq \frac{\mu_{hS}^2}{m_h^2},
\eea
and we can write
\bea
\label{muhS2}
\mu_{hS}^2 &=& 2 \frac{\partial ^2 V_{\text{quad}}}{\partial S \partial h} = 2 \frac{\partial ^2 {\cal V}_{\text{eff}}(h,S)}{\partial S \partial h} \Big |_{h=v, S=0}, \nn \\
&=& v \left [ \hat{\delta}_1(\mu) + \kappa \frac{g_\chi a_2}{4 \pi^2} \left ( \ln \frac{M_\chi^2}{\mu^2} -1 \right )  \right ] + \kappa \frac{a_2^2}{4\pi}\frac{g_\chi v^3}{M_\chi^2},
\eea
where we have defined the renormalized effective potential ${\cal V}_{\text{eff}}$ as
\bea
{\cal V}_{\text{eff}} \equiv V_{\text{eff}} + \delta V_{\text{eff}}.
\eea
The first term with square brackets in Eq.(\ref{muhS2}) corresponds to the renormalized value of $v \delta_1$ and the last term
corresponds to the finite contribution from all non-renormalizable operators. This can be compared to Eq.(\ref{dim5mixing}) where we have included only the contribution from the renormalized $\delta_1$ coupling and the third diagram in Fig.~\ref{fig:singletmixing} whose Wilson coefficent is denoted as $C_2$. The above result, which is given in Eq.(\ref{eq:resum}) of the text, is the generalized result where the contribution of the entire tower of higher dimension operators is resummed.

\OMIT{
\bibliographystyle{h-physrev3.bst}
\bibliography{Higgs}
}

\bibliographystyle{h-physrev3.bst}
\bibliography{Higgs.bib}

\begin{thebibliography}{10}

\bibitem{Perlmutter:1998np}
Supernova Cosmology Project, S.~Perlmutter {\em et~al.},
\newblock Astrophys. J. {\bf 517}, 565 (1999), astro-ph/9812133.

\bibitem{Riess:1998cb}
Supernova Search Team, A.~G. Riess {\em et~al.},
\newblock Astron. J. {\bf 116}, 1009 (1998), astro-ph/9805201.

\bibitem{Schmidt:1998ys}
Supernova Search Team, B.~P. Schmidt {\em et~al.},
\newblock Astrophys. J. {\bf 507}, 46 (1998), astro-ph/9805200.

\bibitem{Garnavich:1998th}
Supernova Search Team, P.~M. Garnavich {\em et~al.},
\newblock Astrophys. J. {\bf 509}, 74 (1998), astro-ph/9806396.

\bibitem{Knop:2003iy}
Supernova Cosmology Project, R.~A. Knop {\em et~al.},
\newblock Astrophys. J. {\bf 598}, 102 (2003), astro-ph/0309368.

\bibitem{Faber:1979pp}
S.~M. Faber and J.~S. Gallagher,
\newblock Ann. Rev. Astron. Astrophys. {\bf 17}, 135 (1979).

\bibitem{Bosma:1981zz}
A.~Bosma,
\newblock Astron. J. {\bf 86}, 1825 (1981).

\bibitem{Rubin:1985ze}
V.~C. Rubin, D.~Burstein, J.~Ford, W.~K., and N.~Thonnard,
\newblock Astrophys. J. {\bf 289}, 81 (1985).

\bibitem{Hu:1994uz}
W.~Hu and N.~Sugiyama,
\newblock Astrophys. J. {\bf 444}, 489 (1995), astro-ph/9407093.

\bibitem{Hu:1994jd}
W.~Hu and N.~Sugiyama,
\newblock Phys. Rev. {\bf D51}, 2599 (1995), astro-ph/9411008.

\bibitem{Jungman:1995bz}
G.~Jungman, M.~Kamionkowski, A.~Kosowsky, and D.~N. Spergel,
\newblock Phys. Rev. {\bf D54}, 1332 (1996), astro-ph/9512139.

\bibitem{Zaldarriaga:1997ch}
M.~Zaldarriaga, D.~N. Spergel, and U.~Seljak,
\newblock Astrophys. J. {\bf 488}, 1 (1997), astro-ph/9702157.

\bibitem{Eisenstein:1997ik}
D.~J. Eisenstein and W.~Hu,
\newblock Astrophys. J. {\bf 496}, 605 (1998), astro-ph/9709112.

\bibitem{Eisenstein:2005su}
SDSS, D.~J. Eisenstein {\em et~al.},
\newblock Astrophys. J. {\bf 633}, 560 (2005), astro-ph/0501171.

\bibitem{Clowe:2006eq}
D.~Clowe {\em et~al.},
\newblock Astrophys. J. {\bf 648}, L109 (2006), astro-ph/0608407.

\bibitem{Zhang:2007nk}
P.~Zhang, M.~Liguori, R.~Bean, and S.~Dodelson,
\newblock Phys. Rev. Lett. {\bf 99}, 141302 (2007), 0704.1932.

\bibitem{Angle:2007uj}
XENON, J.~Angle {\em et~al.},
\newblock Phys. Rev. Lett. {\bf 100}, 021303 (2008), 0706.0039.

\bibitem{Ahmed:2008eu}
CDMS, Z.~Ahmed {\em et~al.},
\newblock (2008), 0802.3530.

\bibitem{Adriani:2008zr}
O.~Adriani {\em et~al.},
\newblock (2008), 0810.4995.

\bibitem{Barwick:1997ig}
HEAT, S.~W. Barwick {\em et~al.},
\newblock Astrophys. J. {\bf 482}, L191 (1997), astro-ph/9703192.

\bibitem{Beatty:2004cy}
J.~J. Beatty {\em et~al.},
\newblock Phys. Rev. Lett. {\bf 93}, 241102 (2004), astro-ph/0412230.

\bibitem{Aguilar:2007yf}
AMS-01, M.~Aguilar {\em et~al.},
\newblock Phys. Lett. {\bf B646}, 145 (2007), astro-ph/0703154.

\bibitem{Gardner:2006za}
S.~Gardner,
\newblock Phys. Rev. Lett. {\bf 100}, 041303 (2008), astro-ph/0611684.

\bibitem{Gardner:2008yn}
S.~Gardner,
\newblock (2008), 0811.0967.

\bibitem{DeRujula:1989fe}
A.~De~Rujula, S.~L. Glashow, and U.~Sarid,
\newblock Nucl. Phys. {\bf B333}, 173 (1990).

\bibitem{Feng:2008mu}
J.~L. Feng, H.~Tu, and H.-B. Yu,
\newblock JCAP {\bf 0810}, 043 (2008), 0808.2318.

\bibitem{ArkaniHamed:2008qn}
N.~Arkani-Hamed, D.~P. Finkbeiner, T.~Slatyer, and N.~Weiner,
\newblock (2008), 0810.0713.

\bibitem{Ackerman:2008gi}
L.~Ackerman, M.~R. Buckley, S.~M. Carroll, and M.~Kamionkowski,
\newblock (2008), 0810.5126.

\bibitem{Baumgart:2009tn}
M.~Baumgart, C.~Cheung, J.~T. Ruderman, L.-T. Wang, and I.~Yavin,
\newblock (2009), 0901.0283.

\bibitem{Aguirre:2001xs}
A.~Aguirre, C.~P. Burgess, A.~Friedland, and D.~Nolte,
\newblock Class. Quant. Grav. {\bf 18}, R223 (2001), hep-ph/0105083.

\bibitem{Zurek:2008qg}
K.~M. Zurek,
\newblock (2008), 0811.4429.

\bibitem{Bai:2009it}
Y.~Bai and Z.~Han,
\newblock (2009), 0902.0006.

\bibitem{Moody:1984ba}
J.~E. Moody and F.~Wilczek,
\newblock Phys. Rev. {\bf D30}, 130 (1984).

\bibitem{Hoyle:2004cw}
C.~D. Hoyle {\em et~al.},
\newblock Phys. Rev. {\bf D70}, 042004 (2004), hep-ph/0405262.

\bibitem{Kaplan:2000hh}
D.~B. Kaplan and M.~B. Wise,
\newblock JHEP {\bf 08}, 037 (2000), hep-ph/0008116.

\bibitem{Bovy:2008gh}
J.~Bovy and G.~R. Farrar,
\newblock (2008), 0807.3060.

\bibitem{Carroll:2008ub}
S.~M. Carroll, S.~Mantry, M.~J. Ramsey-Musolf, and C.~W. Stubbs,
\newblock (2008), 0807.4363.

\bibitem{Damour:1990tw}
T.~Damour, G.~W. Gibbons, and C.~Gundlach,
\newblock Phys. Rev. Lett. {\bf 64}, 123 (1990).

\bibitem{Friedman:1991dj}
J.~A. Frieman and B.-A. Gradwohl,
\newblock Phys. Rev. Lett. {\bf 67}, 2926 (1991).

\bibitem{Gradwohl:1992ue}
B.-A. Gradwohl and J.~A. Frieman,
\newblock Astrophys. J. {\bf 398}, 407 (1992).

\bibitem{Anderson:1997un}
G.~W. Anderson and S.~M. Carroll,
\newblock (1997), astro-ph/9711288.

\bibitem{Carroll:1998zi}
S.~M. Carroll,
\newblock Phys. Rev. Lett. {\bf 81}, 3067 (1998), astro-ph/9806099.

\bibitem{Amendola:2001rc}
L.~Amendola and D.~Tocchini-Valentini,
\newblock Phys. Rev. {\bf D66}, 043528 (2002), astro-ph/0111535.

\bibitem{Farrar:2003uw}
G.~R. Farrar and P.~J.~E. Peebles,
\newblock Astrophys. J. {\bf 604}, 1 (2004), astro-ph/0307316.

\bibitem{Gubser:2004du}
S.~S. Gubser and P.~J.~E. Peebles,
\newblock Phys. Rev. {\bf D70}, 123511 (2004), hep-th/0407097.

\bibitem{Gubser:2004uh}
S.~S. Gubser and P.~J.~E. Peebles,
\newblock Phys. Rev. {\bf D70}, 123510 (2004), hep-th/0402225.

\bibitem{Bertolami:2004nh}
O.~Bertolami and J.~Paramos,
\newblock Phys. Rev. {\bf D71}, 023521 (2005), astro-ph/0408216.

\bibitem{Nusser:2004qu}
A.~Nusser, S.~S. Gubser, and P.~J.~E. Peebles,
\newblock Phys. Rev. {\bf D71}, 083505 (2005), astro-ph/0412586.

\bibitem{Bean:2007ny}
R.~Bean, E.~E. Flanagan, and M.~Trodden,
\newblock Phys. Rev. {\bf D78}, 023009 (2008), 0709.1128.

\bibitem{Alimi:2008ee}
J.~M. Alimi and A.~Fuzfa,
\newblock JCAP {\bf 0809}, 014 (2008), 0804.4100.

\bibitem{Fuzfa:2007sv}
A.~Fuzfa and J.~M. Alimi,
\newblock Phys. Rev. {\bf D75}, 123007 (2007), astro-ph/0702478.

\bibitem{Coc:2008yu}
A.~Coc, K.~A. Olive, J.-P. Uzan, and E.~Vangioni,
\newblock Phys. Rev. {\bf D79}, 103512 (2009), 0811.1845.

\bibitem{Einasto:2006si}
SDSS, J.~Einasto {\em et~al.},
\newblock Astron. Astrophys. {\bf 462}, 397 (2007), astro-ph/0604539.

\bibitem{Kesden:2006vz}
M.~Kesden and M.~Kamionkowski,
\newblock Phys. Rev. {\bf D74}, 083007 (2006), astro-ph/0608095.

\bibitem{Kesden:2006zb}
M.~Kesden and M.~Kamionkowski,
\newblock Phys. Rev. Lett. {\bf 97}, 131303 (2006), astro-ph/0606566.

\bibitem{Chou:2006ia}
M.-Y. Chou {\em et~al.},
\newblock (2006), astro-ph/0605101.

\bibitem{Bean:2008ac}
R.~Bean, E.~E. Flanagan, I.~Laszlo, and M.~Trodden,
\newblock Phys. Rev. {\bf D78}, 123514 (2008), 0808.1105.

\bibitem{Lammerzahl:2001qr}
e.~. Lammerzahl, C., e.~. Everitt, C. W.~F., and e.~. Hehl, F.~W.,
\newblock Prepared for 220th WE-Heraeus Seminar on Gyros, Clocks, and
  Interferometers: Testing General Relativity in Space, Bad Honnef, Germany,
  22-27 Aug 1999.

\bibitem{Lockerbie:1998ar}
N.~A. Lockerbie,
\newblock Nucl. Phys. Proc. Suppl. {\bf 61B}, 3 (1998).

\bibitem{Barger:2007im}
V.~Barger, P.~Langacker, M.~McCaskey, M.~J. Ramsey-Musolf, and G.~Shaughnessy,
\newblock Phys. Rev. {\bf D77}, 035005 (2008), 0706.4311.

\bibitem{He:2008qm}
X.-G. He, T.~Li, X.-Q. Li, J.~Tandean, and H.-C. Tsai,
\newblock (2008), 0811.0658.

\bibitem{Barger:2008jx}
V.~Barger, P.~Langacker, M.~McCaskey, M.~Ramsey-Musolf, and G.~Shaughnessy,
\newblock (2008), 0811.0393.

\bibitem{Schlamminger:2007ht}
S.~Schlamminger, K.~Y. Choi, T.~A. Wagner, J.~H. Gundlach, and E.~G.
  Adelberger,
\newblock Phys. Rev. Lett. {\bf 100}, 041101 (2008), 0712.0607.

\bibitem{Williams:2003wu}
J.~G. Williams, S.~G. Turyshev, and J.~Murphy, Thomas~W.,
\newblock Int. J. Mod. Phys. {\bf D13}, 567 (2004), gr-qc/0311021.

\bibitem{Dimopoulos:2006nk}
S.~Dimopoulos, P.~W. Graham, J.~M. Hogan, and M.~A. Kasevich,
\newblock Phys. Rev. Lett. {\bf 98}, 111102 (2007), gr-qc/0610047.

\bibitem{Peebles:2009th}
J.~A. K. A. N. P. J.~E. Peebles,
\newblock (2009), 0902.3452.

\bibitem{Shifman:1978zn}
M.~A. Shifman, A.~I. Vainshtein, and V.~I. Zakharov,
\newblock Phys. Lett. {\bf B78}, 443 (1978).

\bibitem{Belanger:2008sj}
G.~Belanger, F.~Boudjema, A.~Pukhov, and A.~Semenov,
\newblock (2008), 0803.2360.

\bibitem{Jungman:1995df}
G.~Jungman, M.~Kamionkowski, and K.~Griest,
\newblock Phys. Rept. {\bf 267}, 195 (1996), hep-ph/9506380.

\bibitem{Cheng:1988cz}
T.~P. Cheng,
\newblock Phys. Rev. {\bf D38}, 2869 (1988).

\bibitem{Cheng:1988im}
H.-Y. Cheng,
\newblock Phys. Lett. {\bf B219}, 347 (1989).

\bibitem{Burgess:2000yq}
C.~P. Burgess, M.~Pospelov, and T.~ter Veldhuis,
\newblock Nucl. Phys. {\bf B619}, 709 (2001), hep-ph/0011335.

\bibitem{O'Connell:2006wi}
D.~O'Connell, M.~J. Ramsey-Musolf, and M.~B. Wise,
\newblock Phys. Rev. {\bf D75}, 037701 (2007), hep-ph/0611014.

\bibitem{Cirelli:2005uq}
M.~Cirelli, N.~Fornengo, and A.~Strumia,
\newblock Nucl. Phys. {\bf B753}, 178 (2006), hep-ph/0512090.

\bibitem{FileviezPerez:2008bj}
P.~Fileviez~Perez, H.~H. Patel, M.~J. Ramsey-Musolf, and K.~Wang,
\newblock (2008), 0811.3957.

\bibitem{Akerib:2005za}
CDMS, D.~S. Akerib {\em et~al.},
\newblock Phys. Rev. {\bf D73}, 011102 (2006), astro-ph/0509269.

\bibitem{Ni:2006fh}
K.~Ni and L.~Baudis,
\newblock Adv. Space Res. {\bf 41}, 2019 (2008), astro-ph/0611124.

\bibitem{Brunetti:2004cf}
WARP, R.~Brunetti {\em et~al.},
\newblock New Astron. Rev. {\bf 49}, 265 (2005), astro-ph/0405342.

\bibitem{Akerib:2006rr}
D.~S. Akerib {\em et~al.},
\newblock Nucl. Instrum. Meth. {\bf A559}, 411 (2006).

\bibitem{Aprile:2002ef}
E.~Aprile {\em et~al.},
\newblock (2002), astro-ph/0207670.

\end{thebibliography}

\end{document}